\documentclass[oneside]{article}
\usepackage[a4paper]{geometry}
\usepackage[latin1]{inputenc} 
\usepackage[T1]{fontenc}
\usepackage{RR}
\usepackage{hyperref}

\usepackage{float}
\usepackage{algorithm}
\usepackage[noend]{algorithmic}
\usepackage[geometry]{ifsym}
\usepackage {ifsym}
\usepackage{amssymb}
\usepackage{yhmath}
\usepackage{amscd}
\usepackage{dsfont}
\usepackage{amsmath}
\usepackage{epsfig}
\usepackage{graphics}
\usepackage{psfrag}
\usepackage{rotating}
\usepackage{amsmath}
\usepackage{amsfonts}
\usepackage{url}
\usepackage{color}
\usepackage{float}
\usepackage{balance} 
\usepackage{bbm}

\usepackage{IEEEtrantools}

\usepackage[toc,page]{appendix}

\usepackage{enumitem}

\newtheorem{lemma}{Lemma}

\newtheorem{theorem}{Theorem}
\newtheorem{definition}{Definition}

\newcommand{\bs}{\boldsymbol}

\newcommand{\ds}{\displaystyle}

\newcommand{\pr}[1]{\mathrm{Pr} \left(#1\right)}
\newcommand{\SNR}{\mathrm{SNR}}

\newcommand{\INR}{\mathrm{INR}}

\newcommand{\sfT}{\textsf{T}}

\newcommand{\Cgicnof}{\mathcal{C}_{\mathrm{GIC-NOF}}}
\newcommand{\agicnof}{\underline{\mathcal{C}}_{\mathrm{GIC-NOF}}}
\newcommand{\cgicnof}{\overline{\mathcal{C}}_{\mathrm{GIC-NOF}}}

\RRNo{8861}
\RRdate{March 2016}
\RRauthor{
Victor Quintero
 \and Samir M. Perlaza
\and Iñaki Esnaola 
\and Jean-Marie Gorce 
}
\authorhead{Quintero \& Perlaza \& Esnaola \& Gorce}
\RRtitle{}
\RRetitle{Approximate Capacity of the Two-User Gaussian Interference Channel with Noisy Channel-Output Feedback.}
\titlehead{Approximate Capacity Region of the Two-User G-IC-NOF.}
\RRnote{Victor Quintero, Samir M. Perlaza  and Jean-Marie Gorce are with the CITI Laboratory of the Institut National de Recherche en Informatique et en Automatique (INRIA), Universit\'{e} de Lyon, and Institut National de Sciences Apliqu\'ees (INSA) de Lyon. 6 Av. des Arts 69621 Villeurbanne, France. ($\lbrace$victor.quintero-florez, samir.perlaza, jean-marie.gorce$\rbrace$@inria.fr).

I\~naki Esnaola is with the Department of Automatic Control and Systems Engineering, University of Sheffield, Mappin Street, Sheffield, S1 3JD, United Kingdom. (esnaola@sheffield.ac.uk).

Victor Quintero is also with Universidad del Cauca, Popay\'{a}n, Colombia.

Samir M. Perlaza and I\~naki Esnaola are also with the Department of Electrical Engineering at Princeton University, Princeton, NJ.

This research was supported in part by the European Commission under Marie Sklodowska-Curie Individual Fellowship No. 659316 (CYBERNETS); and the Administrative Department of Science, Technology, and Innovation of Colombia (Colciencias), fellowship No. 617-2013.

Parts of this work were presented at the IEEE International Workshop on Information Theory (ITW), Cambridge, United Kingdom, September, 2016. This work was also submitted to the IEEE Transactions on Information Theory in November 10 2016.}

\RRresume{
Ce rapport pr\'esente une approximation de la capacit\'e du canal \`a interf\'erences Gaussien avec r\'etroalimentation degrad\'ee (G-IC-NOF). \`A cette fin, une r\'egion atteignable et une r\'egion converse du G-IC-NOF sont introduites, permettant de d\'eterminer que la capacit\'e du G-IC-NOF est approch\'ee \`a $4.4$ bits par utilisation de canal.  
}
\RRabstract{
In this research report, an achievability region and a converse region for the two-user Gaussian interference channel with noisy channel-output feedback (G-IC-NOF)  are presented. The achievability region is obtained using a random coding argument and three well-known techniques: rate splitting, superposition coding and backward decoding. The converse region is obtained using some of the existing perfect-output feedback outer-bounds as well as a set of new outer-bounds that are obtained by using  genie-aided models of the original G-IC-NOF. Finally, it is shown that the achievability region and the converse region approximate the capacity region of the G-IC-NOF to within a constant gap in bits. 
}
\RRmotcle{R\'egion atteignable, r\'egion converse, canal \`a interf\'erences Gaussien, r\'etroalimentation d\'egrad\'ee.}
\RRkeyword{Achievable region, converse region, two-user Gaussian interference channel, noisy chanel-output feedback}
\RRprojet{Socrate}  
\RCGrenoble 

\begin{document}
\makeRR   

\clearpage
\tableofcontents
\clearpage

\section{Notation}

Throughout this research report, sets are denoted with uppercase calligraphic letters, e.g. $\mathcal{X}$. Random variables are denoted by uppercase letters, e.g., $X$. The realizations and the set of events from which the random variable $X$ takes values are respectively denoted by  $x$ and $\mathcal{X}$. The probability distribution of $X$ over the set $\mathcal{X}$ is denoted $P_{X}$. Whenever a second random variable $Y$ is involved, $P_{X \, Y}$ and $P_{Y|X}$ denote respectively the joint probability distribution of $(X, Y)$ and the conditional probability distribution of $Y$ given $X$. Let $N$ be a fixed natural number. An $N$-dimensional vector of random variables is denoted by ${\bf X} = (X_{1}, X_{2}, ..., X_{N})^\sfT$ and a corresponding realization is denoted by ${\bf x}= (x_{1}, x_{2}, ..., x_{N})^\sfT \in \mathcal{X}^{N}$.  Given ${\bf X} = (X_{1}, X_{2}, ..., X_{N})^\sfT$ and $(a,b) \in \mathds{N}^2$, with $a < b \leqslant N$,  the $(b-a+1)$-dimensional vector of random variables formed by the components $a$ to $b$ of $\bs{X}$ is denoted by ${{\bf X}_{(a:b)} = (X_a, X_{a+1}, \ldots, X_b)^\sfT}$.  The notation $(\cdot)^+$ denotes the positive part operator, i.e., $(\cdot)^+ = \max(\cdot, 0)$ and $\mathbb{E}_{X}[ \cdot ]$ denotes the expectation with respect to the distribution of the random variable $X$. The logarithm function $\log$ is assumed to be base $2$. 

\clearpage

\section{Problem Formulation}\label{SecProbForm}

\begin{figure}[t!]
 \centerline{\epsfig{figure=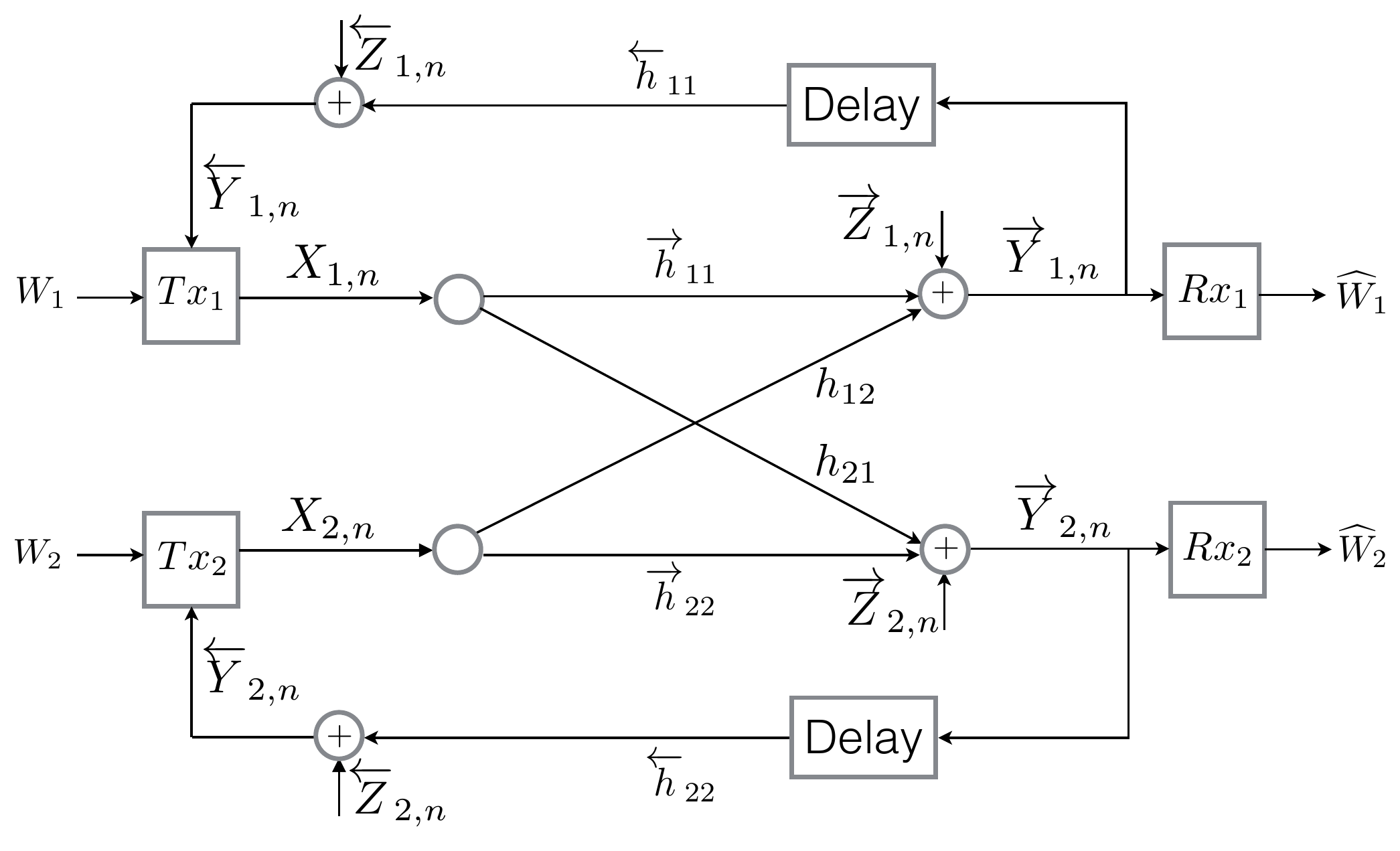,width=0.8\textwidth}}
  \caption{Gaussian interference channel with noisy channel-output feedback at channel use~$n$.}
  \label{Fig:G-IC-NOF}
\end{figure}

\noindent
This section introduces  the two-user Gaussian interference channel with noisy channel-output feedback (G-IC-NOF) and defines an approximation to its corresponding capacity region. 

\noindent
Consider the two-user G-IC-NOF in Figure~\ref{Fig:G-IC-NOF}. Transmitter $i$, with $i \in \{1,2\}$, communicates with receiver $i$ subject to the interference produced by transmitter $j$, with $j \in \{1,2\} \backslash \{i\}$. There are two independent and uniformly distributed messages, $W_i \in \mathcal{W}_i$, with $\mathcal{W}_i=\{1, 2,  \ldots, 2^{NR_i}\}$, where $N$ denotes the fixed block-length in channel uses and $R_i$ is the transmission rate in bits per channel use. At each block, transmitter $i$  sends the codeword ${\bs{X}_{i}=\left(X_{i,1}, X_{i,2}, \ldots, X_{i,N}\right)^\sfT \in \mathcal{X}_i^N}$, where $\mathcal{X}_i$ and $\mathcal{X}_i^N$ are respectively the channel-input alphabet and the codebook of transmitter $i$. 

\begin{figure}[t!]
 \centerline{\epsfig{figure=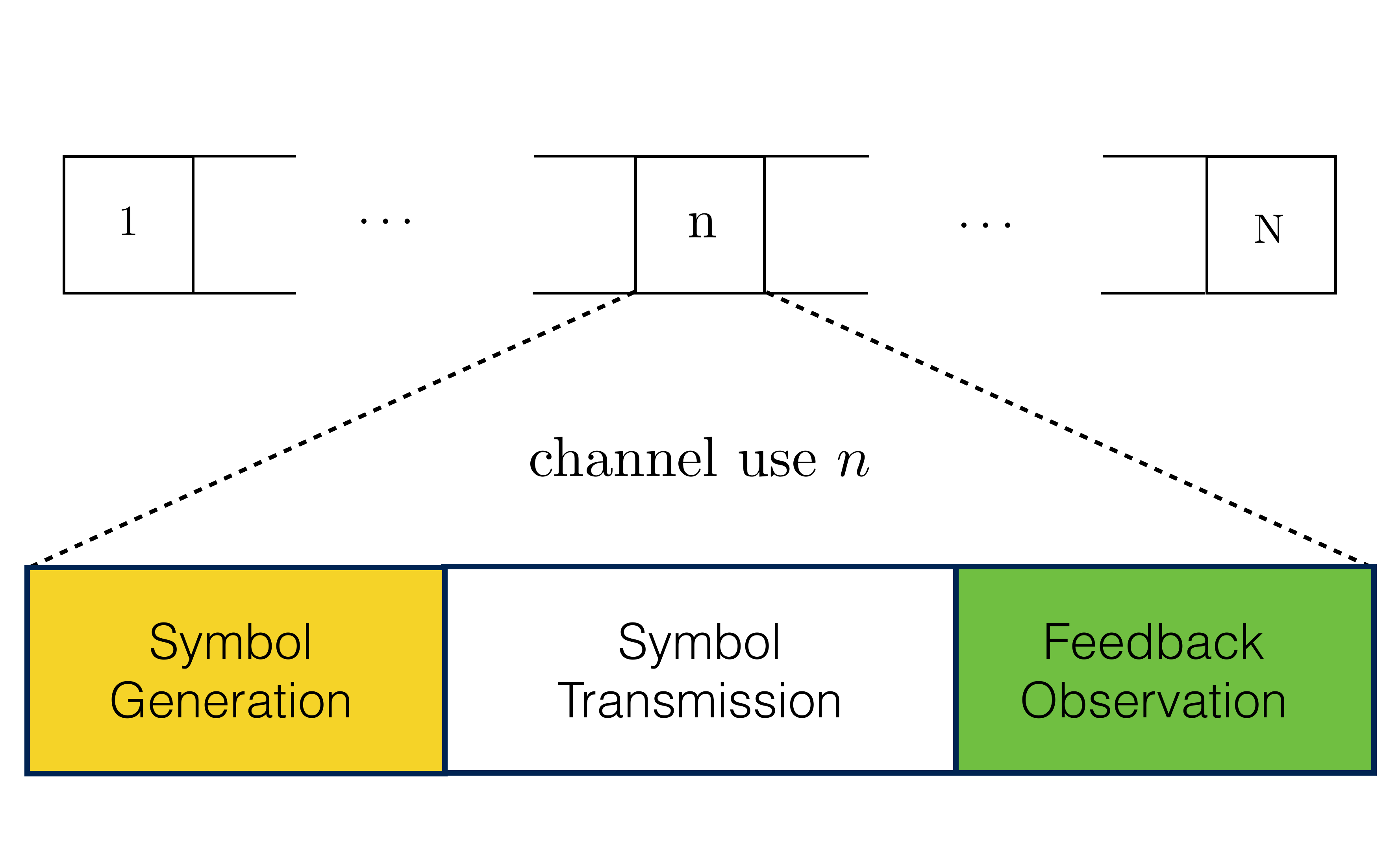,width=0.8\textwidth}}
  \caption{Phases of channel use $n$: Symbol generation phase occurs following \eqref{Eqencod}; Symbol transmission phase occurs following \eqref{Eqsignalyif}; and feedback observation occurs following \eqref{Eqsignalyib}.}
  \label{Fig:channelusen}
\end{figure}

\noindent
The channel coefficient from transmitter $j$ to receiver $i$ is denoted by $h_{ij}$; the channel coefficient from transmitter $i$ to receiver $i$ is denoted by $\overrightarrow{h}_{ii}$; and the channel coefficient from channel-output $i$ to transmitter $i$ is denoted by $\overleftarrow{h}_{ii}$. All channel coefficients are assumed to be non-negative real numbers.
At a given channel use $n \in \{1, 2, \ldots, N\}$, the channel output at receiver $i$ is denoted by $\overrightarrow{Y}_{i,n}$.  
During channel use $n$, the input-output relation of the channel model is given by
\begin{IEEEeqnarray}{lcl}
\label{Eqsignalyif}
\overrightarrow{Y}_{i,n}&=& \overrightarrow{h}_{ii}X_{i,n} + h_{ij}X_{j,n}+\overrightarrow{Z}_{i,n},
\end{IEEEeqnarray}
where $\overrightarrow{Z}_{i,n}$ is a real Gaussian random variable with zero mean and unit variance that represents the noise at the input of receiver $i$.
Let $d>0$ be the finite feedback delay measured in channel uses. At the end of channel use $n$, transmitter $i$ observes $\overleftarrow{Y}_{i,n}$, which consists of a scaled and noisy version of $\overrightarrow{Y}_{i,n-d}$ (see Figure~\ref{Fig:channelusen}). More specifically,
\begin{IEEEeqnarray}{rcl}
\label{Eqsignalyib}
\overleftarrow{Y}_{i,n}  &=& 
\begin{cases}
 \overleftarrow{Z}_{i,n} &  \textrm{for } n \! \in \lbrace \! 1, \! 2,  \ldots, d \rbrace  \\ 
\overleftarrow{h}_{ii}\overrightarrow{Y}_{i,n-d} \! + \! \overleftarrow{Z}_{i,n}, \!  &  \textrm{for } n \! \in \lbrace  d \! + \! 1, \! d \! + \! 2, \ldots, \! N \rbrace,
 \end{cases} \quad
\end{IEEEeqnarray}
where $\overleftarrow{Z}_{i,n}$ is a real Gaussian random variable with zero mean and unit variance that represents the noise in the feedback link of transmitter-receiver pair  $i$. The random variables $\overrightarrow{Z}_{i,n}$ and $\overleftarrow{Z}_{i,n}$ are independent and identically distributed.
In the following, without loss of generality, the feedback delay is assumed to be one channel use, i.e., $d=1$. 
The encoder of transmitter $i$ is defined by a set of deterministic functions $f_i^{(1)}, f_i^{(2)}, \ldots, f_i^{(N)}$, with $f_i^{(1)}:\mathcal{W}_i \rightarrow \mathcal{X}_i$ and for all $n \in \{2, 3, \ldots, N\}$, $f_i^{(n)}:\mathcal{W}_i\times\mathds{R}^{n-1} \rightarrow \mathcal{X}_i$, such that
\begin{subequations}
\label{Eqencod}
\begin{IEEEeqnarray}{lcl}
\label{Eqencodi1}
X_{i,1}&=&f_i^{(1)}\left(W_i\right),  \mbox{ and }  \\
\label{Eqencodit}
X_{i,n}&=&f_i^{(n)}\left(W_i,\overleftarrow{Y}_{i,1}, \overleftarrow{Y}_{i,2}, \ldots,\overleftarrow{Y}_{i,n-1}\right).
\end{IEEEeqnarray}
\end{subequations}
The components of the input vector $\bs{X}_{i}$ are real numbers subject to an average power constraint:
\begin{equation}
\label{Eqconstpow}
\frac{1}{N}\sum_{n=1}^{N}\mathbb{E}\left({X_{i,n}}^2\right) \leq 1,
\end{equation}
where the expectation is taken over the joint distribution of the message indexes $W_1$, $W_2$, and the noise terms, i.e., $\overrightarrow{Z}_{1}$, $\overrightarrow{Z}_{2}$, $\overleftarrow{Z}_{1}$, and $\overleftarrow{Z}_{2}$. The dependence of $X_{i,n}$ on $W_1$, $W_2$, and the previously observed noise realizations is due to the effect of feedback as shown in \eqref{Eqsignalyib} and \eqref{Eqencod}. 

\noindent
Let $T\in \mathds{N}$ be fixed. Assume that during a given communication, $T$ blocks, each of $N$ channel uses, are transmitted. Hence, the decoder of receiver $i$ is defined by a deterministic function ${\psi_i: \mathds{R}_i^{N T} \rightarrow \mathcal{W}_i^{T}}$.
At the end of the communication, receiver $i$ uses the vector $\Big(\overrightarrow{Y}_{i,1}$, $\overrightarrow{Y}_{i,2}$, $\ldots$, $\overrightarrow{Y}_{i,NT}\Big)^\sfT$ to obtain an estimate of the message indices:
\begin{IEEEeqnarray}{rcl}
\label{Eqdecoder}
\left(\widehat{W}_i^{(1)}, \widehat{W}_i^{(2)}, \ldots, \widehat{W}_i^{(T)}\right) &=& \psi_i \left(\overrightarrow{Y}_{i,1}, \overrightarrow{Y}_{i,2}, \ldots, \overrightarrow{Y}_{i,NT}\right), \quad
\end{IEEEeqnarray} 
where $\widehat{W}_i^{(t)}$ is an estimate of the message index sent during block $t \in \lbrace 1, 2, \ldots, T \rbrace$.
The decoding error probability in the two-user G-IC-NOF during block $t$, denoted by $P_{e}^{(t)}(N)$, is given by   
\begin{IEEEeqnarray}{rcl}
\label{EqDecErrorProb}
\! P_{e}^{(t)} \! ( N ) &=&   \max \!  \Bigg(  \pr{\! \widehat{W_1}^{(t)} \!  \neq \!  W_1^{(t)} \! }  ,  \pr{\! \widehat{W_2}^{(t)} \!  \neq \!  W_2^{(t)} }\Bigg). \! \quad
\end{IEEEeqnarray}

\noindent
The definition of an achievable rate pair $(R_1,R_2) \in \mathds{R}_+^{2}$ is given below. 
\begin{definition}[Achievable Rate Pairs]\label{DefAchievableRatePairs}\emph{
A rate pair $(R_1,R_2) \in \mathds{R}_+^{2}$ is achievable if there exists at least one pair of codebooks $\mathcal{X}_1^{N}$ and $\mathcal{X}_2^{N}$ with codewords of length $N$, and the corresponding encoding functions $f_1^{(1)}, f_1^{(2)}, \ldots,f_1^{(N)}$ and $f_2^{(1)}, f_2^{(2)}, \ldots, f_2^{(N)}$ such that the decoding error probability $P_{e}^{(t)}(N)$ can be made arbitrarily small by letting the block-length $N$ grow to infinity, for all blocks $t \in \lbrace 1, 2,  \ldots, T \rbrace$.
 }
\end{definition}
The two-user G-IC-NOF in Figure~\ref{Fig:G-IC-NOF} can be fully described by six parameters: $\overrightarrow{\SNR}_i$, $\overleftarrow{\SNR}_i$, and $\INR_{ij}$, with $i \in \{1,2\}$ and $j \in \{1,2\} \backslash \{i\}$, which are defined as follows:
\begin{IEEEeqnarray}{rcl}
\label{EqSNRifwd}
\overrightarrow{\SNR}_i &=& \overrightarrow{h}_{ii}^2, \\
\label{EqINRij}
\INR_{ij}&=& h_{ij}^2 \mbox{ and } \\
\label{EqSNRibwd}
\overleftarrow{\SNR}_i&=&\overleftarrow{h}_{ii}^2\left(\overrightarrow{h}_{ii}^2 + 2\overrightarrow{h}_{ii}h_{ij}+h_{ij}^2+1\right). \quad
\end{IEEEeqnarray}

\noindent
The analysis presented in this report focuses exclusively on the case in which $\INR_{ij} > 1$ for all $\left(i,j\right) \in \lbrace 1,2 \rbrace \times \lbrace \lbrace1,2 \rbrace \setminus \lbrace i \rbrace \rbrace$. The reason for exclusively considering this case follows from the the fact that when $\INR_{ij} \leqslant 1$, the transmitter-receiver pair $i$ is impaired mainly by noise instead of interference. In this case, treating interference as noise is optimal and feedback does not bring a significant rate improvement. 

\clearpage

\section{Main Results} \label{SectMainRes}

This section introduces an achievable region (Theorem~\ref{TheoremA-G-IC-NOF})  and a converse region (Theorem~\ref{TheoremC-G-IC-NOF}), denoted by $\agicnof$ and $\cgicnof$ respectively, for the two-user G-IC-NOF with fixed parameters $\overrightarrow{\SNR}_{1}$, $\overrightarrow{\SNR}_{2}$, $\INR_{12}$, $\INR_{21}$, $\overleftarrow{\SNR}_{1}$, and $\overleftarrow{\SNR}_{2}$.
In general, the capacity region of a given multi-user channel is said to be approximated to within a constant gap according to the following definition.

\begin{definition}[Approximation to within $\xi$ units]\label{DefGap}\emph{
A closed and convex set $\mathcal{T}\subset\mathbb{R}_{+}^{m}$ is approximated to within $\xi$ units by the sets $\underline{\mathcal{T}}$ and $\overline{\mathcal{T}}$ if  $\underline{\mathcal{T}} \subseteq \mathcal{T} \subseteq \overline{\mathcal{T}}$ and for all $\bs{t}=(t_1, \ldots, t_m)\in \overline{\mathcal{T}}$, $\left(\left(t_1-\xi\right)^+, \ldots, \left(t_m-\xi\right)^+\right) \in \underline{\mathcal{T}}$.}
\end{definition}

\noindent
Denote by $\Cgicnof$ the capacity region of the 2-user G-IC-NOF.  The achievable region $\agicnof$ and the converse region $\cgicnof$ approximate the capacity region $\Cgicnof$ to within $4.4$ bits (Theorem~\ref{TheoremGAP-G-IC-NOF}).

\subsection{An Achievable Region}
The description of the achievable region $\agicnof$ is presented using the constants $a_{1,i}$; the functions $a_{2,i}:[0,1] \rightarrow \mathds{R}_{+}$,  $a_{l,i}:[0,1]^2\rightarrow \mathds{R}_{+}$, with $l \in \lbrace 3, \ldots, 6 \rbrace$; and $a_{7,i}:[0,1]^3\rightarrow \mathds{R}_{+}$, which are defined as follows, for all $i \in \lbrace 1, 2 \rbrace$, with $j \in \lbrace 1, 2 \rbrace \setminus \lbrace i \rbrace$:

\begin{subequations}
\label{Eq-a}
\begin{IEEEeqnarray}{rcl}
\label{Eq-a1}
a_{1,i}  &=&  \frac{1}{2}\log \left(2+\frac{\overrightarrow{\SNR_{i}}}{\INR_{ji}}\right)-\frac{1}{2}, \\
\label{Eq-a2}
a_{2,i}(\rho) &=& \frac{1}{2}\log \Big(b_{1,i}(\rho)+1\Big)-\frac{1}{2}, \\
\label{Eq-a3}
a_{3,i}(\rho,\mu) &=& \frac{1}{2} \log \left( \frac{ \overleftarrow{\SNR}_i \Big(b_{2,i}(\rho)+2\Big)+b_{1,i}(1)+1}{\overleftarrow{\SNR}_i\Big(  \left(1-\mu\right)  b_{2,i}( \rho )+2\Big)\!+ b_{1,i}( 1  ) + 1}  \right), \\ 
\label{Eq-a4}
a_{4,i}(\rho,\mu) &=& \frac{1}{2}\log \bigg(\Big(1-\mu\Big)b_{2,i}(\rho)+2 \bigg)-\frac{1}{2},\\
\label{Eq-a5}
a_{5,i}(\rho,\mu) &=& \frac{1}{2}\log \left(2+\frac{\overrightarrow{\SNR}_{i}}{\INR_{ji}}+\Big(1-\mu\Big)b_{2,i}(\rho)\right)-\frac{1}{2},\\
\label{Eq-a6}
a_{6,i}(\rho,\mu) &=& \frac{1}{2}\!\log\! \left(\!\frac{\overrightarrow{\SNR}_{i}}{\INR_{ji}}\bigg(\Big(1\!-\!\mu\Big)b_{2,j}(\rho)\!+\!1\bigg)\!+\!2\right)\!-\!\frac{1}{2},\\
\label{Eq-a7}
a_{7,i}(\rho,\!\mu_1\!,\!\mu_2\!) &=& \frac{1}{2}\!\log \Bigg(\!\frac{\overrightarrow{\SNR}_{i}}{\INR_{ji}}\bigg(\Big(1\!-\!\mu_i\Big)b_{2,j}(\rho)\!+\!1\bigg)+\Big(1\!-\!\mu_j\Big)b_{2,i}(\rho)+2\Bigg)\!-\!\frac{1}{2},
\end{IEEEeqnarray}
\end{subequations}
where the functions $b_{l,i}:[0,1]\rightarrow \mathds{R}_{+}$, with $(l,i) \in \lbrace1, 2 \rbrace^2$ are defined as follows: 
\begin{subequations}
\label{Eqfnts}
\begin{IEEEeqnarray}{rcl}
\label{Eqb1i}
b_{1,i}(\rho)&=&\overrightarrow{\SNR}_{i}+2\rho\sqrt{\overrightarrow{\SNR}_{i}\INR_{ij}}+\INR_{ij} \mbox{ and } \\
\label{Eqb5i}
b_{2,i}(\rho)&=&\Big(1-\rho\Big)\INR_{ij}-1,
\end{IEEEeqnarray}
\end{subequations}
with $j \in \lbrace 1, 2 \rbrace \setminus \lbrace i \rbrace$.

\noindent
Note that the functions in \eqref{Eq-a} and \eqref{Eqfnts} depend on $\overrightarrow{\SNR}_{1}$, $\overrightarrow{\SNR}_{2}$, $\INR_{12}$, $\INR_{21}$, $\overleftarrow{\SNR}_{1}$, and $\overleftarrow{\SNR}_{2}$, however as these parameters are fixed in this analysis, this dependence is not emphasized in the definition of these functions. Finally, using this notation, Theorem~\ref{TheoremA-G-IC-NOF} is presented on the top of this page.
\begin{figure*}[t]
\begin{theorem} \label{TheoremA-G-IC-NOF} \emph{
The capacity region $\Cgicnof$ contains the region $\agicnof$ given by the closure of the set of all possible non-negative achievable rate pairs $(R_1,R_2)$ that satisfy:
\begin{subequations}
\label{EqRa-G-IC-NOF}
\begin{IEEEeqnarray}{rcl}
\label{EqR1a-G-IC-NOF}
R_{1}  & \leqslant & \min\Big(a_{2,1}(\rho),a_{6,1}(\rho,\mu_1)+a_{3,2}(\rho,\mu_1), a_{1,1}+a_{3,2}(\rho,\mu_1)+a_{4,2}(\rho,\mu_1)\Big),  \\ 
\label{EqR2a-G-IC-NOF}
R_{2}   & \leqslant & \min\Big(a_{2,2}(\rho),a_{3,1}(\rho,\mu_2)+a_{6,2}(\rho,\mu_2), a_{3,1}(\rho,\mu_2)+a_{4,1}(\rho,\mu_2)+a_{1,2}\Big),   \\
\nonumber
R_{1}+R_{2}  & \leqslant & \min\Big(a_{2,1}(\rho)+a_{1,2}, a_{1,1}+a_{2,2}(\rho), a_{3,1}(\rho,\mu_2)+a_{1,1}+a_{3,2}(\rho,\mu_1)+a_{7,2}(\rho,\mu_1,\mu_2), \\
\nonumber
& & a_{3,1}(\rho,\mu_2)+a_{5,1}(\rho,\mu_2)+a_{3,2}(\rho,\mu_1)+a_{5,2}(\rho,\mu_1), \\
\label{EqR1+R2a-G-IC-NOF}
& & a_{3,1}(\rho,\mu_2)+a_{7,1}(\rho,\mu_1,\mu_2)+a_{3,2}(\rho,\mu_1)+a_{1,2}\Big),  \\
\nonumber
2R_{1}+R_{2}  & \leqslant & \min\Big(a_{2,1}(\rho)+a_{1,1}+a_{3,2}(\rho,\mu_1)+a_{7,2}(\rho,\mu_1,\mu_2),  \\
\nonumber
& &  a_{3,1}(\rho,\mu_2)+a_{1,1}+a_{7,1}(\rho,\mu_1,\mu_2)+2a_{3,2}(\rho,\mu_1)+a_{5,2}(\rho,\mu_1), \\
\label{Eq2R1+R2a-G-IC-NOF}
& & a_{2,1}(\rho)+a_{1,1}+a_{3,2}(\rho,\mu_1)+a_{5,2}(\rho,\mu_1)\Big), \\
\nonumber
R_{1}+2R_{2}  & \leqslant & \min\Big(a_{3,1}(\rho,\mu_2)+a_{5,1}(\rho,\mu_2)+a_{2,2}(\rho)+a_{1,2}, \\
\nonumber
& & a_{3,1}(\rho,\mu_2)+a_{7,1}(\rho,\mu_1,\mu_2)+a_{2,2}(\rho)+a_{1,2}, \\
\label{EqR1+2R2a-G-IC-NOF}
& &  2a_{3,1}(\rho,\mu_2)+a_{5,1}(\rho,\mu_2)+a_{3,2}(\rho,\mu_1)+a_{1,2}+a_{7,2}(\rho,\mu_1,\mu_2)\Big),
\end{IEEEeqnarray}
\end{subequations}
with $\left(\rho, \mu_1, \mu_2\right) \in \left[0,\left(1-\max\left(\frac{1}{\INR_{12}},\frac{1}{\INR_{21}}\right) \right)^+\right]\times[0,1]\times[0,1]$.
}
\end{theorem}
\vspace{-5mm}
\end{figure*}

\begin{IEEEproof}
The proof of Theorem~\ref{TheoremA-G-IC-NOF} is presented in Appendix~\ref{AppAch-IC-NOF}.
\end{IEEEproof}
\subsection{Comments on the Achievability}
The achievable region is obtained using a random coding argument and combining three classical tools: rate splitting, superposition coding, and backward decoding. This coding scheme is described in Appendix~\ref{AppAch-IC-NOF} and it is specially designed for the two-user IC-NOF. Consequently, only the strictly needed number of superposition code-layers is used.  Other achievable schemes, as reported in \cite{SyQuoc-TIT-2015}, can also be obtained as special cases of the more general scheme presented in \cite{Tuninetti-ISIT-2007}. However, in this more general case, the resulting code for the IC-NOF contains a handful of unnecessary superposing code-layers, which complicates the error probability analysis.     
\subsection{A Converse Region}

The description of the converse region $\cgicnof$ is determined by the ratios  $\frac{\INR_{ij}}{\overrightarrow{\SNR}_{j}}$, and $\frac{\INR_{ji}}{\overrightarrow{\SNR}_{j}}$, for all $i \in \lbrace 1, 2 \rbrace$, with $j \in \lbrace 1, 2 \rbrace\setminus\lbrace i \rbrace$. All relevant scenarios regarding these ratios are described by two events denoted by $S_{l_{1},1}$ and $S_{l_{2},2}$, where $(l_{1},l_{2}) \in \lbrace 1, \ldots, 5 \rbrace^2$. The events are defined as follows:
 \begin{subequations}
\label{EqSi}
\begin{IEEEeqnarray}{rcl}
\label{EqS1i}  
 S_{1,i}&: \quad & \overrightarrow{\SNR}_{j} < \min\left(\INR_{ij},\INR_{ji}\right), \\ 
\label{EqS2i}
 S_{2,i}&: \quad & \INR_{ji} \leqslant  \overrightarrow{\SNR}_{j} < \INR_{ij}, \\
  \label{EqS3i}
 S_{3,i}&: \quad & \INR_{ij} \leqslant \overrightarrow{\SNR}_{j} < \INR_{ji},  \\
 \label{EqS4i}
 S_{4,i}&: \quad & \max\left(\INR_{ij}, \INR_{ji}\right) \leqslant \overrightarrow{\SNR}_{j} < \INR_{ij}\INR_{ji}, \\
 \label{EqS5i}
 S_{5,i}&: \quad & \overrightarrow{\SNR}_{j} \geqslant  \INR_{ij}\INR_{ji}.
 \end{IEEEeqnarray}
\end{subequations}
Note that for all $i \in \lbrace 1, 2 \rbrace$, the events $S_{1,i}$, $S_{2,i} $, $S_{3,i}$, $S_{4,i}$, and $S_{5,i}$ are mutually exclusive. This observation shows that given any $4$-tuple $(\overrightarrow{\SNR}_{1}, \overrightarrow{\SNR}_{2}, \INR_{12}, \INR_{21})$, there always exists one and only one pair of events  $(S_{l_{1},1}, S_{l_{2},2})$, with $(l_{1},l_{2}) \in \lbrace 1, \ldots, 5 \rbrace^2$, that identifies a unique scenario. Note also that the pairs of events $(S_{2,1}, S_{2,2})$ and $(S_{3,1}, S_{3,2})$ are not feasible. In view of this, twenty-three different scenarios can be identified using the events in \eqref{EqSi}.
Once the exact scenario is identified, the converse region is described using the functions $\kappa_{l,i}: [0,1]\rightarrow \mathds{R}_{+}$, with $l \in \lbrace 1, \ldots, 3\rbrace$; $\kappa_{l}: [0,1]\rightarrow \mathds{R}_{+}$, with $l \in \lbrace 4, 5 \rbrace$; $\kappa_{6,l}: [0,1]\rightarrow \mathds{R}_{+}$, with $l \in \lbrace 1, \ldots, 4\rbrace$; and $\kappa_{7,i,l}:[0,1]\rightarrow \mathds{R}_{+}$, with $l \in \lbrace 1, 2 \rbrace$. These functions are defined as follows for all $i \in \lbrace 1, 2 \rbrace$, with $j \in \lbrace 1, 2 \rbrace\setminus\lbrace i \rbrace$:
\begin{subequations}
\label{Eqconv}
\begin{IEEEeqnarray}{rcl}
\label{Eqconv1}
\kappa_{1,i}(\rho)  & = & \frac{1}{2}\log \Big(b_{1,i}(\rho)+1\Big), \\ 
\label{Eqconv2}
\kappa_{2,i}(\rho)  & = & \frac{1}{2}\log \Big(1+b_{5,j}(\rho)\Big) \!+\! \frac{1}{2}\log \Bigg(1\!+\! \frac{b_{4,i}(\rho)}{1+b_{5,j}(\rho)}\Bigg)\!, \, \qquad \\
\label{Eqconv3}
\kappa_{3,i}(\rho)  & = & \frac{1}{2} \! \log \! \left(\frac{\bigg(b_{4,i}(\rho)+b_{5,j}(\rho)+1\bigg)\overleftarrow{\SNR}_j}{\! \bigg(b_{1,j}(1) \! + \! 1\bigg)\bigg(b_{4,i}(\rho) \! + 1 \! \bigg)} \! + \! 1 \! \right) +\frac{1}{2}\log\Big(b_{4,i}(\rho)+1\Big),\\
\label{Eqconv4}
\kappa_{4}(\rho)  & = & \frac{1}{2}\log \Bigg(1+\frac{b_{4,1}(\rho)}{1+b_{5,2}(\rho)}\Bigg) \! + \! \frac{1}{2}\log \Big(b_{1,2}(\rho)+1\Big)\!, \\
\label{Eqconv5}
\kappa_{5}(\rho)  & = & \frac{1}{2}\log \Bigg(1 \! + \! \frac{b_{4,2}(\rho)}{1 \!+ \! b_{5,1}(\rho)}\Bigg) \! + \! \frac{1}{2}\log \Big(b_{1,1}(\rho) \! + \! 1\Big),\\
\label{Eqk6} 
\kappa_{6}(\rho) &=& \begin{cases} \kappa_{6,1}(\rho) & \textrm{ if } (S_{1,2} \lor S_{2,2} \lor S_{5,2} )  \land (S_{1,1} \lor S_{2,1} \lor S_{5,1} )\\
                                                \kappa_{6,2} (\rho) & \textrm{if }(S_{1,2} \lor S_{2,2} \lor S_{5,2} ) \land (S_{3,1} \lor S_{4,1} )\\
                                                \kappa_{6,3}(\rho) & \textrm{if } (S_{3,2} \lor S_{4,2}) \land (S_{1,1} \lor S_{2,1} \lor S_{5,1})\\
                                                \kappa_{6,4}(\rho) & \textrm{if } (S_{3,2} \lor S_{4,2} ) \land ( S_{3,1} \lor S_{4,1})
                                           \end{cases} \\
\label{Eqk77}
\kappa_{7,i}(\rho) &=& \begin{cases} \kappa_{7,i,1}(\rho) & \textrm{if } (S_{1,i} \lor S_{2,i} \lor S_{5,i})\\
                                                  \kappa_{7,i,2}(\rho) & \textrm{if } (S_{3,i} \lor S_{4,i})
                                                \end{cases}\qquad\qquad\qquad\qquad
\end{IEEEeqnarray}
\end{subequations}
\begin{figure*}[t!]
 \centerline{\epsfig{figure=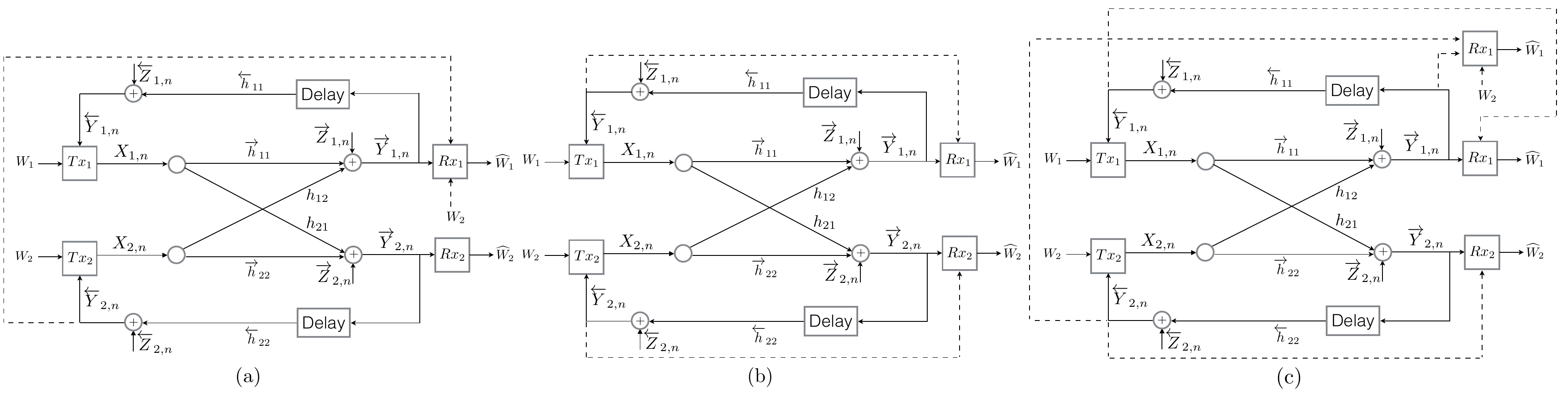,width=1.0\textwidth}}
  \caption{Genie-Aided G-IC-NOF models for channel use~$n$. $(a)$ Model used to calculate the outer-bound on $R_1$; $(b)$ Model used to calculate the outer-bound on $R_1+R_2$; and  $(c)$ Model used to calculate the outer-bound on $2 R_1+R_2$}
  \label{Fig:G-IC-NOF-Conv}
\vspace{-6mm}
\end{figure*}
where
\begin{subequations}
\label{Eqconv6}
\begin{IEEEeqnarray}{rcl}
\nonumber
\kappa_{6,1}(\rho)  &=&  \frac{1}{2} \! \log \! \Big(b_{1,1}(\rho) \! + \! b_{5,1}(\rho)\INR_{21}\Big) \! - \! \frac{1}{2}\log\Big(1 \! + \! \INR_{12}\Big) +\frac{1}{2}\log\left(1+\frac{b_{5,2}(\rho)\overleftarrow{\SNR}_2}{b_{1,2}(1)+1}\right)\\
\nonumber
& & +\frac{1}{2}\log\Big(b_{1,2}(\rho)+b_{5,1}(\rho)\INR_{21}\Big) -\frac{1}{2}\log\Big(1 \! + \! \INR_{21}\Big) \! + \! \frac{1}{2}\log\left(\! 1 \! + \! \frac{b_{5,1}(\rho)\overleftarrow{\SNR}_1}{b_{1,1}(1)+1}\right)\\
\label{Eqconv61}
& & +\log(2 \pi e),
\end{IEEEeqnarray}
\begin{IEEEeqnarray}{rcl}
\label{Eqconv62}   
\kappa_{6,2}(\rho) & = &   \frac{1}{2}\log\left(b_{6,2}(\rho)+\frac{b_{5,1}(\rho)\INR_{21}}{\overrightarrow{\SNR}_2}\Big(\overrightarrow{\SNR}_2+b_{3,2}\Big)\right) - \frac{1}{2}\log\Big(1 \!+ \! \INR_{12}\Big)  \\
\nonumber
& & +\frac{1}{2}\log\left(1+\frac{b_{5,1}(\rho)\overleftarrow{\SNR}_1}{b_{1,1}(1)+1}\right)+ \frac{1}{2} \! \log\Big(b_{1,1}(\rho) \! + \! b_{5,1}(\rho)\INR_{21}\Big) \! -\frac{1}{2}\log\Big(1+\INR_{21}\Big)\\
\nonumber
& & +\frac{1}{2}\log\Bigg(1+\frac{b_{5,2}(\rho)}{\overrightarrow{\SNR}_2}\left(\INR_{12}+\frac{b_{3,2} \overleftarrow{\SNR}_2}{b_{1,2}(1)+1}\right)\Bigg)-\frac{1}{2}\log\left(1+\frac{b_{5,1}(\rho)\INR_{21}}{\overrightarrow{\SNR}_2}\right)+\log(2 \pi e),\\
\label{Eqconv63}
\kappa_{6,3}(\rho) & = & \frac{1}{2}\log\Bigg(b_{6,1}(\rho)+\frac{b_{5,1}(\rho)\INR_{21}}{\overrightarrow{\SNR}_1}\Big(\overrightarrow{\SNR}_1+b_{3,1}\Big)\Bigg)-\frac{1}{2}\log\Big(1+\INR_{12}\Big)\\
\nonumber
& & +\frac{1}{2}\log\left(1+\frac{b_{5,2}(\rho)\overleftarrow{\SNR}_2}{b_{1,2}(1)+1}\right) +\frac{1}{2}\log\Big(b_{1,2}(\rho) \! + \! b_{5,1}(\rho)\INR_{21}\Big) \! - \! \frac{1}{2}\log\Big(1 \! + \! \INR_{21}\Big)\\
\nonumber
& & +\frac{1}{2}\log\Bigg(1+\frac{b_{5,1}(\rho)}{\overrightarrow{\SNR}_1}\left(\INR_{21}+\frac{b_{3,1} \overleftarrow{\SNR}_1}{b_{1,1}(1)+1}\right)\Bigg)-\frac{1}{2}\log\left(1+\frac{b_{5,1}(\rho)\INR_{21}}{\overrightarrow{\SNR}_1}\right) +\log(2 \pi e), \\
\nonumber
\kappa_{6,4}(\rho) &=& \frac{1}{2}\log\left(b_{6,1}(\rho)+\frac{b_{5,1}(\rho)\INR_{21}}{\overrightarrow{\SNR}_1}\Big(\overrightarrow{\SNR}_1+b_{3,1}\Big)\right)-\frac{1}{2}\log\Big(1+\INR_{12}\Big)-\frac{1}{2}\log\Big(1+\INR_{21}\Big)\\
\nonumber
& & +\frac{1}{2}\log\left(1+\frac{b_{5,2}(\rho)}{\overrightarrow{\SNR}_2}\left(\INR_{12}+\frac{b_{3,2}\overleftarrow{\SNR}_2}{b_{1,2}(1)+1}\right)\right)-\frac{1}{2}\log\left(1+\frac{b_{5,1}(\rho)\INR_{21}}{\overrightarrow{\SNR}_2}\right)\\
\nonumber
& & -\frac{1}{2}\log\left(1+\frac{b_{5,1}(\rho)\INR_{21}}{\overrightarrow{\SNR}_1}\right)+\frac{1}{2}\log\left(b_{6,2}(\rho)+\frac{b_{5,1}(\rho)\INR_{21}}{\overrightarrow{\SNR}_2}\Big(\overrightarrow{\SNR}_2+b_{3,2}\Big)\right)\\
\label{Eqconv64}
& & +\frac{1}{2}\log\Bigg(1+\frac{b_{5,1}(\rho)}{\overrightarrow{\SNR}_1}\left(\INR_{21}+\frac{b_{3,1} \overleftarrow{\SNR}_1}{b_{1,1}(1)+1}\right)\Bigg)+\log(2 \pi e),
\end{IEEEeqnarray}
\end{subequations}
and 
\vspace{-2mm}
\begin{subequations}
\label{Eqconv7i}
\begin{IEEEeqnarray}{rcl}
\nonumber
\kappa_{7,i,1}(\rho) &=& \frac{1}{2}\log\Big(b_{1,i}(\rho)+1\Big)-\frac{1}{2}\log\Big(1+\INR_{ij}\Big)+\frac{1}{2}\log\left(1+\frac{b_{5,j}(\rho)\overleftarrow{\SNR}_j}{b_{1,j}(1)+1}\right)\\
\nonumber
& & +\frac{1}{2}\log\Big(b_{1,j}(\rho)+b_{5,i}(\rho)\INR_{ji}\Big)+\frac{1}{2}\log\Big(1 \! + \! b_{4,i}(\rho) \!+ \! b_{5,j}(\rho) \Big) \! - \! \frac{1}{2}\log\Big(1 \! + \! b_{5,j}(\rho)\Big)\\
\label{Eqconv7i1}
& & +2\log(2 \pi e), 
\end{IEEEeqnarray}
\vspace{-4mm}
\begin{IEEEeqnarray}{rcl}
\nonumber
\kappa_{7,i,2}(\rho) &=& \frac{1}{2}\log\Big(b_{1,i}(\rho)+1\Big)-\frac{1}{2}\log\Big(1+\INR_{ij}\Big)-\frac{1}{2}\log\Big(1+b_{5,j}(\rho)\Big)\\
\nonumber
& & +\frac{1}{2}\log\Big(1+b_{4,i}(\rho)+b_{5,j}(\rho) \Big)+\frac{1}{2}\log\Bigg(1+\Big(1-\rho^2\Big)\frac{\INR_{ji}}{\overrightarrow{\SNR}_j}\Bigg(\INR_{ij}+\frac{b_{3,j}\overleftarrow{\SNR}_j}{b_{1,j}(1)+1}\Bigg)\Bigg)\\
\nonumber
& & -\frac{1}{2}\log\left(1+\frac{b_{5,i}(\rho)\INR_{ji}}{\overrightarrow{\SNR}_j}\right)+\frac{1}{2}\!\log\!\left(\!b_{6,j}(\rho)\!+\!\frac{b_{5,i}(\rho)\INR_{ji}}{\overrightarrow{\SNR}_j}\Big(\overrightarrow{\SNR}_j+b_{3,j}\Big)\right)\\
\label{Eqconv7i2}
& & +2\log(2 \pi e), \quad
\end{IEEEeqnarray}
\end{subequations}
where the functions $b_{l,i}$, with $(l,i) \in \lbrace1, 2 \rbrace^2$ are defined in \eqref{Eqfnts}; $b_{3,i}$ are constants; and the functions $b_{l,i}:[0,1]\rightarrow \mathds{R}_{+}$, with $(l,i) \in \lbrace 4, 5, 6 \rbrace\times\lbrace1, 2 \rbrace$ are defined as follows, with $j \in \lbrace 1, 2 \rbrace \setminus \lbrace i \rbrace$:
\begin{subequations}
\label{Eqfnts2}
\begin{IEEEeqnarray}{rcl}
\label{Eqb2i}
b_{3,i}&=&\overrightarrow{\SNR}_i-2\sqrt{\overrightarrow{\SNR}_i\INR_{ji}}+\INR_{ji}, \\
\label{Eqb3i}
b_{4,i}(\rho)&=&\Big(1-\rho^2\Big)\overrightarrow{\SNR}_{i}, \\
\label{Eqb4i}
b_{5,i}(\rho)&=&\Big(1-\rho^2\Big)\INR_{ij},\\
\nonumber
b_{6,i}(\rho)&=&\overrightarrow{\SNR}_i \! + \! \INR_{ij} \! + \!2\rho\sqrt{\INR_{ij}}\left(\sqrt{\overrightarrow{\SNR}_i} \! - \! \sqrt{\INR_{ji}}\right) \! + \! \frac{\INR_{ij}\sqrt{\INR_{ji}}}{\overrightarrow{\SNR}_i} \left(\sqrt{\INR_{ji}}\! - \! 2\sqrt{\overrightarrow{\SNR}_i}\right).\\
\label{Eqb6i}
\end{IEEEeqnarray}
\end{subequations}

\noindent
Note that the functions in \eqref{Eqconv}, \eqref{Eqconv6}, \eqref{Eqconv7i} and \eqref{Eqfnts2} depend on $\overrightarrow{\SNR}_{1}$, $\overrightarrow{\SNR}_{2}$, $\INR_{12}$, $\INR_{21}$, $\overleftarrow{\SNR}_{1}$, and $\overleftarrow{\SNR}_{2}$. However, these parameters are fixed in this analysis, and therefore, this dependence is not emphasized in the definition of these functions.
Finally, using this notation, Theorem~\ref{TheoremC-G-IC-NOF} is presented below.
\begin{theorem} \label{TheoremC-G-IC-NOF} \emph{
The capacity region $\Cgicnof$ is contained within the region $\cgicnof$ given by the closure of the set of non-negative rate pairs  $(R_1,R_2)$ that for all $i \in \lbrace 1, 2 \rbrace$, with $j\in\lbrace 1, 2 \rbrace\setminus\lbrace i \rbrace$ satisfy:
\begin{subequations}
\label{EqRic-G-IC-NOF}
\begin{IEEEeqnarray}{rcl}
\label{EqRic-12-G-IC-NOF}
R_{i}  & \leqslant & \min\left(\kappa_{1,i}(\rho), \kappa_{2,i}(\rho)\right), \\ 
\label{EqRic-3-G-IC-NOF}
R_{i}  & \leqslant & \kappa_{3,i}(\rho), \\
\label{EqR1+R2c-12-G-IC-NOF}
R_{1}+R_{2}  & \leqslant & \min\left(\kappa_{4}(\rho), \kappa_{5}(\rho)\right),\\
\label{EqR1+R2c-3g-G-IC-NOF}
R_{1}+R_{2}  & \leqslant &\kappa_{6}(\rho),\\
\label{Eq2Ri+Rjc-g-G-IC-NOF}
2R_i+R_j&\leqslant&  \kappa_{7,i}(\rho),
\end{IEEEeqnarray}
\end{subequations}
with $\rho \in [0,1]$.
}
\end{theorem} 
\begin{IEEEproof}
The proof of Theorem~\ref{TheoremC-G-IC-NOF} is presented in Appendix~\ref{App-C-G-IC-NOF}.
\end{IEEEproof}

\subsection{Comments on the Converse Region}

The outer bounds \eqref{EqRic-12-G-IC-NOF} and \eqref{EqR1+R2c-12-G-IC-NOF} correspond to the outer bounds for the case of perfect channel-output feedback \cite{Suh-TIT-2011}. 
The bounds \eqref{EqRic-3-G-IC-NOF}, \eqref{EqR1+R2c-3g-G-IC-NOF} and \eqref{Eq2Ri+Rjc-g-G-IC-NOF} correspond to new outer bounds that generalize those presented in \cite{SyQuoc-TIT-2015} for the two-user symmetric G-IC-NOF. These new outer-bounds were obtained using the genie-aided models shown in Figure~\ref{Fig:G-IC-NOF-Conv}.

\subsection{A Gap Between the Achievable Region and the Converse Region}
\balance

\begin{figure}[t]
\centerline{\epsfig{figure=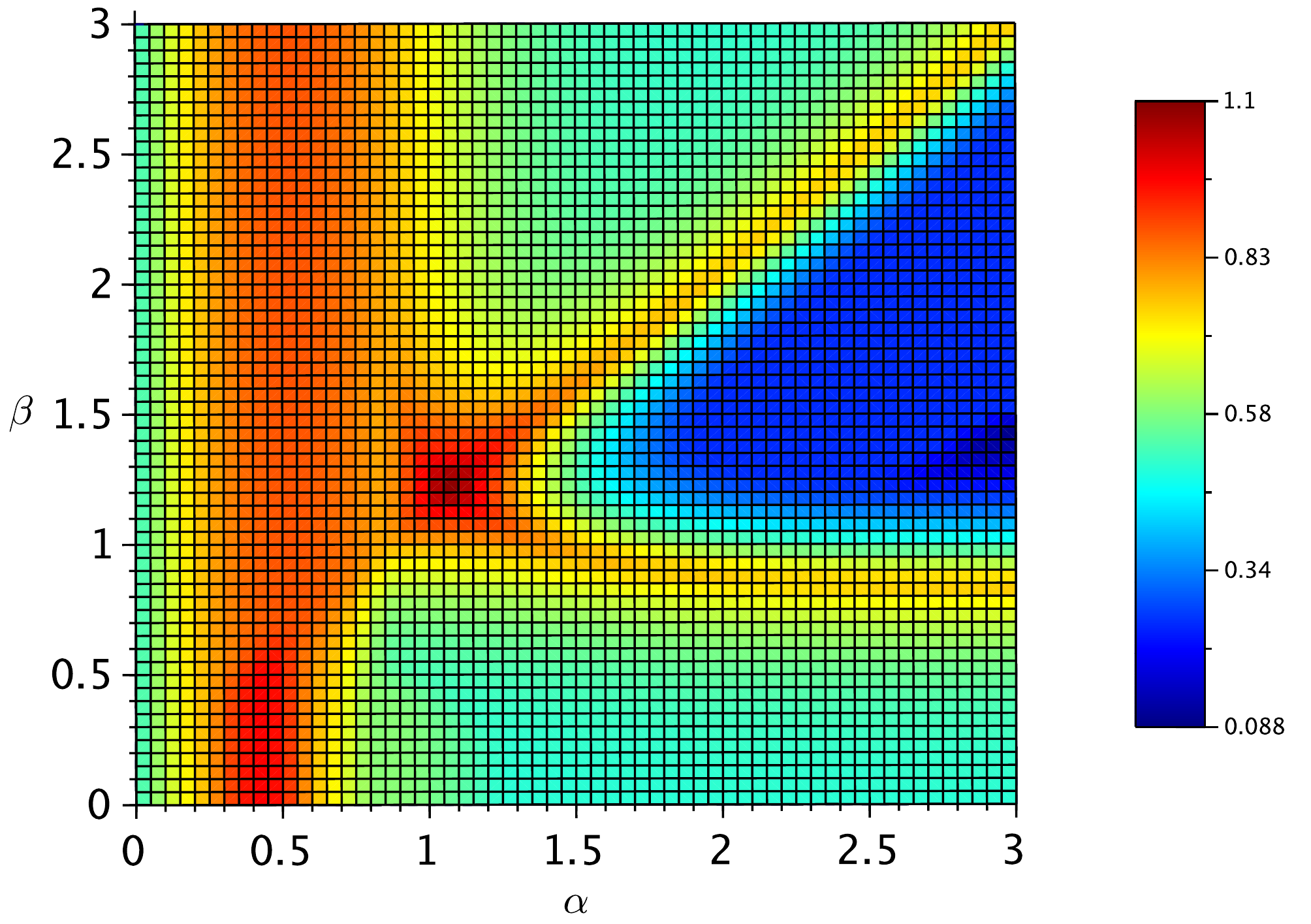,width=0.8\textwidth}}
\caption{Gap between the converse region $\cgicnof$ and the achievable region $\agicnof$ of the two-user G-IC-NOF, under symmetric channel conditions, i.e., $\protect\overrightarrow{\SNR}_1=\protect\overrightarrow{\SNR}_2=\protect\overrightarrow{\SNR}$, $\INR_{12}=\INR_{21}=\INR$, and $\protect\overleftarrow{\SNR}_1=\protect\overleftarrow{\SNR}_2=\protect\overleftarrow{\SNR}$, as a function of $\alpha=\frac{\log \INR}{\log{\protect\overrightarrow{\SNR}}}$ and $\beta=\frac{\log\protect\overleftarrow{\SNR}}{\log\protect\overrightarrow{\SNR}}$.}
\label{FigGapGICNOF}
\end{figure}

Theorem~\ref{TheoremGAP-G-IC-NOF} describes the gap between the achievable region $\agicnof$ and the converse region $\cgicnof$ (Definition~\ref{DefGap}).

\begin{theorem} \label{TheoremGAP-G-IC-NOF} \emph{The capacity region of the two-user G-IC-NOF is approximated to within $4.4$ bits by the achievable region $\agicnof$ and the converse region $\cgicnof$.
}
\end{theorem} 

\begin{IEEEproof}
The proof of Theorem~\ref{TheoremGAP-G-IC-NOF} is presented in Appendix~\ref{AppG-Gap}.
\end{IEEEproof}

\noindent
Figure~\ref{FigGapGICNOF} presents the exact gap existing between the achievable region $\agicnof$ and the converse region $\cgicnof$ for the case in which $\overrightarrow{\SNR}_1=\overrightarrow{\SNR}_2=\overrightarrow{\SNR}$, $\INR_{12}=\INR_{21}=\INR$, and $\overleftarrow{\SNR}_1=\overleftarrow{\SNR}_2=\overleftarrow{\SNR}$ as a function of $\alpha=\frac{\log \INR}{\log{\overrightarrow{\SNR}}}$ and $\beta=\frac{\log\overleftarrow{\SNR}}{\log\overrightarrow{\SNR}}$. Note that in this case, the maximum gap is $1.1$ bits and occurs when $\alpha=1.05$ and $\beta=1.2$.

\clearpage

\section{Conclusions}
An achievable region and a converse region for the two-user G-IC-NOF have been introduced. It has been shown that these regions approximate the capacity region of the two-user G-IC-NOF to within $4.4$ bits.

\clearpage

\begin{appendices}

\section{Proof of Achievability} \label{AppAch-IC-NOF}

This appendix describes an achievability scheme for the IC-NOF based on a three-part message splitting, superposition coding, and backward decoding. 

\noindent
\textbf{Codebook Generation}: Fix a strictly positive joint probability distribution  
\begin{IEEEeqnarray}{rcl}
\nonumber
&P&_{U\, U_1\,U_2\, V_1\,V_2\, X_{1,P}\, X_{2,P}}(u, u_1,u_2, v_1,v_2, x_{1,P}, x_{2,P})=P_U(u) P_{U_1|U}(u_1|u) P_{U_2|U}(u_2|u) \\
\label{Eqprobdist}
& & P_{V_1|U\,U_1}(v_1|u,u_1)P_{V_2|U\,U_2}(v_2|u,u_2) P_{X_{1,P}|U\,U_1\,V_1}(x_{1,P}|u,u_1,v_1) P_{X_{2,P}|U\,U_2\,V_2}(x_{2,P}|u,u_2,v_2), \qquad
\end{IEEEeqnarray}
for all $\left(u, u_1, u_2, v_1, v_2, x_{1,P}, x_{2,P}\right) \in \left(\mathcal{X}_1\cup \mathcal{X}_2\right) \times \mathcal{X}_1 \times \mathcal{X}_2 \times \mathcal{X}_1 \times \mathcal{X}_2 \times \mathcal{X}_1 \times \mathcal{X}_2$.

\noindent
Let $R_{1,C1}$, $R_{1,C2}$, $R_{2,C1}$, $R_{2,C2}$, $R_{1,P}$, and $R_{2,P}$ be non-negative real numbers. Let also $R_{1,C}=R_{1,C1}$ $+$ $R_{1,C2}$, ${R_{2,C}=R_{2,C1}+R_{2,C2}}$, $R_{1}=R_{1,C}+R_{1,P}$, and ${R_{2}=R_{2,C}+R_{2,P}}$. 

\noindent
Generate $2^{N(R_{1,C1} + R_{2,C1})}$ i.i.d. $N$-length codewords ${\bs{u}(s,r) = \big(u_{1}(s,r), u_{2}(s,r), \ldots, u_{N}(s,r)\big)}$ according to 

\begin{equation}
P_{\bs{U}}\big(\bs{u}(s,r)\big) = \ds\prod_{i =1}^N P_{U}(u_{i}(s,r)),
\end{equation}
with $s \in \lbrace 1, 2,  \ldots, 2^{NR_{1,C1}}\rbrace$ and $r \in \lbrace 1, 2,  \ldots, 2^{NR_{2,C1}}\rbrace$. 

\noindent
For encoder $1$, generate for each codeword $\bs{u}(s,r)$, $2^{NR_{1,C1}}$ i.i.d. $N$-length codewords $\bs{u}_1(s,r,k) = \big(u_{1,1}(s,r,k), u_{1,2}(s,r,k), \ldots, u_{1,N}(s,r,k)\big)$  according to 
\begin{equation}
P_{\bs{U}_1|\bs{U}}\big(\bs{u}_1(s,r,k)|\bs{u}(s,r)\big) = \ds\prod_{i =1}^N P_{U_{1}|U}\big(u_{1,i}(s,r,k)|u_{i}(s,r)\big), 
\end{equation}
with $k \in \lbrace 1, 2,  \ldots, 2^{NR_{1,C1}}\rbrace$.  For each pair of codewords $\big(\bs{u}(s,r),\bs{u}_{1}(s,r,k)\big)$, generate $2^{NR_{1,C2}}$ i.i.d. $N$-length codewords $\bs{v}_1(s,r,k,l) = \big(v_{1,1}(s,r,k,l), v_{1,2}(s,r,k,l), \ldots, v_{1,N}(s,r,k,l)\big)$  according to 
\begin{IEEEeqnarray}{rcl}
P_{\bs{V}_1 | \bs{U}\,\bs{U}_1}\big(\bs{v}_1(s,r,k,l)|\bs{u}(s,r) ,\bs{u}_1(s,r,k)\big) = \ds\prod_{i =1}^N P_{V_{1}|U\,U_{1}}\big(v_{1,i}(s,r,k,l)|u_{i}(s,r),u_{1,i}(s,r,k)\big), \quad
\end{IEEEeqnarray}
with $l \in \lbrace 1, 2, \ldots, 2^{NR_{1,C2}}\rbrace$. For each tuple of codewords $\big(\bs{u}(s,r)$, $\bs{u}_{1}(s,r,k)$, $\bs{v}_{1}(s,r,k,l)\big)$, generate  $2^{NR_{1,P}}$ i.i.d. $N$-length codewords $\bs{x}_{1,P}(s,r,k,l,q) = \big(x_{1,P,1}(s,r,k,l,q), x_{1,P,2}(s,r,k,l,q), \ldots$, $x_{1,P,N}(s,r,k,l,q)\big)$ according to 
\begin{IEEEeqnarray}{rcl}
\nonumber
&P&_{\bs{X}_{1,P} | \bs{U}\, \bs{U}_{1} \!\, \!\bs{V}_{1}}\!\big(\bs{x}_{1,P}(s,r,k,l,q) |  \bs{u}(s,r),\!\bs{u}_{1}(s,r,k),\!\bs{v}_{1}(s,r,k,l)\!\big)\!\\
& & =\ds\prod_{i =1}^N P_{X_{1,P}|U\,U_{1}\,V_{1}}\big(x_{1,P,i}(s,r,k,l,q)|u_{i}(s,r),u_{1,i}(s,r,k),v_{1,i}(s,r,k,l)\big), 
\end{IEEEeqnarray}
with $q \in \lbrace 1, 2, \ldots, 2^{NR_{1,P}}\rbrace$. 

\begin{figure*}[t!]
 \centerline{\epsfig{figure=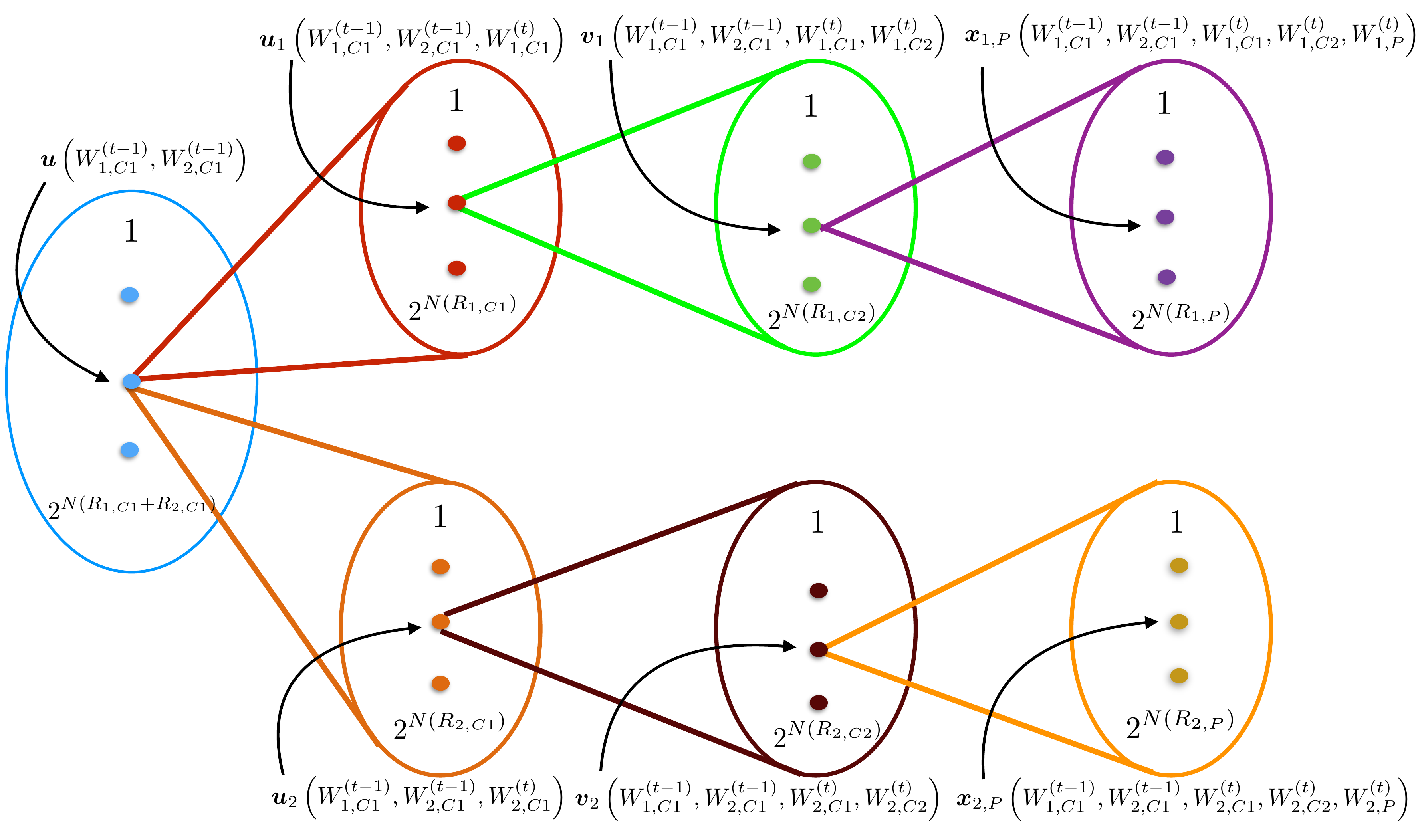,width=1\textwidth}}
 \caption{Structure of the superposition code. The codewords corresponding to the message indices $W_{1,C1}^{(t-1)}, W_{2,C1}^{(t-1)},W_{i,C1}^{(t)},W_{i,C2}^{(t)},W_{i,P}^{(t)}$ with $i \in \lbrace 1, 2 \rbrace$ as well as the block index $t$ are both highlighted. The (approximate) number of codewords for each code layer is also highlighted.} 
\label{FigSuperpos}
\end{figure*}

\noindent
For encoder $2$, generate for each codeword $\bs{u}(s,r)$,  $2^{NR_{2,C1}}$ i.i.d. $N$-length codewords  $\bs{u}_2(s,r,j) = \big(u_{2,1}(s,r,j), u_{2,2}(s,r,j), \ldots, u_{2,N}(s,r,j)\big)$  according to 
\begin{equation}
P_{\bs{U}_2|\bs{U}}\big(\bs{u}_2(s,r,j)|\bs{u}(s,r)\big) = \ds\prod_{i =1}^N P_{U_{2}|U}\big(u_{2,i}(s,r,j)|u_{i}(s,r)\big), 
\end{equation}
with $j \in \lbrace 1, 2,  \ldots, 2^{NR_{2,C1}}\rbrace$. For each pair of codewords $\big(\bs{u}(s,r),\bs{u}_{2}(s,r,j)\big)$, generate $2^{NR_{2,C2}}$ i.i.d. length-$N$ codewords $\bs{v}_2(s,r,j,m)=\big(v_{2,1}(s,r,j,m), v_{2,2}(s,r,j,m),  \ldots, v_{2,N}(s,r,j,m)\big)$ according to 
\begin{IEEEeqnarray}{rcl}
P_{\bs{V}_2 | \bs{U}\, \bs{U}_2}\big(\bs{v}_2(s,r,j,m) | \bs{u}(s,r), \bs{u}_2(s,r,j)\big) = \ds\prod_{i =1}^N P_{V_{2} |  U\, U_{2}}(v_{2,i}(s,r,j,m) |  u_{i}(s,r), u_{2,i}(s,r,j)),  \qquad
\end{IEEEeqnarray}
with $m \in \lbrace 1, 2,  \ldots, 2^{NR_{2,C2}}\rbrace$. For each tuple of codewords $\big(\bs{u}(s,r)$, $\bs{u}_{2}(s,r,j),\bs{v}_{2}(s,r,j,m)\big)$, generate  $2^{NR_{2,P}}$ i.i.d. $N$-length codewords $\bs{x}_{2,P}(s,r,j,m,b) \! = \! \big(\! x_{2,P,1}(s,r,j,m,b) \!, \! x_{2,P,2}(s,r,j,m,b) \!, \! \ldots$,  $ x_{2,P,N}(s,r,j,m,b) \!\big)$ according to 
\begin{IEEEeqnarray}{rcl}
\nonumber
&P&_{\bs{X}_{2,P}  |  \bs{U}\,\bs{U}_{2}\! \, \! \bs{V}_{2}} \!\big( \! \bs{x}_{2,P}(s,r,j,m,b)|  \bs{u}(s,r),  \bs{u}_{2}(s,r,j), \! \bs{v}_{2}(s,r,j,m) \! \big) \!  \\
& & = \ds\prod_{i =1}^N P_{X_{2,P} | U\, U_{2}\, V_{2}}\big(x_{2,P,i}(s,r,j,m,b) | u_{i}(s,r), u_{2,i}(s,r,j), v_{2,i}(s,r,j,m,b)\big), 
\end{IEEEeqnarray}
with $b \in \lbrace 1, 2,  \ldots, 2^{NR_{2,P}}\rbrace$.  
The resulting code structure is shown in Figure~\ref{FigSuperpos}.

\noindent
\textbf{Encoding}: Denote by $W_{i}^{(t)} \in \lbrace 1, 2, \ldots, 2^{NR_{i}} \rbrace$  the message index of transmitter $i \in \lbrace 1,2 \rbrace$ during block $t \in \lbrace 1, 2,  \ldots, T \rbrace$, with $T$ the total number of blocks. Let $W_{i}^{(t)}$ be composed by the message index $W_{i,C}^{(t)} \in \lbrace 1, 2,  \ldots, 2^{NR_{i,C}} \rbrace$ and message index $W_{i,P}^{(t)} \in \lbrace 1$, $2$, $  \ldots, 2^{NR_{i,P}} \rbrace$. That is, $W_{i}^{(t)}=\left(W_{i,C}^{(t)},W_{i,P}^{(t)}\right)$.  The message index $W_{i,P}^{(t)}$ must be reliably decoded at receiver $i$. Let also $W_{i,C}^{(t)}$ be composed by the message indices $W_{i,C1}^{(t)} \in \lbrace 1, 2,  \ldots, 2^{NR_{i,C1}} \rbrace$ and $W_{i,C2}^{(t)} \in \lbrace 1, 2,  \ldots, 2^{NR_{i,C2}} \rbrace$. That is, $W_{i,C}^{(t)}=\Big(W_{i,C1}^{(t)}$,$W_{i,C2}^{(t)}\Big)$.  The message index $W_{i,C1}^{(t)}$ must be reliably decoded by the other transmitter (via feedback) and by the non-intended receiver, but not necessarily by the intended receiver. The message index $W_{i,C2}^{(t)}$ must be reliably decoded by the non-intended receiver, but not necessarily by the intended receiver. 

\noindent
Consider Markov encoding over $T$ blocks. At encoding step $t$, with $t \in \lbrace 1, 2,  \ldots, T \rbrace$, transmitter $1$ sends the codeword:
 \begin{IEEEeqnarray}{rcl}
 \nonumber
 \bs{x}_1^{(t)} &=& \Theta_1 \Bigg(\! \bs{u}\Big(\!W_{1,C1}^{(t-1)}, W_{2,C1}^{(t-1)} \!\Big), \! \bs{u}_1\Big(\! W_{1,C1}^{(t-1)}, W_{2,C1}^{(t-1)},W_{1,C1}^{(t)} \!\Big), \bs{v}_1\Big(W_{1,C1}^{(t-1)}, W_{2,C1}^{(t-1)},W_{1,C1}^{(t)}, W_{1,C2}^{(t)}\Big), \\
\label{Eqtransmittercodeword}
 & & \bs{x}_{1,P}\Big(W_{1,C1}^{(t-1)}, W_{2,C1}^{(t-1)}, W_{1,C1}^{(t)}, W_{1,C2}^{(t)},W_{1,P}^{(t)}\Big)\Bigg), 
 \end{IEEEeqnarray}
 where,  $\Theta_1: \left(\mathcal{X}_1\cup \mathcal{X}_2\right)^{N} \times \mathcal{X}_1^{N} \times \mathcal{X}_1^{N} \times \mathcal{X}_1^{N} \rightarrow \mathcal{X}_1^{N}$ is a function that transforms the codewords $\bs{u}\Big(W_{1,C1}^{(t-1)}$, $W_{2,C1}^{(t-1)}\Big)$, $\bs{u}_1\Big(W_{1,C1}^{(t-1)}, W_{2,C1}^{(t-1)},W_{1,C1}^{(t)}\Big)$, $\bs{v}_1\Big(W_{1,C1}^{(t-1)}, W_{2,C1}^{(t-1)},W_{1,C1}^{(t)}, W_{1,C2}^{(t)}\Big) \ $, and $\bs{x}_{1,P}\Big(W_{1,C1}^{(t-1)}$, $W_{2,C1}^{(t-1)}$, $W_{1,C1}^{(t)}$, $W_{1,C2}^{(t)}$, $W_{1,P}^{(t)}\Big)$ into the N-dimensional vector $\bs{x}_1^{(t)}$ of channel inputs. The indices $W_{1,C1}^{(0)} = W_{1,C1}^{(T)} = s^*$ and  $W_{2,C1}^{(0)} = W_{2,C1}^{(T)} = r^*$, and the pair $(s^*,r^*) \in  \lbrace 1, 2,  \ldots, 2^{N \, R_{1,C1}} \rbrace \times \lbrace 1, 2,  \ldots, 2^{NR_{2,C1}} \rbrace$  are pre-defined and known by both receivers and transmitters. It is worth noting that the message index  $W_{2,C1}^{(t-1)}$ is obtained by transmitter $1$ from the feedback signal $\overleftarrow{\bs{y}}_{1}^{(t-1)}$ at the end of the previous encoding step $t-1$ (see Figure~\ref{Fig:channelusen}). 

\noindent
Transmitter $2$ follows a similar encoding scheme.

\noindent
\textbf{Decoding}: Both receivers decode their message indices at the end of block $T$ in a backward decoding fashion. At each decoding step $t$, with $t \in \lbrace 1, 2,  \ldots, T \rbrace$, receiver $1$ obtains the message indices $\big(\widehat{W}_{1,C1}^{(T-t)}$, $\widehat{W}_{2,C1}^{(T-t)}$, $\widehat{W}_{1,C2}^{(T-(t-1))}$, $\widehat{W}_{1,P}^{(T-(t-1))}$, $\widehat{W}_{2,C2}^{(T-(t-1))}\big) \in \lbrace 1$, $2,  \ldots $, $ 2^{NR_{1,C1}}\rbrace \times \lbrace 1$, $ 2,  \ldots, 2^{NR_{2,C1}} \rbrace  \times  \lbrace 1$, $ 2,  \ldots, 2^{NR_{1,C2}} \rbrace \times  \lbrace 1$, $ 2,  \ldots, 2^{NR_{1,P}}\rbrace \times  \lbrace 1$, $2,  \ldots, 2^{NR_{2,C2}} \rbrace $ from the channel output $\overrightarrow{\bs{y}}_1^{(T-(t-1))}$. The tuple $\Big(\widehat{W}_{1,C1}^{(T-t)}$, $\widehat{W}_{2,C1}^{(T-t)}$, $\widehat{W}_{1,C2}^{(T-(t-1))}$, $\widehat{W}_{1,P}^{(T-(t-1))}$, $\widehat{W}_{2,C2}^{(T-(t-1))}\Big)$ is the unique tuple that satisfies
\begin{IEEEeqnarray}{ll}
\nonumber
\Big( & \bs{u}\left(\widehat{W}_{1,C1}^{(T-t)}, \widehat{W}_{2,C1}^{(T-t)}\right), \bs{u}_1\left(\widehat{W}_{1,C1}^{(T-t)}, \widehat{W}_{2,C1}^{(T-t)}, W_{1,C1}^{(T-(t-1))}\right),  \\
\nonumber
& \bs{v}_1 \left(\widehat{W}_{1,C1}^{(T-t)}, \widehat{W}_{2,C1}^{(T-t)}, W_{1,C1}^{(T-(t-1))}, \widehat{W}_{1,C2}^{(T-(t-1))} \right), \\
\nonumber
& \bs{x}_{1,P}\Big(\widehat{W}_{1,C1}^{(T-t)}, \widehat{W}_{2,C1}^{(T-t)}, W_{1,C1}^{(T-(t-1))}, \widehat{W}_{1,C2}^{(T-(t-1))}, \widehat{W}_{1,P}^{(T-(t-1))}\Big),\\
\nonumber
&  \bs{u}_2\left(\widehat{W}_{1,C1}^{(T-t)}, \widehat{W}_{2,C1}^{(T-t)}, W_{2,C1}^{(T-(t-1))}\right), \bs{v}_2\left(\widehat{W}_{1,C1}^{(T-t)}, \widehat{W}_{2,C1}^{(T-t)}, W_{2,C1}^{(T-(t-1))},\widehat{W}_{2,C2}^{(T-(t-1))}\right), \\
\label{EqDecodingW1cW2cW1p}
& \overrightarrow{\bs{y}}_1^{(T-(t-1))} \Big) \in \mathcal{T}_{\big[U \ U_1 \ V_1  \ X_{1,P} \ U_2 \ V_2  \ \overrightarrow{Y}_1\big]}^{(N, e)}, 
\end{IEEEeqnarray}
where $W_{1,C1}^{(T-(t-1))}$ and $W_{2,C1}^{(T-(t-1))}$ are assumed to be perfectly decoded in the previous decoding step $t-1$. The set $\mathcal{T}_{\big[U \ U_1 \ V_1  \ X_{1,P} \ U_2 \ V_2  \ \overrightarrow{Y}_1\big]}^{(N, e)}$ represents the set of jointly typical sequences of the random variables $U, U_1, V_1, X_{1,P}, U_2, V_2$, and $\overrightarrow{Y}_1$, with $e>0$.
Receiver $2$ follows a similar decoding scheme.

\noindent
\textbf{Probability of Error Analysis}: An error might occur during encoding step $t$ if the message index $W_{2,C1}^{(t-1)}$ is not correctly decoded at transmitter $1$. From the asymptotic equipartion property (AEP) \cite{Cover-Book-1991}, it follows that the message index $W_{2,C1}^{(t-1)}$ can be reliably decoded at transmitter $1$ during encoding step $t$, under the condition:
\begin{eqnarray}
\nonumber
R_{2,C1} & \leqslant & I\left( \overleftarrow{Y}_1 ; U_2  | U, U_1, V_1, X_1 \right) \\
\label{EqConditionNoError1}
&=& I\left( \overleftarrow{Y}_1 ; U_2  | U, X_1  \right).
\end{eqnarray} 
An error might occur during the (backward) decoding step $t$ if the message indices $W_{1,C1}^{(T-t)}$, $W_{2,C1}^{(T-t)}$, $W_{1,C2}^{(T-(t-1))}, W_{1,P}^{(T-(t-1))}$, and $W_{2,C2}^{(T-(t-1))}$ are not decoded correctly given that the message indices $W_{1,C1}^{(T-(t-1))}$ and $W_{2,C1}^{(T-(t-1))}$ were correctly decoded in the previous decoding step $t-1$. 
These errors might arise for two reasons: $(i)$ there does not exist a tuple $\Big(\widehat{W}_{1,C1}^{(T-t)}$, $\widehat{W}_{2,C1}^{(T-t)}, \widehat{W}_{1,C2}^{(T-(t-1))},\widehat{W}_{1,P}^{(T-(t-1))},\widehat{W}_{2,C2}^{(T-(t-1))}\Big)$  that satisfies \eqref{EqDecodingW1cW2cW1p}, or $(ii)$ there exist several tuples $\Big(\widehat{W}_{1,C1}^{(T-t)}, \widehat{W}_{2,C1}^{(T-t)}, \widehat{W}_{1,C2}^{(T-(t-1))},\widehat{W}_{1,P}^{(T-(t-1))}, \widehat{W}_{2,C2}^{(T-(t-1))}\Big)$ that simultaneously satisfy \eqref{EqDecodingW1cW2cW1p}. 
From the asymptotic equipartion property (AEP) \cite{Cover-Book-1991}, the probability of an error due to $(i)$ tends to zero when $N$ grows to infinity. Consider the error due to $(ii)$ and define the event $E_{(s, r, l, q, m)}$ that describes the case in which the codewords $\big(\bs{u}(s,r)$, $\bs{u}_1(s,r,W_{1,C1}^{(T-(t-1))})$, $\bs{v}_1(s,r,W_{1,C1}^{(T-(t-1))},l)$, $\bs{x}_{1,P}(s,r,W_{1,C1}^{(T-(t-1))},l,q)$, $\bs{u}_2(s,r,W_{2,C1}^{(T-(t-1))})$, and $\bs{v}_2(s,r,W_{2,C1}^{(T-(t-1))},m)\big)$ are jointly typical with $ \overrightarrow{\bs{y}}_1^{(T-(t-1))}$ during decoding step $t$. 
Assume now that the codeword to be decoded at decoding step $t$ corresponds to the indices $(s,r,l,q,m) = (1,1,1,1,1)$ without loss of generality due to the symmetry of the code. Then, the probability of error due to $(ii)$ during decoding step $t$, can be  bounded as follows:
\begin{IEEEeqnarray}{lcl}
\nonumber
P_e & = &\pr{\ds\bigcup_{(s,r,l,q,m) \neq (1,1,1,1,1)} E_{(s,r,l,q,m)} }\\
\label{EqConditionNoError2a}
& \leqslant & \ds\sum_{\scriptscriptstyle (s, r, l, q, m) \in \mathcal{T}} \pr{ E_{(s,r,l,q,m)}}, 
\end{IEEEeqnarray}
with $\mathcal{T}=\Big\lbrace \lbrace 1$,$ 2,  \ldots 2^{NR_{1,C1}} \rbrace \times \lbrace 1$,$ 2,  \ldots 2^{NR_{2,C1}} \rbrace \times \lbrace 1$,$ 2,  \ldots 2^{NR_{1,C2}} \rbrace \times \lbrace 1$,$ 2,  \ldots 2^{NR_{1,P}} \rbrace \times \lbrace 1$ , $ 2,  \ldots 2^{NR_{2,C2}} \rbrace \Big\rbrace \setminus \lbrace (1,1,1,1,1) \rbrace$.

\noindent
From AEP \cite{Cover-Book-1991}, it follows that
\begin{IEEEeqnarray}{lcl}
\nonumber
P_e& \leqslant & 2^{N (R_{2,C2} - I(\overrightarrow{Y}_1;V_2 | U, U_1, U_2, V_1, X_1) + 2\epsilon) }+2^{N (R_{1,P} - I(\overrightarrow{Y}_1;X_1 | U, U_1, U_2, V_1, V_2) + 2\epsilon) }\\
 \nonumber
 & & +2^{N (R_{2,C2} +R_{1,P} - I(\overrightarrow{Y}_1;V_2, X_1 | U, U_1, U_2, V_1) + 2\epsilon) }+2^{N (R_{1,C2} - I(\overrightarrow{Y}_1; V_1, X_1 | U, U_1, U_2, V_2) + 2\epsilon) } \\
 \nonumber
 & & +2^{N (R_{1,C2} +R_{2,C2} - I(\overrightarrow{Y}_1;V_1,V_2, X_1 | U, U_1, U_2) + 2\epsilon) }+2^{N (R_{1,C2} +R_{1,P} - I(\overrightarrow{Y}_1; V_1, X_1 | U, U_1, U_2, V_2) + 2\epsilon) }\\
 \nonumber
 & & +2^{N (R_{1,C2} + R_{1,P}+R_{2,C2} - I(\overrightarrow{Y}_1;V_1, V_2, X_1 | U, U_1, U_2) + 2\epsilon) }+2^{N (R_{2,C1}  - I(\overrightarrow{Y}_1; U, U_1, U_2, V_1, V_2, X_1) + 2\epsilon) }\\ 
\nonumber
 & & +2^{N (R_{2,C1} + R_{2,C2} - I(\overrightarrow{Y}_1; U, U_1, U_2, V_1, V_2, X_1) + 2\epsilon) }+2^{N (R_{2,C1} +R_{1,P} - I(\overrightarrow{Y}_1; U, U_1, U_2, V_1, V_2, X_1) + 2\epsilon) } \\
  \nonumber
& & +2^{N (R_{2,C1} + R_{1,P}+R_{2,C2} - I(\overrightarrow{Y}_1; U, U_1, U_2, V_1, V_2, X_1) + 2\epsilon) } +2^{N (R_{2,C1} +R_{1,C2}  - I(\overrightarrow{Y}_1; U, U_1, U_2, V_1, V_2, X_1) + 2\epsilon) }\\
\nonumber
& & +2^{N (R_{2,C1} +R_{1,C2}+R_{2,C2} - I(\overrightarrow{Y}_1; U, U_1, U_2, V_1, V_2, X_1) + 2\epsilon) }\\
\nonumber
& & +2^{N (R_{2,C1} +R_{1,C2}+R_{1,P}  - I(\overrightarrow{Y}_1; U, U_1, U_2, V_1, V_2, X_1) + 2\epsilon) }\\
\nonumber
& & +2^{N (R_{2,C} +R_{1,C2}+R_{1,P} - I(\overrightarrow{Y}_1; U, U_1, U_2, V_1, V_2, X_1) + 2\epsilon) }+2^{N (R_{1,C1} - I(\overrightarrow{Y}_1; U, U_1, U_2, V_1, V_2, X_1) + 2\epsilon) }\\
\nonumber
& & +2^{N (R_{1,C1} +R_{2,C2} - I(\overrightarrow{Y}_1; U, U_1, U_2, V_1, V_2, X_1) + 2\epsilon) } +2^{N (R_{1,C1} +R_{1,P} - I(\overrightarrow{Y}_1; U, U_1, U_2, V_1, V_2, X_1) + 2\epsilon) }
\end{IEEEeqnarray}
\begin{IEEEeqnarray}{lcl}
\nonumber
& & +2^{N (R_{1,C1} +R_{1,P}+R_{2,C2} - I(\overrightarrow{Y}_1; U, U_1, U_2, V_1, V_2, X_1) + 2\epsilon) } +2^{N (R_{1,C1} +R_{1,C2} - I(\overrightarrow{Y}_1; U, U_1, U_2, V_1, V_2, X_1) + 2\epsilon) } \\
\nonumber
& & +2^{N (R_{1,C1} +R_{1,C2}+R_{2,C2} - I(\overrightarrow{Y}_1; U, U_1, U_2, V_1, V_2, X_1) + 2\epsilon) } \\
\nonumber
& & +2^{N (R_{1,C1} +R_{1,C2}+R_{1,P} - I(\overrightarrow{Y}_1; U, U_1, U_2, V_1, V_2, X_1) + 2\epsilon) } \\
\nonumber
& & +2^{N (R_{1,C1} +R_{1,C2}+R_{1,P}+R_{2,C2} - I(\overrightarrow{Y}_1; U, U_1, U_2, V_1, V_2, X_1) + 2\epsilon) } \\
\nonumber
& & +2^{N (R_{1,C1} +R_{2,C1} - I(\overrightarrow{Y}_1; U, U_1, U_2, V_1, V_2, X_1) + 2\epsilon) }+2^{N (R_{1,C1} +R_{2,C1}+R_{2,C2}  - I(\overrightarrow{Y}_1; U, U_1, U_2, V_1, V_2, X_1) + 2\epsilon) }\\
\nonumber
& & +2^{N (R_{1,C1} +R_{2,C1}+R_{1,P} - I(\overrightarrow{Y}_1; U, U_1, U_2, V_1, V_2, X_1) + 2\epsilon) } \\
\nonumber
& & +2^{N (R_{1,C1} +R_{2,C1}+R_{1,P}+R_{2,C2} - I(\overrightarrow{Y}_1; U, U_1, U_2, V_1, V_2, X_1) + 2\epsilon) }\\
\nonumber
& & +2^{N (R_{1,C1} +R_{2,C1}+R_{1,C2} - I(\overrightarrow{Y}_1; U, U_1, U_2, V_1, V_2, X_1) + 2\epsilon) }\\
\nonumber
& & +2^{N (R_{1,C1} +R_{2,C1}+R_{1,C2} +R_{2,C2}  - I(\overrightarrow{Y}_1; U, U_1, U_2, V_1, V_2, X_1) + 2\epsilon) }\\
\nonumber
& & +2^{N (R_{1,C1} +R_{2,C1}+R_{1,C2} +R_{1,P} - I(\overrightarrow{Y}_1; U, U_1, U_2, V_1, V_2, X_1) + 2\epsilon) }+2^{N (R_{1} +R_{2,C} - I(\overrightarrow{Y}_1; U, U_1, U_2, V_1, V_2, X_1) + 2\epsilon) }.\\
\label{EqConditionNoError2b}
\end{IEEEeqnarray}
The same analysis of the probability of error holds for transmitter-receiver pair $2$.
Hence, in general, from \eqref{EqConditionNoError1} and \eqref{EqConditionNoError2b}, reliable decoding holds under the following conditions for transmitter $i \in \lbrace1,2 \rbrace$, with $j \in  \lbrace1,2 \rbrace\setminus\lbrace i \rbrace$:

\begin{subequations}
\label{EqRateRegion-z}
\begin{IEEEeqnarray}{rcl}
\nonumber
R_{j,C1}  & \leqslant &  I\left( \overleftarrow{Y}_i ; U_j  | U, U_i, V_i,X_i  \right) \\ 
\nonumber
&=& I\left( \overleftarrow{Y}_i ; U_j  | U, X_i  \right)\\
\label{EqRateRegion0}
& \triangleq &  \theta_{1,i}, \\
\nonumber
R_{i} + R_{j,C}  & \leqslant &  I(\overrightarrow{Y}_i; U,U_i, U_j,V_i, V_j, X_i) \\
\nonumber
&=& I(\overrightarrow{Y}_i; U, U_j,V_j, X_i) \\
\label{EqRateRegion1}
&\triangleq&  \theta_{2,i}, \\
\nonumber
R_{j,C2}  & \leqslant & I(\overrightarrow{Y}_i; V_j | U, U_i, U_j, V_i, X_i)\\
\nonumber
&=& I(\overrightarrow{Y}_i; V_j | U, U_j, X_i) \\
\label{EqRateRegion2}
&\triangleq&  \theta_{3,i}, \\
\nonumber
R_{i,P}    &\leqslant &  I(\overrightarrow{Y}_i; X_i | U, U_i, U_j,V_i, V_j) \\
\label{EqRateRegion3}
&\triangleq& \theta_{4,i}, \\
\nonumber
R_{i,P}+R_{j,C2}  & \leqslant & I(\overrightarrow{Y}_i; V_j, X_i | U, U_i, U_j, V_i) \\
\label{EqRateRegion4}
&\triangleq& \theta_{5,i}, \\
\nonumber
R_{i,C2}+R_{i,P}  & \leqslant & I(\overrightarrow{Y}_i; V_i,X_i | U, U_i, U_j, V_j) \\
\nonumber
&=& I(\overrightarrow{Y}_i; X_i | U, U_i, U_j, V_j) \\
\label{EqRateRegion5}
&\triangleq& \theta_{6,i},  \mbox{ and }\\
\nonumber
R_{i,C2}+R_{i,P}+R_{j,C2} & \leqslant & I(\overrightarrow{Y}_i; V_i, V_j, X_i | U, U_i, U_j) \\
\nonumber
&=& I(\overrightarrow{Y}_i; V_j, X_i | U, U_i, U_j) \\
\label{EqRateRegion6}
&\triangleq& \theta_{7,i}.
\end{IEEEeqnarray}
\end{subequations}
Taking into account that $R_i=R_{i,C1}+R_{i,C2}+R_{i,P}$, a Fourier-Motzkin elimination process in \eqref{EqRateRegion-z} yields:
\begin{subequations}
\label{EqRateRegion2b}
\begin{IEEEeqnarray}{rcl}
\label{EqRateRegion21}
R_{1}  & \leqslant & \min\left(\theta_{2,1},\theta_{6,1}+\theta_{1,2},\theta_{4,1}+\theta_{1,2}+\theta_{3,2}\right),  \\ 
\label{EqRateRegion22}
R_{2}   & \leqslant & \min\left(\theta_{2,2},\theta_{1,1}+a_{6,2},\theta_{1,1}+\theta_{3,1}+\theta_{4,2}\right), \\
\nonumber
R_{1}+R_{2}  & \leqslant & \min(\theta_{2,1}+\theta_{4,2}, \theta_{2,1}+a_{6,2}, \theta_{4,1}+\theta_{2,2}, \theta_{6,1}+\theta_{2,2}, \theta_{1,1}+\theta_{3,1}+\theta_{4,1}+\theta_{1,2}+\theta_{5,2},  \\
\nonumber
& & \theta_{1,1}+\theta_{7,1}+\theta_{1,2}+\theta_{5,2}, \theta_{1,1}+\theta_{4,1}+\theta_{1,2}+\theta_{7,2}, \theta_{1,1}+\theta_{5,1}+\theta_{1,2}+\theta_{3,2}+\theta_{4,2}, \\
\label{EqRateRegion23}
& & \theta_{1,1}+\theta_{5,1}+\theta_{1,2}+\theta_{5,2}, \theta_{1,1}+\theta_{7,1}+\theta_{1,2}+\theta_{4,2}), \\
\nonumber
2R_{1}+R_{2}  & \leqslant & \min(\theta_{2,1}+\theta_{4,1}+\theta_{1,2}+\theta_{7,2}, \theta_{1,1}+\theta_{4,1}+\theta_{7,1}+2\theta_{1,2}+\theta_{5,2}, \theta_{2,1}+\theta_{4,1}+\theta_{1,2}+\theta_{5,2}), \\
\label{EqRateRegion24}\\
\nonumber
R_{1}+2R_{2}  & \leqslant & \min(\theta_{1,1}+\theta_{5,1}+\theta_{2,2}+\theta_{4,2}, \theta_{1,1}+\theta_{7,1}+\theta_{2,2}+\theta_{4,2}, 2\theta_{1,1}+\theta_{5,1}+\theta_{1,2}+\theta_{4,2}+\theta_{7,2}), \\
\label{EqRateRegion25}
\end{IEEEeqnarray}
\end{subequations}
where $\theta_{l,i}$ are defined in \eqref{EqRateRegion-z} with $(l,i) \in \lbrace1,  \ldots, 7 \rbrace \times \lbrace 1, 2 \rbrace$.

\noindent
Consider that transmitter $i$ uses the following Gaussian input distribution: 
\begin{equation}
\label{EqXia}
X_i=U+U_i+V_i+X_{i,P}, 
\end{equation}
where $U$, $U_1$, $U_2$, $V_1$, $V_2$, $X_{1,P}$, and $X_{2,P}$ in \eqref{Eqprobdist} are mutually independent and distributed as follows:  
\begin{subequations}
\label{EqAchievGaussDef}
\begin{IEEEeqnarray}{rcl}
\label{EqU}
U&\sim&\mathcal{N}\left(0,\rho \right), \\
\label{EqUi}
U_i&\sim&\mathcal{N}\left(0,\mu_i \lambda_{i,C}\right), \\
\label{EqVi}
V_i&\sim&\mathcal{N}\left(0,(1-\mu_i)\lambda_{i,C}\right), \\
\label{EqXip}
X_{i,P}&\sim&\mathcal{N}\left(0,\lambda_{i,P}\right),
\end{IEEEeqnarray}
\end{subequations}
with
\begin{subequations}
\label{EqAchievGaussDef2}
\begin{equation}
\label{Eqpwrallc}
\rho+\lambda_{i,C}+\lambda_{i,P} = 1 \mbox{ and}
\end{equation}
\begin{IEEEeqnarray}{rcl}
\label{Eqpwrp}
\lambda_{i,P}&=&\min\left(\frac{1}{\INR_{ji}},1\right), 
\end{IEEEeqnarray}
\end{subequations}
where $\mu_i \in \left[0,1\right]$ and $\rho \in \left[0,\left(1-\max\left(\frac{1}{\INR_{12}},\frac{1}{\INR_{21}}\right) \right)^+\right]$. 

\noindent
The parameters $\rho$, $\mu_i$, and $\lambda_{i,P}$ define a particular coding scheme for transmitter $i$.
The assignment in \eqref{Eqpwrp} is based on the intuition obtained from the linear deterministic model, in which the power of the signal $X_{i,P}$ from transmitter $i$ to receiver $j$ must be observed at the noise level.  
From \eqref{Eqsignalyif}, \eqref{Eqsignalyib}, and \eqref{EqXia}, the right-hand side of the inequalities in \eqref{EqRateRegion-z} can be written in terms of $\overrightarrow{\SNR}_1$, $\overrightarrow{\SNR}_2$, $\INR_{12}$, $\INR_{21}$, $\overleftarrow{\SNR}_1$, $\overleftarrow{\SNR}_2$, $\rho$, $\mu_1$, and $\mu_2$ as follows:
\begin{subequations}
\label{EqIab}
\begin{IEEEeqnarray}{rcl}
\nonumber
\theta_{1,i} &=& I\left(\overleftarrow{Y}_i ; U_j  | U, X_i  \right) \\
\nonumber
&=& \frac{1}{2}\log \left(\frac{\overleftarrow{\SNR}_i\Big(b_{2,i}(\rho)+2\Big)+b_{1,i}(1)+1}{\overleftarrow{\SNR}_i\Big(\left(1\!-\!\mu_j\right)b_{2,i}(\rho)\!+\!2\Big)\!+\!b_{1,i}(1)\!+1}\right) \\
\label{EqIab8}
&=& a_{3,i}(\rho,\mu_j),
\end{IEEEeqnarray}
\begin{IEEEeqnarray}{rcl}
\nonumber
\theta_{2,i}&=& I\left(\overrightarrow{Y}_i; U, U_j,V_j, X_i \right) \\
\nonumber
&=&  \frac{1}{2}\log \Big(b_{1,i}(\rho)+1\Big)-\frac{1}{2} \\
\label{EqIab9}
&=& a_{2,i}(\rho), \\
\nonumber
\theta_{3,i}&=& I\left(\overrightarrow{Y}_i; V_j | U, U_j, X_i \right) \\
\nonumber
&=&  \frac{1}{2}\log \bigg(\Big(1-\mu_j\Big)b_{2,i}(\rho)+2\bigg)-\frac{1}{2} \\
\label{EqIab10}
&=& a_{4,i}(\rho,\mu_j), \\
\nonumber
\theta_{4,i}&=& I\left(\overrightarrow{Y}_i; X_i | U, U_i, U_j,V_i, V_j   \right) \\
\nonumber
&=&  \frac{1}{2}\log\left(\frac{\overrightarrow{\SNR}_{i}}{\INR_{ji}}+2\right)-\frac{1}{2} \\
\label{EqIab11}
&=& a_{1,i},\\
\nonumber
\theta_{5,i}&=& I\left(\overrightarrow{Y}_i; V_j, X_i | U, U_i, U_j, V_i   \right) \\
\nonumber
& & = \frac{1}{2}\log \left(2+\frac{\overrightarrow{\SNR}_{i}}{\INR_{ji}}+\Big(1-\mu_j\Big)b_{2,i}(\rho)\right)-\frac{1}{2} \\
\label{EqIab12}
&=& a_{5,i}(\rho,\mu_j),\\
\nonumber
\theta_{6,i} &=& I\left(\overrightarrow{Y}_i; X_i | U, U_i, U_j, V_j   \right) \\
\nonumber
& & = \frac{1}{2}\log \left(\frac{\overrightarrow{\SNR}_{i}}{\INR_{ji}}\bigg(\Big(1-\mu_i\Big)b_{2,j}(\rho)+1\right)+2\bigg)-\frac{1}{2} \\
\label{EqIab13}
&=& a_{6,i}(\rho,\mu_i),\\
\nonumber
\theta_{7,i}&=& I\left( \overrightarrow{Y}_i; V_j, X_i | U, U_i, U_j  \right)  \\
\nonumber
&=&  \frac{1}{2} \! \log\! \left(\!\frac{\overrightarrow{\SNR}_{i}}{\INR_{ji}}\!\bigg(\!\Big(1\!-\!\mu_i\Big)b_{2,j}(\rho)\!+\!1\bigg)\!+\!\Big(1\!-\!\mu_j\Big)b_{2,i}(\rho)\!+\!2\!\right)\!-\frac{1}{2}\\
\label{EqIab14}
&=& a_{7,i}(\rho,\mu_1, \mu_2).
\end{IEEEeqnarray}
\end{subequations}
Finally, plugging \eqref{EqIab} into  \eqref{EqRateRegion2b}  (after some trivial manipulations) yields the system of inequalities in Theorem~\ref{TheoremA-G-IC-NOF}.  
The sum-rate bound in \eqref{EqRateRegion23} can be simplified as follows:
\begin{IEEEeqnarray}{rcl}
\nonumber
R_{1}+R_{2}   &\leqslant&  \min \Big(a_{2,1}(\rho)+a_{1,2}, a_{1,1}+a_{2,2}(\rho), a_{3,1}(\rho,\mu_2)+a_{1,1}+a_{3,2}(\rho,\mu_1)+a_{7,2}(\rho,\mu_1,\mu_2),   \\
\nonumber
& & a_{3,1}(\rho,\mu_2)\!+\!a_{5,1}(\rho,\mu_2)\!+\!a_{3,2}(\rho,\mu_1)\!+\!a_{5,2}(\rho,\mu_1), \\
\label{EqRateRegion22simplified} 
& & a_{3,1}(\rho,\mu_2)+a_{7,1}(\rho,\mu_1,\mu_2)+a_{3,2}(\rho,\mu_1)+a_{1,2} \Big).
\end{IEEEeqnarray}
Note that this follows from the realization that $\max( a_{2,1}(\rho)+a_{1,2}, a_{1,1}+a_{2,2}(\rho)$ , $a_{3,1}(\rho,\mu_2)+a_{1,1}+a_{3,2}(\rho,\mu_1)$ $+$ $a_{7,2}(\rho,\mu_1,\mu_2)$, $a_{3,1}(\rho,\mu_2)+a_{5,1}(\rho,\mu_2)+a_{3,2}(\rho,\mu_1)+a_{5,2}(\rho,\mu_1), a_{3,1}(\rho,\mu_2)+a_{7,1}(\rho,\mu_1,\mu_2)+a_{3,2}(\rho,\mu_1)+a_{1,2})  \leqslant \min(a_{2,1}+a_{6,2}(\rho,\mu_2),a_{6,1}(\rho,\mu_1)+a_{2,2}(\rho), a_{3,1}(\rho,\mu_2)+a_{4,1}(\rho,\mu_2)+a_{1,1}+a_{3,2}(\rho,\mu_1)+a_{5,2}(\rho,\mu_1), a_{3,1}(\rho,\mu_2)+a_{7,1}(\rho,\mu_1,\mu_2)+a_{3,2}(\rho,\mu_1)+a_{5,2}(\rho,\mu_1)$, $a_{3,1}(\rho,\mu_2)$ $+$ $a_{5,1}(\rho,\mu_2)$ $+$ $a_{3,2}(\rho,\mu_1)$ $+$ $\theta_{3,2}$ $+$ $a_{1,2})$.
Therefore, the inequalities in \eqref{EqRateRegion2b} simplify into \eqref{EqRa-G-IC-NOF} and this completes the proof of Theorem~\ref{TheoremA-G-IC-NOF}. 

\newpage

\section{Proof of Converse} \label{App-C-G-IC-NOF}

This appendix provides a proof of the Theorem~\ref{TheoremC-G-IC-NOF}. The outer bounds \eqref{EqRic-12-G-IC-NOF} and \eqref{EqR1+R2c-12-G-IC-NOF} correspond to the outer bounds of the case of perfect channel-output feedback \cite{Suh-TIT-2011}. 
The bounds \eqref{EqRic-3-G-IC-NOF}, \eqref{EqR1+R2c-3g-G-IC-NOF} and \eqref{Eq2Ri+Rjc-g-G-IC-NOF} correspond to new outer bounds.
Before presenting the proof, consider the parameter $h_{ji,U}$, with $i \in \lbrace 1, 2 \rbrace$ and $j \in \lbrace 1, 2 \rbrace\setminus\lbrace i \rbrace$, defined as follows:
\begin{equation}
h_{ji,U}=\left\{
    \begin{array}{ll}
      0 & \textrm{if } (S_{1,i} \lor S_{2,i} \lor S_{3,i}) \\
    \sqrt{\frac{\INR_{ij}\INR_{ji}}{\overrightarrow{\SNR}_{j}}} & \textrm{if } (S_{4,i} \lor S_{5,i}),                                                                                                
  \end{array} \right.
\label{Eqdefhjiu}
\end{equation} 
where the events $S_{1,i}$, $S_{2,i}$, $S_{3,i}$, $S_{4,i}$, and $S_{5,i}$ are defined in \eqref{EqSi}. 
Consider also the following signals:
\begin{IEEEeqnarray}{rcl}
\label{EqsignalXiCG}
X_{i,C,n}&=& \sqrt{\INR_{ji}}X_{i,n}+\overrightarrow{Z}_{j,n} \mbox{ and }\\
\label{EqsignalXiUG}
X_{i,U,n}&=&h_{ji,U}X_{i,n}+\overrightarrow{Z}_{j,n},
\end{IEEEeqnarray} 
where,  $X_{i,n}$ and $\overrightarrow{Z}_{j,n}$ are the channel input of transmitter $i$ and the noise observed at receiver $j$ during a given channel use $n \in \lbrace1, 2,  \ldots, N \rbrace$, as described by \eqref{Eqsignalyif}. 
The following lemma is also fundamental in the present proof of Theorem~\ref{TheoremC-G-IC-NOF}.
\begin{lemma} \label{Lemmahelpsumandweightedrates} \emph{
For all $i \in \lbrace 1, 2 \rbrace$, with $j \in \lbrace 1, 2 \rbrace\setminus\lbrace i \rbrace$,  the following holds: 
\begin{IEEEeqnarray}{lcl}
\nonumber
 I\Big(\bs{X}_{i,C}, \bs{X}_{j,U}, \overleftarrow{\bs{Y}}_i, W_i; \overleftarrow{\bs{Y}}_j, W_j\Big) &\leqslant& h\left(\overleftarrow{\bs{Y}}_j| W_j \right) +\sum_{n=1}^{N}\Big[h\left(X_{j,U,n}|X_{i,C,n} \right)+h\left(\overleftarrow{Y}_{i,n}|X_{i,n},X_{j,U,n}\right)\\
\label{Eqhelpc1}
& & -\frac{3}{2}\log\left(2\pi e\right) \Big]. 
\end{IEEEeqnarray}
}
\end{lemma}
\begin{IEEEproof}
The proof of Lemma~\ref{Lemmahelpsumandweightedrates} is presented at the end of this Appendix. 
\end{IEEEproof}

\noindent
\textbf{Proof of \eqref{EqRic-3-G-IC-NOF}}:  From the assumption that the message index $W_i$ is i.i.d. following a uniform distribution over the set $\mathcal{W}_i$, the following holds for any $k \in \lbrace 1, 2, \ldots, N \rbrace$: 
\begin{IEEEeqnarray}{lcl}
\nonumber
NR_i &=& H\left(W_i\right)\\
\nonumber
&=& H\left(W_i|W_j\right)\\
\nonumber
&\stackrel{(a)}{\leqslant}& I\left(W_i;\overrightarrow{\bs{Y}}_i, \overleftarrow{\bs{Y}}_j |W_j\right)+N\delta(N)\\
\nonumber
&\leqslant& \sum_{n=1}^{N}\Big[h\Big(\overrightarrow{Y}_{i,n},\overleftarrow{Y}_{j,n} |W_j, \overrightarrow{\bs{Y}}_{i,(1:n-1)},\overleftarrow{\bs{Y}}_{j,(1:n-1)}, \bs{X}_{j,(1:n)}\Big)-h\left(\overrightarrow{Z}_{i,n} \right)-h\left(\overleftarrow{Z}_{j,n} \right)\Big]+N\delta(N)\\
\nonumber
&\leqslant& \sum_{n=1}^{N}\Big[h\left(\overrightarrow{Y}_{i,n},\overleftarrow{Y}_{j,n} | X_{j,n}\right)-h\left(\overrightarrow{Z}_{i,n} \right)-h\left(\overleftarrow{Z}_{j,n} \right)\Big]+N\delta(N) \\
\label{EqproofRi-1gc1}
&=& N\left[h\left(\overrightarrow{Y}_{i,k},\overleftarrow{Y}_{j,k} | X_{j,k}\right)-\log\left(2 \pi e\right)\right]+N\delta(N),
\end{IEEEeqnarray}
where (a) follows from Fano's inequality (see Figure~\ref{Fig:G-IC-NOF-Conv}a). 

\noindent
From \eqref{EqproofRi-1gc1}, the following holds in the asymptotic regime: 
\begin{IEEEeqnarray} {lcl}
\label{EqproofRi-1gc2}
\nonumber
R_i&\leqslant& h\left(\overrightarrow{Y}_{i,k},\overleftarrow{Y}_{j,k} | X_{j,k}\right)-\log\left(2 \pi e\right) \\
\label{EqproofRi-1gc3}
&\leqslant& \frac{1}{2}\log\Big(b_{3,i}+1\Big)+\frac{1}{2} \! \log \! \left( \! \frac{\Big(b_{3,i}+b_{4,j}(\rho)+1\Big)\overleftarrow{\SNR}_j}{\Big(b_{1,j}(\rho)+1\Big)\Big(b_{3,i}+\left(1-\rho^2\right)\Big)} \!+ \! 1 \! \right).
\end{IEEEeqnarray}

\noindent
This completes the proof of \eqref{EqRic-3-G-IC-NOF}. 

\noindent
\textbf{Proof of \eqref{EqR1+R2c-3g-G-IC-NOF}}: 

\noindent
From the assumption that the message indices $W_1$ and $W_2$  are i.i.d. following a uniform distribution over the sets $\mathcal{W}_1$ and $\mathcal{W}_2$ respectively, the following holds for any $k \in \lbrace 1, 2, \ldots, N \rbrace$:
\begin{IEEEeqnarray}{lcl}
\nonumber
N&\Big(& \! R_1+R_2\Big) = H\left(W_1\right)+H\left(W_2\right)\\
\nonumber
&\stackrel{(a)}{\leqslant}& I\left(W_1;\overrightarrow{\bs{Y}}_1,\overleftarrow{\bs{Y}}_1\right)+I\left(W_2;\overrightarrow{\bs{Y}}_2,\overleftarrow{\bs{Y}}_2\right)+N\delta(N)\\
\nonumber
&=& h\left(\overrightarrow{\bs{Y}}_1\right)+h\left(\overleftarrow{\bs{Z}}_1|\overrightarrow{\bs{Y}}_1\right)-h\left(\overleftarrow{\bs{Y}}_1|W_1\right)-h\left(\overrightarrow{\bs{Y}}_1|W_1,\overleftarrow{\bs{Y}}_1, \bs{X}_1\right)+h\left(\overrightarrow{\bs{Y}}_2\right)+h\left(\overleftarrow{\bs{Z}}_2|\overrightarrow{\bs{Y}}_2\right)\\
\nonumber
& & -h\left(\overleftarrow{\bs{Y}}_2|W_2\right)-h\left(\overrightarrow{\bs{Y}}_2|W_2,\overleftarrow{\bs{Y}}_2, \bs{X}_2\right)+N\delta(N)\\
\nonumber
&\leqslant& h\left(\overrightarrow{\bs{Y}}_1\right)+h\left(\overleftarrow{\bs{Z}}_1\right)-h\left(\overleftarrow{\bs{Y}}_1|W_1\right)-h\left(\bs{X}_{2,C}|W_1,\overleftarrow{\bs{Y}}_1, \bs{X}_1\right)+h\left(\overrightarrow{\bs{Y}}_2\right)+h\left(\overleftarrow{\bs{Z}}_2\right)-h\left(\overleftarrow{\bs{Y}}_2|W_2\right)\\
\nonumber
& & -h\left(\bs{X}_{1,C}|W_2,\overleftarrow{\bs{Y}}_2, \bs{X}_2\right)+N\delta(N)\\
\nonumber
&=& h\left(\overrightarrow{\bs{Y}}_1\right)-h\left(\overleftarrow{\bs{Y}}_1|W_1\right)-h\left(\bs{X}_{2,C},\overrightarrow{\bs{Z}}_{2} |W_1,\overleftarrow{\bs{Y}}_1, \bs{X}_1\right)+h\left(\overrightarrow{\bs{Z}}_{2} |W_1,\overleftarrow{\bs{Y}}_1, \bs{X}_1, \bs{X}_{2,C}\right)+h\left(\overrightarrow{\bs{Y}}_2\right)\\
\nonumber
& & -h\left(\overleftarrow{\bs{Y}}_2|W_2\right)-h\left(\bs{X}_{1,C}, \overrightarrow{\bs{Z}}_{1}|W_2,\overleftarrow{\bs{Y}}_2, \bs{X}_2\right)+h\left(\overrightarrow{\bs{Z}}_{1}|W_2,\overleftarrow{\bs{Y}}_2, \bs{X}_2, \bs{X}_{1,C}\right)+N\log\left(2\pi e\right)\\
\nonumber
& & +N\delta(N)\\
\nonumber
&=& h\left(\overrightarrow{\bs{Y}}_1\right)-h\left(\overleftarrow{\bs{Y}}_1|W_1\right)-h\left(\bs{X}_{2,C},\bs{X}_{1,U} |W_1,\overleftarrow{\bs{Y}}_1, \bs{X}_1\right)+h\left(\overrightarrow{\bs{Z}}_{2} |W_1,\overleftarrow{\bs{Y}}_1, \bs{X}_1, \bs{X}_{2,C}\right)\\
\nonumber
& & +h\left(\overrightarrow{\bs{Y}}_2\right)-h\left(\overleftarrow{\bs{Y}}_2|W_2\right)-h\left(\bs{X}_{1,C}, \bs{X}_{2,U}|W_2,\overleftarrow{\bs{Y}}_2, \bs{X}_2\right)+h\left(\overrightarrow{\bs{Z}}_{1}|W_2,\overleftarrow{\bs{Y}}_2, \bs{X}_2, \bs{X}_{1,C}\right)\\
\nonumber
& & +N\log\left(2\pi e\right)+N\delta(N)\\
\nonumber
&=& h\left(\overrightarrow{\bs{Y}}_1\right) \! - \! h\left(\overleftarrow{\bs{Y}}_1|W_1\right) \!+ \! \Big[I\left(\bs{X}_{2,C},\bs{X}_{1,U} ; W_1,\overleftarrow{\bs{Y}}_1\right)-h\left(\bs{X}_{2,C},\bs{X}_{1,U}\right)\Big]+h\left(\overrightarrow{\bs{Y}}_2\right)\\
\nonumber
& & -h\left(\overleftarrow{\bs{Y}}_2|W_2\right)+\left[I\left(\bs{X}_{1,C}, \bs{X}_{2,U};W_2,\overleftarrow{\bs{Y}}_2\right)-h\left(\bs{X}_{1,C}, \bs{X}_{2,U}\right)\right]\\
\nonumber
& &  +h\left(\overrightarrow{\bs{Z}}_{1}|W_2,\overleftarrow{\bs{Y}}_2, \bs{X}_2, \bs{X}_{1,C}\right)+h\left(\overrightarrow{\bs{Z}}_{2} |W_1,\overleftarrow{\bs{Y}}_1, \bs{X}_1, \bs{X}_{2,C}\right)+N\log\left(2\pi e\right)+N\delta(N) \\ 
\nonumber
&\leqslant& h\left(\overrightarrow{\bs{Y}}_1\right) \! - \! h\left(\overleftarrow{\bs{Y}}_1|W_1\right) \! + \! \Big[I\left(\bs{X}_{2,C},\bs{X}_{1,U} ; W_1,\overleftarrow{\bs{Y}}_1\right)-h\left(\bs{X}_{2,C},\bs{X}_{1,U}\right)\Big]+h\left(\overrightarrow{\bs{Y}}_2\right)\\
\nonumber
& & -h\left(\overleftarrow{\bs{Y}}_2|W_2\right)+\left[I\left(\bs{X}_{1,C}, \bs{X}_{2,U};W_2,\overleftarrow{\bs{Y}}_2\right)-h\left(\bs{X}_{1,C}, \bs{X}_{2,U}\right)\right] + \!   \Big[h\left(\bs{X}_{2,C},\bs{X}_{1,U}| \overrightarrow{\bs{Y}}_{2}\right) \! -\\
\nonumber
& &    \! h\left(\bs{X}_{2,C},\bs{X}_{1,U}| \overrightarrow{\bs{Y}}_{2}, \bs{X}_{1}, \bs{X}_{2}\right) \!  \Big] + \!  \Big[h\left(\bs{X}_{1,C},\bs{X}_{2,U}| \overrightarrow{\bs{Y}}_{1}\right)  \! - \!  h\left(\bs{X}_{1,C},\bs{X}_{2,U}| \overrightarrow{\bs{Y}}_{1}, \bs{X}_{2}, \bs{X}_{1}\right) \! \Big]\\
\nonumber
& & +h\left(\overrightarrow{\bs{Z}}_{1}|W_2,\overleftarrow{\bs{Y}}_2, \bs{X}_2, \bs{X}_{1,C}\right) +h\left(\overrightarrow{\bs{Z}}_{2} |W_1,\overleftarrow{\bs{Y}}_1, \bs{X}_1, \bs{X}_{2,C}\right)+N\log\left(2\pi e\right)+N\delta(N) \\
\nonumber
&\stackrel{(b)}{=}& h\left(\overrightarrow{\bs{Y}}_1|\bs{X}_{1,C},\bs{X}_{2,U}\right)-h\left(\overleftarrow{\bs{Y}}_1|W_1\right)+I\left(\bs{X}_{2,C},\bs{X}_{1,U} ; W_1,\overleftarrow{\bs{Y}}_1\right)+h\left(\overrightarrow{\bs{Y}}_2|\bs{X}_{2,C},\bs{X}_{1,U}\right)\\
\nonumber
& & -h\left(\overleftarrow{\bs{Y}}_2|W_2\right)+I\left(\bs{X}_{1,C}, \bs{X}_{2,U};W_2,\overleftarrow{\bs{Y}}_2\right)-h\left(\overrightarrow{\bs{Z}}_1,\overrightarrow{\bs{Z}}_2| \overrightarrow{\bs{Y}}_{2}, \bs{X}_{1}, \bs{X}_{2}\right)\\
\nonumber
& & -h\left(\overrightarrow{\bs{Z}}_2,\overrightarrow{\bs{Z}}_1| \overrightarrow{\bs{Y}}_{1}, \bs{X}_{2}, \bs{X}_{1}\right)+h\left(\overrightarrow{\bs{Z}}_{1}|W_2,\overleftarrow{\bs{Y}}_2, \bs{X}_2, \bs{X}_{1,C}\right)+h\left(\overrightarrow{\bs{Z}}_{2} |W_1,\overleftarrow{\bs{Y}}_1, \bs{X}_1, \bs{X}_{2,C}\right) \\
\nonumber
& & +N\log\left(2\pi e\right)+N\delta(N)\\
\nonumber
&\stackrel{(c)}{\leqslant}& h\left(\overrightarrow{\bs{Y}}_1|\bs{X}_{1,C},\bs{X}_{2,U}\right)-h\left(\overleftarrow{\bs{Y}}_1|W_1\right)+I\left(\bs{X}_{2,C},\bs{X}_{1,U}; W_1,\overleftarrow{\bs{Y}}_1\right)+h\left(\overrightarrow{\bs{Y}}_2|\bs{X}_{2,C},\bs{X}_{1,U}\right)\\
\nonumber
& & -h\left(\overleftarrow{\bs{Y}}_2|W_2\right)+I\left(\bs{X}_{1,C}, \bs{X}_{2,U};W_2,\overleftarrow{\bs{Y}}_2\right)+N\log\left(2\pi e\right)+N\delta(N)\\
\nonumber
&\leqslant& h\left(\overrightarrow{\bs{Y}}_1|\bs{X}_{1,C},\bs{X}_{2,U}\right)-h\left(\overleftarrow{\bs{Y}}_1|W_1\right)+I\left(\bs{X}_{2,C},\bs{X}_{1,U}, W_2,\overleftarrow{\bs{Y}}_2; W_1,\overleftarrow{\bs{Y}}_1\right)\\
\nonumber
& & +h\left(\overrightarrow{\bs{Y}}_2|\bs{X}_{2,C},\bs{X}_{1,U}\right)-h\left(\overleftarrow{\bs{Y}}_2|W_2\right)+I\left(\bs{X}_{1,C}, \bs{X}_{2,U}, W_1,\overleftarrow{\bs{Y}}_1;W_2,\overleftarrow{\bs{Y}}_2\right)+N\log\left(2\pi e\right)\\
\nonumber
& & +N\delta(N)
\end{IEEEeqnarray}
\begin{IEEEeqnarray}{lcl}
\nonumber
&\stackrel{(d)}{\leqslant}& \sum_{n=1}^{N}\Big[h\left(\overrightarrow{Y}_{1,n}|\bs{X}_{1,C},\bs{X}_{2,U}, \overrightarrow{\bs{Y}}_{1,(1:n-1)}\right)+h\left(X_{1,U,n}|X_{2,C,n} \right)+h\left(\overleftarrow{Y}_{2,n}|X_{2,n},X_{1,U,n}\right)\\
\nonumber
& & +h\left(\overrightarrow{Y}_{2,n}|\bs{X}_{2,C},\bs{X}_{1,U} \overrightarrow{\bs{Y}}_{2,(1:n-1)}\right)+h\left(X_{2,U,n}|X_{1,C,n} \right)+h\left(\overleftarrow{Y}_{1,n}|X_{1,n},X_{2,U,n}\right)\\
\nonumber
& & -3\log\left(2\pi e\right)\Big]+N\log\left(2\pi e\right)+N\delta(N)\\
\nonumber
&\leqslant& \sum_{n=1}^{N}\Big[h\left(\overrightarrow{Y}_{1,n}|X_{1,C,n},X_{2,U,n}\right)+h\left(X_{1,U,n}|X_{2,C,n} \right)+h\left(\overleftarrow{Y}_{2,n}|X_{2,n},X_{1,U,n}\right)\\
\nonumber
& & +h\left(\overrightarrow{Y}_{2,n}|X_{2,C,n},X_{1,U,n}\right)+h\left(X_{2,U,n}|X_{1,C,n} \right)+h\left(\overleftarrow{Y}_{1,n}|X_{1,n},X_{2,U,n}\right)-3\log\left(2\pi e\right)\Big]\\
\nonumber
& & +N\log\left(2\pi e\right)+N\delta(N)\\
\nonumber
&=& N\Big[h\left(\overrightarrow{Y}_{1,k}|X_{1,C,k},X_{2,U,k}\right)+h\left(X_{1,U,k}|X_{2,C,k} \right)+h\left(\overleftarrow{Y}_{2,k}|X_{2,k},X_{1,U,k}\right)\\
\nonumber
& & +h\left(\overrightarrow{Y}_{2,k}|X_{2,C,k},X_{1,U,k}\right)+h\left(X_{2,U,k}|X_{1,C,k} \right)+h\left(\overleftarrow{Y}_{1,k}|X_{1,k},X_{2,U,k}\right)-3\log\left(2 \pi e\right)\Big]\\
\label{EqproofR1+R2gc1}
& & +N\log\left(2\pi e\right)+N\delta(N), 
\end{IEEEeqnarray}
where
(a) follows from Fano's inequality (see Figure~\ref{Fig:G-IC-NOF-Conv}b); 
(b) follows from the fact that $h\left(\overrightarrow{\bs{Y}}_i\right)-h\left(\bs{X}_{i,C},\bs{X}_{j,U} \right)+h\left(\bs{X}_{i,C},\bs{X}_{j,U}|\overrightarrow{\bs{Y}}_i \right)=h\left(\overrightarrow{\bs{Y}}_i|\bs{X}_{i,C},\bs{X}_{j,U}\right)$; 
(c) follows from the fact that $\qquad$ $h\Big(\overrightarrow{\bs{Z}}_{i}|W_j$, $\overleftarrow{\bs{Y}}_j, \bs{X}_j, \bs{X}_{i,C}\Big) -h\left(\overrightarrow{\bs{Z}}_i,\overrightarrow{\bs{Z}}_j| \overrightarrow{\bs{Y}}_{j}, \bs{X}_{i}, \bs{X}_{j}\right) \leqslant 0$; and 
(d) follows from Lemma~\ref{Lemmahelpsumandweightedrates}.

\noindent
From \eqref{EqproofR1+R2gc1},  the following holds in the asymptotic regime for any $k \in \lbrace 1, 2, \ldots, N \rbrace$:   
\begin{IEEEeqnarray} {lcl}
\nonumber
R_1+R_2 &\leqslant& h\left(\overrightarrow{Y}_{1,k}|X_{1,C,k},X_{2,U,k}\right)+h\left(X_{1,U,k}|X_{2,C,k} \right)+h\left(\overleftarrow{Y}_{2,k}|X_{2,k},X_{1,U,k}\right)\\
\nonumber
& & +h\left(\overrightarrow{Y}_{2,k}|X_{2,C,k},X_{1,U,k}\right)+h\left(X_{2,U,k}|X_{1,C,k} \right)+h\left(\overleftarrow{Y}_{1,k}|X_{1,k},X_{2,U,k}\right)-2\log\left(2 \pi e\right) \\
\nonumber
&\leqslant&\frac{1}{2}\log\left(\det\left(\textrm{Var}\left(\overrightarrow{Y}_{1,k},X_{1,C,k},X_{2,U,k}\right)\right)\right)-\frac{1}{2}\log\left(\INR_{12}+1\right)\\
\nonumber
& & +\frac{1}{2}\log\left(\det\left(\textrm{Var}\left(\overleftarrow{Y}_{2,k},X_{2,k},X_{1,U,k}\right)\right)\right)-\frac{1}{2}\log\left(\det\left(\textrm{Var}\left(X_{2,k},X_{1,U,k}\right)\right)\right)\\
\nonumber
& & +\frac{1}{2}\log\left(\det\left(\textrm{Var}\left(\overrightarrow{Y}_{2,k},X_{2,C,k},X_{1,U,k}\right)\right)\right)-\frac{1}{2}\log\left(\INR_{21}+1\right)\\
\nonumber
& & +\frac{1}{2}\log\left(\det\left(\textrm{Var}\left(\overleftarrow{Y}_{1,k},X_{1,k},X_{2,U,k}\right)\right)\right)-\frac{1}{2}\log\left(\det\left(\textrm{Var}\left(X_{1,k},X_{2,U,k}\right)\right)\right)+\log\left(2\pi e\right),\\
\label{EqproofR1+R2gc3}
\end{IEEEeqnarray}
where for all $i \in \lbrace 1, 2 \rbrace$, with $j \in {\lbrace 1, 2 \rbrace\setminus\lbrace i \rbrace}$ the following holds:
\begin{subequations}
\label{EqVarAux1}
\begin{IEEEeqnarray} {rcl}
\nonumber
\det \Big(\textrm{Var}\Big(\overrightarrow{Y}_{j,k},X_{j,C,k}&,&X_{i,U,k}\Big)\Big) = \overrightarrow{\SNR}_{j}+\INR_{ji}+h_{ji,U}^2-2h_{ji,U}\sqrt{\INR_{ji}}\\
\nonumber
& & +\left(1-\rho^2\right)\Big(\INR_{ij}\INR_{ji}+h_{ji,U}^2\left(\overrightarrow{\SNR}_{j}+\INR_{ij} \right)-2h_{ji,U}\INR_{ij}\sqrt{\INR_{ji}} \Big)\\
\label{EqvarYjXjcgXiug}
& & +2\rho\sqrt{\overrightarrow{\SNR}_{j}}\left(\sqrt{\INR_{ji}}-h_{ji,U}\right),\\
\nonumber
\det \Big(\textrm{Var}\Big(\overleftarrow{Y}_{j,k},X_{j,k}&,&X_{i,U,k}\Big)\Big) = 1+h_{ji,U}^2\left(1-\rho^2\right) \\
\label{EqvarYjXjXiug}
& & +\frac{\overleftarrow{\SNR}_j\left(1-\rho^2\right)\left(h_{ji,U}^2-2h_{ji,U}\sqrt{\INR_{ji}}+\INR_{ji} \right )}{\left(\overrightarrow{\SNR}_j + 2\rho\sqrt{\overrightarrow{\SNR}_j\INR_{ji}}+\INR_{ji}+1\right)}, \mbox{and }  \\
\label{EqvarXjXiug}
\det \Big(\textrm{Var}\Big(X_{j,k}&,&X_{i,U,k}\Big)\Big) = 1+\left(1-\rho^2 \right)h_{ji,U}^2. 
\end{IEEEeqnarray}
\end{subequations}
The expressions in \eqref{EqVarAux1} depend on $S_{1,i}$, $S_{2,i}$, $S_{3,i}$, $S_{4,i}$, and $S_{5,i}$ via the parameter $h_{ji,U}$ in  \eqref{Eqdefhjiu}.
Hence, the following cases are identified: 

\noindent
\textbf{Case 1:  $(S_{1,2} \lor S_{2,2} \lor S_{5,2})\land(S_{1,1} \lor S_{2,1} \lor S_{5,1})$}.  From \eqref{Eqdefhjiu}, it follows that $h_{12,U}=0$ and $h_{21,U}=0$. Therefore, plugging the expression \eqref{EqVarAux1} into \eqref{EqproofR1+R2gc3} yields \eqref{Eqconv61}. 

\noindent
\textbf{Case 2: $(S_{1,2} \lor S_{2,2} \lor S_{5,2})\land(S_{3,1} \lor S_{4,1})$}.  From \eqref{Eqdefhjiu}, it follows that $h_{12,U}=0$ and $h_{21,U}= \sqrt{\frac{\INR_{12}\INR_{21}}{\overrightarrow{\SNR}_{2}}}$. Therefore, plugging the expression \eqref{EqVarAux1} into \eqref{EqproofR1+R2gc3} yields \eqref{Eqconv62}. 

\noindent
\textbf{Case 3: $(S_{3,2} \lor S_{4,2})\land(S_{1,1} \lor S_{2,1} \lor S_{5,1})$}.  From \eqref{Eqdefhjiu}, it follows that $h_{12,U}=\sqrt{\frac{\INR_{12}\INR_{21}}{\overrightarrow{\SNR}_{1}}}$ and $h_{21,U}=0$. Therefore, plugging the expression \eqref{EqVarAux1} into \eqref{EqproofR1+R2gc3} yields \eqref{Eqconv63}. 

\noindent
\textbf {Case 4: $(S_{3,2} \lor S_{4,2})\land(S_{3,1} \lor S_{4,1})$}. From \eqref{Eqdefhjiu}, it follows that $h_{12,U}= \sqrt{\frac{\INR_{12}\INR_{21}}{\overrightarrow{\SNR}_{1}}}$ and $h_{21,U}=\sqrt{\frac{\INR_{12}\INR_{21}}{\overrightarrow{\SNR}_{2}}}$. Therefore, plugging the expression \eqref{EqVarAux1} into \eqref{EqproofR1+R2gc3} yields \eqref{Eqconv64}.

\noindent
This completes the proof of \eqref{EqR1+R2c-3g-G-IC-NOF}. 

\noindent
\textbf{Proof of \eqref{Eq2Ri+Rjc-g-G-IC-NOF}}:  From the assumption that the message indices $W_i$ and $W_j$  are i.i.d. following a uniform distribution over the sets $\mathcal{W}_i$ and $\mathcal{W}_j$ respectively, for all $i \in \lbrace1,2\rbrace$, with $j \in \lbrace1,2 \rbrace \setminus\lbrace i \rbrace$, the following holds for any $k \in \lbrace 1, 2, \ldots, N \rbrace$:
\begin{IEEEeqnarray}{lcl}
\nonumber
N &\Big(&2 R_i+R_j\Big) = 2H\left(W_i\right)+H\left(W_j\right)\\
\nonumber
&\stackrel{(a)}{=}& H\left(W_i\right)+H\left(W_i|W_j\right)+H\left(W_j\right)\\
\nonumber
&\stackrel{(b)}{\leqslant}& I\left(W_i;\overrightarrow{\bs{Y}}_i,\overleftarrow{\bs{Y}}_i\right)+I\left(W_i;\overrightarrow{\bs{Y}}_i,\overleftarrow{\bs{Y}}_j|W_j\right)+I\left(W_j;\overrightarrow{\bs{Y}}_j,\overleftarrow{\bs{Y}}_j\right)+N\delta(N)\\
\nonumber
&\leqslant& h\left(\overrightarrow{\bs{Y}}_i\right)+h\left(\overleftarrow{\bs{Z}}_i\right)-h\left(\overleftarrow{\bs{Y}}_i|W_i\right)-h\left(\overrightarrow{\bs{Y}}_i|W_i,\overleftarrow{\bs{Y}}_i\right)+h\left(\overleftarrow{\bs{Y}}_j|W_j\right)-h\left(\overleftarrow{\bs{Y}}_j|W_i,W_j\right)\\
\nonumber
& & +I\left(W_i;\overrightarrow{\bs{Y}}_i|W_j,\overleftarrow{\bs{Y}}_j\right)+h\left(\overrightarrow{\bs{Y}}_j\right)+h\left(\overleftarrow{\bs{Z}}_j\right)-h\left(\overleftarrow{\bs{Y}}_j|W_j\right)-h\left(\overrightarrow{\bs{Y}}_j|W_j,\overleftarrow{\bs{Y}}_j\right)+N\delta(N)\\
\nonumber
&=& h\left(\overrightarrow{\bs{Y}}_i\right)-h\left(\overleftarrow{\bs{Y}}_i|W_i\right)-h\left(\overrightarrow{\bs{Y}}_i|W_i,\overleftarrow{\bs{Y}}_i, \bs{X}_i\right)-h\left(\overleftarrow{\bs{Y}}_j|W_i,W_j\right)+I\left(W_i;\overrightarrow{\bs{Y}}_i|W_j,\overleftarrow{\bs{Y}}_j\right)\\
\nonumber
& & +h\left(\overrightarrow{\bs{Y}}_j\right)-h\left(\overrightarrow{\bs{Y}}_j|W_j,\overleftarrow{\bs{Y}}_j, \bs{X}_j\right)+N\log\left(2\pi e\right)+N\delta(N)\\
\nonumber
&\leqslant& h\left(\overrightarrow{\bs{Y}}_i\right)-h\left(\overleftarrow{\bs{Y}}_i|W_i\right)-h\left(\overrightarrow{\bs{Y}}_i|W_i,\overleftarrow{\bs{Y}}_i, \bs{X}_i\right)+I\left(W_i;\overrightarrow{\bs{Y}}_i|W_j,\overleftarrow{\bs{Y}}_j\right)+h\left(\overrightarrow{\bs{Y}}_j\right)\\
\nonumber
& & -h\left(\overrightarrow{\bs{Y}}_j|W_j,\overleftarrow{\bs{Y}}_j, \bs{X}_j\right)+N\log\left(2\pi e\right)+N\delta(N)\\ 
\nonumber
&\stackrel{(c)}{=}& h\left(\overrightarrow{\bs{Y}}_i\right)-h\left(\overleftarrow{\bs{Y}}_i|W_i\right)-h\left(\bs{X}_{j,C}|W_i,\overleftarrow{\bs{Y}}_i, \bs{X}_i\right)+I\left(W_i;\overrightarrow{\bs{Y}}_i|W_j,\overleftarrow{\bs{Y}}_j\right)+h\left(\overrightarrow{\bs{Y}}_j\right)\\
\nonumber
& & -h\left(\bs{X}_{i,C}|W_j,\overleftarrow{\bs{Y}}_j, \bs{X}_j\right)+N\log\left(2\pi e\right)+N\delta(N)\\
\nonumber
&=& h\left(\overrightarrow{\bs{Y}}_i\right)-h\left(\overleftarrow{\bs{Y}}_i|W_i\right)-h\left(\bs{X}_{j,C}, \overrightarrow{\bs{Z}}_j|W_i,\overleftarrow{\bs{Y}}_i, \bs{X}_i\right)+h\left(\overrightarrow{\bs{Z}}_j|W_i,\overleftarrow{\bs{Y}}_i, \bs{X}_i,\bs{X}_{j,C}\right)\\
\nonumber
& & +I\left(W_i;\overrightarrow{\bs{Y}}_i|W_j,\overleftarrow{\bs{Y}}_j\right)+h\left(\overrightarrow{\bs{Y}}_j\right)-h\left(\bs{X}_{i,C}|W_j,\overleftarrow{\bs{Y}}_j, \bs{X}_j\right)+N\log\left(2\pi e\right)+N\delta(N)\\
\nonumber
&\stackrel{(d)}{=}& h\left(\overrightarrow{\bs{Y}}_i\right)-h\left(\overleftarrow{\bs{Y}}_i|W_i\right)-h\left(\bs{X}_{j,C}, \bs{X}_{i,U}|W_i,\overleftarrow{\bs{Y}}_i, \bs{X}_i\right)+h\left(\overrightarrow{\bs{Z}}_j|W_i,\overleftarrow{\bs{Y}}_i, \bs{X}_i,\bs{X}_{j,C}\right)\\
\nonumber
& & +I\left(W_i;\overrightarrow{\bs{Y}}_i|W_j,\overleftarrow{\bs{Y}}_j\right)+h\left(\overrightarrow{\bs{Y}}_j\right)-h\left(\bs{X}_{i,C}|W_j,\overleftarrow{\bs{Y}}_j, \bs{X}_j\right)+N\log\left(2\pi e\right)+N\delta(N)\\
\nonumber
&\leqslant& h\left(\overrightarrow{\bs{Y}}_i\right)-h\left(\overleftarrow{\bs{Y}}_i|W_i\right)-h\left(\bs{X}_{j,C}, \bs{X}_{i,U}|W_i,\overleftarrow{\bs{Y}}_i\right)+h\left(\overrightarrow{\bs{Z}}_j|W_i,\overleftarrow{\bs{Y}}_i, \bs{X}_i,\bs{X}_{j,C}\right)\\
\nonumber
& & +I\left(W_i;\overrightarrow{\bs{Y}}_i, \bs{X}_{i,C}|W_j,\overleftarrow{\bs{Y}}_j\right)+h\left(\overrightarrow{\bs{Y}}_j\right)-h\left(\bs{X}_{i,C}|W_j,\overleftarrow{\bs{Y}}_j\right)+N\log\left(2\pi e\right)+N\delta(N)\\ 
\nonumber
&=& h\left(\overrightarrow{\bs{Y}}_i\right)-h\left(\overleftarrow{\bs{Y}}_i|W_i\right)-h\left(\bs{X}_{j,C}, \bs{X}_{i,U}|W_i,\overleftarrow{\bs{Y}}_i\right)+h\left(\overrightarrow{\bs{Z}}_j|W_i,\overleftarrow{\bs{Y}}_i, \bs{X}_i,\bs{X}_{j,C}\right)\\
\nonumber
& & +h\left(\overrightarrow{\bs{Y}}_i|W_j,\overleftarrow{\bs{Y}}_j, \bs{X}_{i,C}\right)-h\left(\overrightarrow{\bs{Y}}_i, \bs{X}_{i,C}|W_i, W_j,\overleftarrow{\bs{Y}}_j\right)+h\left(\overrightarrow{\bs{Y}}_j\right)+N\log\left(2\pi e\right)+N\delta(N)
\end{IEEEeqnarray}
\begin{IEEEeqnarray}{lcl}
\nonumber
&\stackrel{(e)}{\leqslant}& h\left(\overrightarrow{\bs{Y}}_i\right)-h\left(\overleftarrow{\bs{Y}}_i|W_i\right)-h\left(\bs{X}_{j,C}, \bs{X}_{i,U}|W_i,\overleftarrow{\bs{Y}}_i\right)+h\left(\overrightarrow{\bs{Z}}_j|W_i,\overleftarrow{\bs{Y}}_i, \bs{X}_i,\bs{X}_{j,C}\right)\\
\nonumber
& & +h\left(\overrightarrow{\bs{Y}}_i|W_j,\overleftarrow{\bs{Y}}_j, \bs{X}_{i,C}\right)-h\left(\overrightarrow{\bs{Y}}_i, \bs{X}_{i,C}|W_i, W_j,\overleftarrow{\bs{Y}}_j, \bs{X}_{i}, \bs{X}_{j}\right)+h\left(\overrightarrow{\bs{Y}}_j\right)+N\log\left(2\pi e\right)\\
\nonumber
& & +N\delta(N)\\
\nonumber
&=& h\left(\overrightarrow{\bs{Y}}_i\right)-h\left(\overleftarrow{\bs{Y}}_i|W_i\right)-h\left(\bs{X}_{j,C}, \bs{X}_{i,U}|W_i,\overleftarrow{\bs{Y}}_i\right)+h\left(\overrightarrow{\bs{Z}}_j|W_i,\overleftarrow{\bs{Y}}_i, \bs{X}_i,\bs{X}_{j,C}\right)\\
\nonumber
& & +h\left(\overrightarrow{\bs{Y}}_i|W_j,\overleftarrow{\bs{Y}}_j, \bs{X}_{i,C}\right)-h\left(\overrightarrow{\bs{Z}}_i, \overrightarrow{\bs{Z}}_j|W_i, W_j,\overleftarrow{\bs{Y}}_j, \bs{X}_{i}, \bs{X}_{j}\right)+h\left(\overrightarrow{\bs{Y}}_j\right)+N\log\left(2\pi e\right)\\
\nonumber
& & +N\delta(N)\\
\nonumber
&\stackrel{(f)}{\leqslant}& h\left(\overrightarrow{\bs{Y}}_i\right)-h\left(\overleftarrow{\bs{Y}}_i|W_i\right)-h\left(\bs{X}_{j,C}, \bs{X}_{i,U}|W_i,\overleftarrow{\bs{Y}}_i\right)+h\left(\overrightarrow{\bs{Y}}_i|W_j,\overleftarrow{\bs{Y}}_j, \bs{X}_{i,C}\right)+h\left(\overrightarrow{\bs{Y}}_j\right)\\
\nonumber
& & +N\log\left(2\pi e\right)+N\delta(N)\\
\nonumber
&\leqslant& h\left(\overrightarrow{\bs{Y}}_i\right)-h\left(\overleftarrow{\bs{Y}}_i|W_i\right)+I\left(\bs{X}_{j,C}, \bs{X}_{i,U};W_i,\overleftarrow{\bs{Y}}_i\right)-h\left(\bs{X}_{j,C}, \bs{X}_{i,U}\right)+h\left(\overrightarrow{\bs{Y}}_i|W_j,\overleftarrow{\bs{Y}}_j, \bs{X}_{i,C}\right)\\
\nonumber
& & +h\left(\overrightarrow{\bs{Y}}_j\right)+h\left(\bs{X}_{j,C}, \bs{X}_{i,U}|\overrightarrow{\bs{Y}}_j\right)+N\log\left(2\pi e\right)+N\delta(N)\\
\nonumber
&\stackrel{(g)}{=}& h\left(\overrightarrow{\bs{Y}}_i\right)-h\left(\overleftarrow{\bs{Y}}_i|W_i\right)+I\left(\bs{X}_{j,C}, \bs{X}_{i,U};W_i,\overleftarrow{\bs{Y}}_i\right)+h\left(\overrightarrow{\bs{Y}}_i|W_j,\overleftarrow{\bs{Y}}_j, \bs{X}_{i,C}\right)+h\left(\overrightarrow{\bs{Y}}_j|\bs{X}_{j,C}, \bs{X}_{i,U}\right)\\
\nonumber
& & +N\log\left(2\pi e\right)+N\delta(N)\\
\nonumber
&\leqslant& h\left(\overrightarrow{\bs{Y}}_i\right)-h\left(\overleftarrow{\bs{Y}}_i|W_i\right)+I\left(\bs{X}_{j,C}, \bs{X}_{i,U}, W_j,\overleftarrow{\bs{Y}}_j;W_i,\overleftarrow{\bs{Y}}_i\right)+h\left(\overrightarrow{\bs{Y}}_i|W_j,\overleftarrow{\bs{Y}}_j, \bs{X}_{i,C}\right)\\
\nonumber
& & +h\left(\overrightarrow{\bs{Y}}_j|\bs{X}_{j,C}, \bs{X}_{i,U}\right)+N\log\left(2\pi e\right)+N\delta(N)\\
\nonumber
&\stackrel{(h)}{\leqslant}& h\left(\overrightarrow{\bs{Y}}_i\right)+\sum_{n=1}^{N}\Big[h\left(X_{i,U,n}|X_{j,C,n} \right)+h\left(\overleftarrow{\bs{Y}}_{j,n}|X_{j,n},X_{i,U,n}\right)-\frac{3}{2}\log\left(2\pi e\right) \Big]\\
\nonumber
& & +h\left(\overrightarrow{\bs{Y}}_i|W_j,\overleftarrow{\bs{Y}}_j, \bs{X}_{i,C}\right)+h\left(\overrightarrow{\bs{Y}}_j|\bs{X}_{j,C}, \bs{X}_{i,U}\right)+N\log\left(2\pi e\right)+N\delta(N)\\
\nonumber
&\stackrel{(i)}{\leqslant}& h\left(\overrightarrow{\bs{Y}}_i\right)+\sum_{n=1}^{N}\Big[h\left(X_{i,U,n}|X_{j,C,n} \right)+h\left(\overleftarrow{\bs{Y}}_{j,n}|X_{j,n},X_{i,U,n}\right)-\frac{3}{2}\log\left(2\pi e\right) \Big]+h\left(\overrightarrow{\bs{Y}}_i|\bs{X}_{i,C}, \bs{X}_j \right)\\
\nonumber
& & +h\left(\overrightarrow{\bs{Y}}_j|\bs{X}_{j,C}, \bs{X}_{i,U}\right)+N\log\left(2\pi e\right)+N\delta(N)\\
\nonumber
&\leqslant& \sum_{n=1}^{N}\Big[h\left(\overrightarrow{Y}_{i,n}\right)+h\left(X_{i,U,n}|X_{j,C,n} \right)+h\left(\overleftarrow{Y}_{j,n}|X_{j,n},X_{i,U,n}\right)-\frac{3}{2}\log\left(2\pi e\right)\\
\nonumber
& & +h\left(\overrightarrow{Y}_{i,n}|X_{i,C,n}, X_{j,n} \right)+h\left(\overrightarrow{Y}_{j,n}|X_{j,C,n}, X_{i,U,n}\right)\Big]+N\log\left(2\pi e\right)+N\delta(N)\\
\nonumber
&=& N\Big[h\left(\overrightarrow{Y}_{i,k}\right)+h\left(X_{i,U,k}|X_{j,C,k} \right)+h\left(\overleftarrow{Y}_{j,k}|X_{j,k},X_{i,U,j}\right)-\frac{5}{2}\log\left(2 \pi e\right)+h\left(\overrightarrow{Y}_{i,k}|X_{i,C,k}, X_{j,k} \right)\\
\label{Eqproof2Ri+Rjgc1}
& & +h\left(\overrightarrow{Y}_{j,k}|X_{j,C,k}, X_{i,U,k}\right)+2\log\left(2\pi e\right)+\delta(N)\Big],
\end{IEEEeqnarray}

\noindent
where,
(a) follows from the fact that $W_1$ and $W_2$ are mutually independent; 
(b) follows from Fano's inequality (see Figure~\ref{Fig:G-IC-NOF-Conv}c); 
(c) follows from \eqref{Eqsignalyif} and \eqref{EqsignalXiCG}; 
(d) follows from \eqref{EqsignalXiUG}; 
(e) follows from \eqref{Eqencod} and the fact that conditioning reduces the entropy; 
(f) follows from the fact that $h\left(\overrightarrow{\bs{Z}}_{j}|W_j,\overleftarrow{\bs{Y}}_i, \bs{X}_i, \bs{X}_{j,C}\right)-h\left(\overrightarrow{\bs{Z}}_i,\overrightarrow{\bs{Z}}_j| W_i, W_j, \overleftarrow{\bs{Y}}_{j}, \bs{X}_{i}, \bs{X}_{j}\right) \leqslant 0$;
(g) follows from the fact that $h\left(\overrightarrow{\bs{Y}}_j\right)-h\left(\bs{X}_{j,C},\bs{X}_{i,U} \right)+h\left(\bs{X}_{j,C},\bs{X}_{i,U}|\overrightarrow{\bs{Y}}_j \right)=h\left(\overrightarrow{\bs{Y}}_j|\bs{X}_{j,C},\bs{X}_{i,U}\right)$; 
(h) follows from Lemma~\ref{Lemmahelpsumandweightedrates}; and
(i) follows from the fact that conditioning reduces the entropy.

\noindent
From \eqref{Eqproof2Ri+Rjgc1},  the following holds in the asymptotic regime:  
\begin{IEEEeqnarray} {lcl}
\nonumber
2R_i+R_j &\leqslant& h\left(\overrightarrow{Y}_{i,k}\right)+h\left(X_{i,U,k}|X_{j,C,k} \right)+h\left(\overleftarrow{Y}_{j,k}|X_{j,k},X_{i,U,k}\right) \! + \! h\left(\overrightarrow{Y}_{i,k}|X_{i,C,k}, X_{j,k} \right) \! \\
\nonumber
& & + \! h\left(\overrightarrow{Y}_{j,k}|X_{j,C,k}, X_{i,U,k}\right) \! - \! \frac{1}{2}\log\left(2 \pi e\right)\\
\nonumber
&\leqslant& \frac{1}{2}\log\left(\overrightarrow{\SNR}_{i}+2\rho\sqrt{\overrightarrow{\SNR}_{i}\INR_{ij}}+\INR_{ij}+1\right)-\frac{1}{2}\log\left(\INR_{ij}+1 \right)\\
\nonumber
& & +\frac{1}{2}\log\left(\det\left(\textrm{Var}\left(\overleftarrow{Y}_{j,k},X_{j,k},X_{i,U,k}\right)\right)\right)-\frac{1}{2}\log\left(\det\left(\textrm{Var}\left(X_{j,k},X_{i,U,k}\right)\right)\right)\\
\nonumber
& & +\frac{1}{2}\log\left(1+\left(1-\rho^2\right)\left(\overrightarrow{\SNR}_{i}+\INR_{ji}\right)\right)-\frac{1}{2}\log\left(1+\left(1-\rho^2 \right)\INR_{ji}\right)\\
\label{Eqproof2Ri+Rjgc3}
& & \! +  \frac{1}{2} \! \log \! \left( \! \det \! \left( \! \textrm{Var}\left(\overrightarrow{Y}_{j,k},X_{j,C,k}, X_{i,U,k}\right)\right)\right) \! + \! 2\log\left(2 \pi e\right).
\end{IEEEeqnarray}

\noindent
The outer bound  on  \eqref{Eqproof2Ri+Rjgc3} depends on $S_{1,i}$, $S_{2,i}$, $S_{3,i}$, $S_{4,i}$, and $S_{5,i}$ via the parameter $h_{ji,U}$ in  \eqref{Eqdefhjiu}. Hence, as in the previous part, the following cases are identified: 

\noindent
\textbf{Case 1: $(S_{1,i} \lor S_{2,i} \lor S_{5,i})$}. From \eqref{Eqdefhjiu}, it follows that $h_{ji,U}=0$. Then, under these conditions,  plugging the expressions \eqref{EqVarAux1} into \eqref{Eqproof2Ri+Rjgc3} yields: \eqref{Eqconv7i1}. 

\noindent
\textbf{Case 2: $(S_{3,i} \lor S_{4,i})$}. From \eqref{Eqdefhjiu}, it follows that $h_{ji,U}=\sqrt{\frac{\INR_{ij}\INR_{ji}}{\overrightarrow{\SNR}_{j}}}$. Then, under these conditions,  plugging the expressions \eqref{EqVarAux1} into \eqref{Eqproof2Ri+Rjgc3} yields \eqref{Eqconv7i2}. 

\noindent
This completes the proof of \eqref{Eq2Ri+Rjc-g-G-IC-NOF} and the proof of Theorem~\ref{TheoremC-G-IC-NOF}. 

\subsection*{Proof of Lemma~\ref{Lemmahelpsumandweightedrates}}\label{SectProofLemma1App2}

Lemma~\ref{Lemmahelpsumandweightedrates} is proved as follows:

\begin{IEEEeqnarray}{lcl}
\nonumber
& & I\left(\bs{X}_{i,C}, \bs{X}_{j,U}, \overleftarrow{\bs{Y}}_i, W_i; \overleftarrow{\bs{Y}}_j, W_j\right)  \\
\nonumber
&=& I\left( W_i; \overleftarrow{\bs{Y}}_j, W_j\right) + I\left(\bs{X}_{i,C}, \bs{X}_{j,U}, \overleftarrow{\bs{Y}}_i; \overleftarrow{\bs{Y}}_j, W_j| W_i\right) \\
\nonumber
&=& h\left(\overleftarrow{\bs{Y}}_j, W_j\right)-h\left(\overleftarrow{\bs{Y}}_j, W_j| W_i\right) + h\left(\bs{X}_{i,C}, \bs{X}_{j,U}, \overleftarrow{\bs{Y}}_i| W_i\right)-h\left(\bs{X}_{i,C}, \bs{X}_{j,U}, \overleftarrow{\bs{Y}}_i | W_i, W_j, \overleftarrow{\bs{Y}}_j\right)\\
\nonumber
&=& h\left(\overleftarrow{\bs{Y}}_j|W_j\right)-h\left(\overleftarrow{\bs{Y}}_j| W_i, W_j\right) + h\left(\bs{X}_{i,C}, \bs{X}_{j,U}, \overleftarrow{\bs{Y}}_i| W_i\right)-h\left(\bs{X}_{i,C}, \bs{X}_{j,U}, \overleftarrow{\bs{Y}}_i | W_i, W_j, \overleftarrow{\bs{Y}}_j \right)\\
\nonumber
&=& h\left(\overleftarrow{\bs{Y}}_j|W_j\right) + h\left(\bs{X}_{i,C}, \bs{X}_{j,U}, \overleftarrow{\bs{Y}}_i| W_i\right)-h\left(\bs{X}_{i,C}, \bs{X}_{j,U}, \overleftarrow{\bs{Y}}_i, \overleftarrow{\bs{Y}}_j | W_i, W_j\right)\\
\nonumber
&=& h\left(\overleftarrow{\bs{Y}}_j|W_j\right) + \sum_{n=1}^{N}\Bigg[h\Big(X_{i,C,n}, X_{j,U,n}, \overleftarrow{Y}_{i,n}| W_i, \bs{X}_{i,C,(1:n-1)}, \bs{X}_{j,U,(1:n-1)}, \overleftarrow{\bs{Y}}_{i,(1:n-1)}, \bs{X}_{i,(1:n)}\Big)\\
\nonumber
& &  -h\Big(X_{i,C,n}, X_{j,U,n}, \overleftarrow{Y}_{i,n}, \overleftarrow{Y}_{j,n} | W_i, W_j,  \bs{X}_{i,C,(1:n-1)}, \bs{X}_{j,U,(1:n-1)}, \overleftarrow{\bs{Y}}_{i,(1:n-1)}, \overleftarrow{\bs{Y}}_{j,(1:n-1)},  \\
\nonumber
& & \bs{X}_{i,(1:n)}, \bs{X}_{j,(1:n)}  \Big)\Bigg] \\
\nonumber
&\leqslant& h\left(\overleftarrow{\bs{Y}}_j|W_j\right) + \sum_{n=1}^{N}\Bigg[h\Big(X_{i,C,n}, X_{j,U,n}, \overleftarrow{Y}_{i,n}| X_{i,n}\Big)-h\Big(\overrightarrow{Z}_{j,n}, \overrightarrow{Z}_{i,n}, \overleftarrow{Y}_{i,n}, \overleftarrow{Y}_{j,n} | W_i, W_j,    \\
\nonumber
& & \bs{X}_{i,C,(1:n-1)}, \bs{X}_{j,U,(1:n-1)}, \overleftarrow{\bs{Y}}_{i,(1:n-1)}, \overleftarrow{\bs{Y}}_{j,(1:n-1)}, \bs{X}_{i,(1:n)}, \bs{X}_{j,(1:n)}  \Big)\Bigg] 
\end{IEEEeqnarray}
\begin{IEEEeqnarray}{lcl}
\nonumber
&=& h\left(\overleftarrow{\bs{Y}}_j|W_j\right) + \sum_{n=1}^{N}\Bigg[h\Big(X_{i,C,n}| X_{i,n}\Big)+h\Big(X_{j,U,n}| X_{i,n}, X_{i,C,n} \Big)+h\Big(\overleftarrow{Y}_{i,n}| X_{i,n}, X_{i,C,n}, X_{j,U,n} \Big) \\
\nonumber
& &  -h\left(\overrightarrow{Z}_{j,n}\right)-h\left(\overrightarrow{Z}_{i,n}\right)-h\Big(\overleftarrow{Y}_{i,n}, \overleftarrow{Y}_{j,n} | W_i, W_j, \bs{X}_{i,C,(1:n-1)}, \bs{X}_{j,U,(1:n-1)}, \overleftarrow{\bs{Y}}_{i,(1:n-1)},   \\
\nonumber
& & \overleftarrow{\bs{Y}}_{j,(1:n-1)}, \bs{X}_{i,(1:n)}, \bs{X}_{j,(1:n)}, \overrightarrow{Z}_{j,n}, \overrightarrow{Z}_{i,n}  \Big)\Bigg] \\
\nonumber
&\leqslant& h\left(\overleftarrow{\bs{Y}}_j|W_j\right) + \sum_{n=1}^{N}\Bigg[h\Big(\overrightarrow{Z}_{j,n}| X_{i,n}\Big)+h\Big(X_{j,U,n}| X_{i,C,n} \Big)+h\Big(\overleftarrow{Y}_{i,n}| X_{i,n}, X_{j,U,n} \Big)-h\left(\overrightarrow{Z}_{j,n}\right) \\
\nonumber
& &  -h\left(\overrightarrow{Z}_{i,n}\right)-h\Big(\overleftarrow{Z}_{i,n}, \overleftarrow{Z}_{j,n} | W_i, W_j, \bs{X}_{i,C,(1:n-1)}, \bs{X}_{j,U,(1:n-1)}, \overleftarrow{\bs{Y}}_{i,(1:n-1)}, \overleftarrow{\bs{Y}}_{j,(1:n-1)},  \bs{X}_{i,(1:n)}, \\
\nonumber
& & \bs{X}_{j,(1:n)}, \overrightarrow{Z}_{j,n}, \overrightarrow{Z}_{i,n}  \Big)\Bigg] \\
\nonumber
&\stackrel{(a)}{=}& h\left(\overleftarrow{\bs{Y}}_j|W_j\right) + \sum_{n=1}^{N}\Bigg[h\Big(X_{j,U,n}| X_{i,C,n} \Big)+h\Big(\overleftarrow{Y}_{i,n}| X_{i,n}, X_{j,U,n} \Big)-h\left(\overrightarrow{Z}_{i,n}\right)-h\Big(\overleftarrow{Z}_{i,n}\Big)\\
\nonumber
& & -h\Big(\overleftarrow{Z}_{j,n}\Big)\Bigg] \\
\nonumber
&=& h\left(\overleftarrow{\bs{Y}}_j|W_j\right) + \sum_{n=1}^{N}\Bigg[h\left(X_{j,U,n}| X_{i,C,n} \right)+h\left(\overleftarrow{Y}_{i,n}| X_{i,n}, X_{j,U,n} \right)-\frac{3}{2}\log\left(2\pi e\right)\Bigg],
\end{IEEEeqnarray}
where
(a) follows from the fact that $\overleftarrow{Z}_{i,n}$ and $\overleftarrow{Z}_{j,n}$ are independent of $W_i$, $W_j$, $\bs{X}_{i,C,(1:n-1)}$, $\bs{X}_{j,U,(1:n-1)}$, $\overleftarrow{\bs{Y}}_{i,(1:n-1)}$, $\overleftarrow{\bs{Y}}_{j,(1:n-1)}$,  $\bs{X}_{i,(1:n)}$, $\bs{X}_{j,(1:n)}$, $\overrightarrow{Z}_{j,n}$, and $\overrightarrow{Z}_{i,n}$.

\noindent
This completes the proof of  Lemma~\ref{Lemmahelpsumandweightedrates}.

\section{Proof of the Gap between the Converse Region and the Achievable Region} \label{AppG-Gap}

This appendix presents a proof of the Theorem~\ref{TheoremGAP-G-IC-NOF}. The  gap, denoted by $\delta$, between the sets $\cgicnof$ and $\agicnof$ (Def.~\ref{DefGap}) is approximated as follows: 
\begin{IEEEeqnarray} {ll}
\label{Eqdelta}
\delta&=\max\left(\delta_{R_1},\delta_{R_2},\frac{\delta_{2R}}{2},\frac{\delta_{3R_1}}{3}, \frac{\delta_{3R_2}}{3}\right),
\end{IEEEeqnarray}      
where 
\begin{subequations}
\label{EqDeltas}
\begin{IEEEeqnarray} {ll}
\nonumber
\delta_{R_1}&=\min\Big(\kappa_{1,1}(\rho'),\kappa_{2,1}(\rho'),\kappa_{3,1}(\rho')\Big)-\min\Big(a_{2,1}(\rho), \! a_{6,1}(\rho,\mu_1) \! + \!  a_{3,2}(\rho,\mu_1), \\
\label{EqdeltaR1}
& a_{1,1} \! + \! a_{3,2}(\rho,\mu_1) \! + \! a_{4,2}(\rho,\mu_1)\Big), \\
\nonumber
\delta_{R_2}&=\min\Big(\kappa_{1,2}(\rho'),\kappa_{2,2}(\rho'),\kappa_{3,2}(\rho')\Big)-\min\Big(a_{2,2}(\rho), \!  a_{3,1}(\rho,\mu_2) \! + \!  a_{6,2}(\rho,\mu_2), \\
\label{EqdeltaR2}
& a_{3,1}(\rho,\mu_2) \! + \! a_{4,1}(\rho,\mu_2) \! + \! a_{1,2}\Big), 
\end{IEEEeqnarray} 
\begin{IEEEeqnarray} {ll}
\nonumber
\delta_{2R}&=\min\Big(\kappa_{4}(\rho'),\kappa_{5}(\rho'),\kappa_{6}(\rho')\Big)-\min\Big(a_{2,1}(\rho)+a_{1,2}, a_{1,1}+a_{2,2}(\rho), \\
\nonumber
& a_{3,1}(\rho,\mu_2)+a_{1,1}+a_{3,2}(\rho,\mu_1)+a_{7,2}(\rho,\mu_1,\mu_2), \\
\nonumber
& a_{3,1}(\rho,\mu_2)+a_{5,1}(\rho,\mu_2)+a_{3,2}(\rho,\mu_1)+a_{5,2}(\rho,\mu_1),  \\
\label{Eqdelta2R}
&  a_{3,1}(\rho,\mu_2)+a_{7,1}(\rho,\mu_1,\mu_2)+a_{3,2}(\rho,\mu_1)+a_{1,2} \Big), \\
\nonumber
\delta_{3R_1}&=\kappa_{7,1}(\rho')-\min\Big(a_{2,1}(\rho)+a_{1,1}+a_{3,2}(\rho,\mu_1)+a_{7,2}(\rho,\mu_1,\mu_2),\\
\nonumber
&  a_{3,1}(\rho,\mu_2)+a_{1,1}+a_{7,1}(\rho,\mu_1,\mu_2)+2a_{3,2}(\rho,\mu_1)+a_{5,2}(\rho,\mu_1),\\
\label{Eqdelta3R1}
&  a_{2,1}(\rho)+a_{1,1}+a_{3,2}(\rho,\mu_1)+a_{5,2}(\rho,\mu_1)\Big), \\
\nonumber
\delta_{3R_2}&=\kappa_{7,2}(\rho')-\min\Big(a_{3,1}(\rho,\mu_2)+a_{5,1}(\rho,\mu_2)+a_{2,2}(\rho)+a_{1,2}, \\
\nonumber
&  a_{3,1}(\rho,\mu_2)+a_{7,1}(\rho,\mu_1,\mu_2)+a_{2,2}(\rho)+a_{1,2}, \\
\label{Eqdelta3R2}
& 2a_{3,1}(\rho,\mu_2)+a_{5,1}(\rho,\mu_2)+a_{3,2}(\rho,\mu_1)+a_{1,2}+a_{7,2}(\rho,\mu_1,\mu_2)\Big),
\end{IEEEeqnarray} 
\end{subequations}
where, $\rho' \in [0,1]$ and $\left(\rho, \mu_1, \mu_2\right) \in \big[0$, $\big(1-\max\big(\frac{1}{\INR_{12}}$, $\frac{1}{\INR_{21}}\big) \big)^+\big]\times[0,1]\times[0,1]$.

\noindent
Note that $\delta_{R_1}$ and $\delta_{R_2}$ represent the gap between the active achievable single-rate bound and the active converse single-rate bound; $\delta_{2R}$ represents the gap between the active achievable sum-rate bound and the active converse sum-rate bound; and, $\delta_{3R_1}$ and $\delta_{3R_2}$ represent the gap between the active achievable weighted sum-rate bound and the active converse weighted sum-rate bound. 

\noindent
It is important to highlight that, as suggested in \cite{Suh-TIT-2011, SyQuoc-TIT-2015}, and \cite{Etkin-TIT-2008}, the gap between $\agicnof$ and $\cgicnof$ can be calculated more precisely. However, the choice in \eqref{Eqdelta} eases the calculations at the expense of less precision. 
Note also that whether the bounds are active (achievable or converse) in either of the equalities in \eqref{EqDeltas} depend on the exact values of $\overrightarrow{\SNR}_{1}$, $\overrightarrow{\SNR}_{2}$, $\INR_{12}$, $\INR_{21}$, $\overleftarrow{\SNR}_{1}$, and $\overleftarrow{\SNR}_{2}$. 
Hence a key point in order to find the gap between the achievable region and the converse region is to choose a convenient coding scheme for the achievable region, i.e., the values of $\rho$, $\mu_1$, and $\mu_2$, according to the definitions in \eqref{EqDeltas} for all $i \in \lbrace 1, 2 \rbrace$. This particular coding scheme is chosen such that the expressions in \eqref{EqDeltas} become simpler to upper bound at the expense of a looser outer bound. This particular coding scheme is different for each interference regime. The following describes all the key cases and the corresponding coding schemes.
 
\noindent
\underline{Case 1}: $\INR_{12} > \overrightarrow{\SNR}_1$ and $\INR_{21}>\overrightarrow{\SNR}_2$.  This case corresponds to the scenario in which both transmitter-receiver pairs are in high interference regime (HIR). Three subcases follow considering the SNR in the feedback links.  

\hangindent=0.7cm \underline{Case 1.1}: $\overleftarrow{\SNR}_2 \leqslant \overrightarrow{\SNR}_1$ and $\overleftarrow{\SNR}_1 \leqslant \overrightarrow{\SNR}_2$. In this case the coding scheme is: $\rho=0$, $\mu_1=0$ and $\mu_2=0$. 

\hangindent=0.7cm \underline{Case 1.2}: $\overleftarrow{\SNR}_2 > \overrightarrow{\SNR}_1$ and $\overleftarrow{\SNR}_1 > \overrightarrow{\SNR}_2$. In this case the coding scheme is: $\rho=0$, $\mu_1=1$, and $\mu_2=1$.

\hangindent=0.7cm \underline{Case 1.3}: $\overleftarrow{\SNR}_2 \leqslant \overrightarrow{\SNR}_1$ and $\overleftarrow{\SNR}_1 > \overrightarrow{\SNR}_2$. In this case the coding scheme is: $\rho=0$, $\mu_1=0$, and $\mu_2=1$.

\noindent
\underline{Case 2}: $\INR_{12} \leqslant \overrightarrow{\SNR}_1$ and $\INR_{21} \leqslant \overrightarrow{\SNR}_2$. This case corresponds to the scenario in which both transmitter-receiver pairs are in low interference regime (LIR). There are twelve subcases that must be studied separately. 

\noindent
In the following four subcases, the achievability scheme presented above is used considering the following coding scheme: $\rho=0$, $\mu_1=0$, and $\mu_2=0$. 

\hangindent=0.7cm \underline{Case 2.1}: $\overleftarrow{\SNR}_1 \leqslant  \INR_{21}$, $\overleftarrow{\SNR}_2 \leqslant \INR_{12}$, $\INR_{12}\INR_{21} >  \overrightarrow{\SNR}_1$ and $\INR_{12}\INR_{21} >  \overrightarrow{\SNR}_2$.

\hangindent=0.7cm \underline{Case 2.2}: $\overleftarrow{\SNR}_1 \leqslant  \INR_{21}$, $\overleftarrow{\SNR}_2\INR_{21} \leqslant \overrightarrow{\SNR}_2$, $\INR_{12}\INR_{21} >  \overrightarrow{\SNR}_1$ and $\INR_{12}\INR_{21} <  \overrightarrow{\SNR}_2$.

\hangindent=0.7cm \underline{Case 2.3}: $\overleftarrow{\SNR}_1\INR_{12} \leqslant \overrightarrow{\SNR}_1$, $\overleftarrow{\SNR}_2 \leqslant \INR_{12}$, $\INR_{12}\INR_{21} <  \overrightarrow{\SNR}_1$ and $\INR_{12}\INR_{21} >  \overrightarrow{\SNR}_2$.

\hangindent=0.7cm \underline{Case 2.4}: $\overleftarrow{\SNR}_1\INR_{12} \leqslant \overrightarrow{\SNR}_1$, $\overleftarrow{\SNR}_2\INR_{21} \leqslant \overrightarrow{\SNR}_2$, $\INR_{12}\INR_{21} <  \overrightarrow{\SNR}_1$ and $\INR_{12}\INR_{21}$ $<$ $\overrightarrow{\SNR}_2$.

\noindent
In the following four subcases, the achievability scheme presented above is used considering the following coding scheme: \indent $\rho=0$,  \indent $\mu_1=\frac{\INR_{21}^2\overleftarrow{\SNR}_2}{\left(\INR_{21}-1\right)\left(\INR_{21}\overleftarrow{\SNR}_2+\overrightarrow{\SNR}_2\right)}$, and  \indent $\mu_2= $ $\frac{\INR_{12}^2\overleftarrow{\SNR}_1}{\left(\INR_{12}-1\right)\left(\INR_{12}\overleftarrow{\SNR}_1+\overrightarrow{\SNR}_1\right)}$. 

\hangindent=0.7cm \underline{Case 2.5}: $\overleftarrow{\SNR}_1 >  \INR_{21}$, $\overleftarrow{\SNR}_2 > \INR_{12}$, $\INR_{12}\INR_{21} >  \overrightarrow{\SNR}_1$ and $\INR_{12}\INR_{21} >  \overrightarrow{\SNR}_2$.

\hangindent=0.7cm \underline{Case 2.6}: $\overleftarrow{\SNR}_1 >  \INR_{21}$, $\overleftarrow{\SNR}_2\INR_{21} > \overrightarrow{\SNR}_2$, $\INR_{12}\INR_{21} >  \overrightarrow{\SNR}_1$ and $\INR_{12}\INR_{21} <  \overrightarrow{\SNR}_2$.

\hangindent=0.7cm \underline{Case 2.7}: $\overleftarrow{\SNR}_1\INR_{12} > \overrightarrow{\SNR}_1$, $\overleftarrow{\SNR}_2 > \INR_{12}$, $\INR_{12}\INR_{21} <  \overrightarrow{\SNR}_1$ and $\INR_{12}\INR_{21} >  \overrightarrow{\SNR}_2$.

\hangindent=0.7cm \underline{Case 2.8}: $\overleftarrow{\SNR}_1\INR_{12} > \overrightarrow{\SNR}_1$, $\overleftarrow{\SNR}_2\INR_{21} > \overrightarrow{\SNR}_2$, $\INR_{12}\INR_{21} <  \overrightarrow{\SNR}_1$ and $\INR_{12}\INR_{21}$ $<$ $\overrightarrow{\SNR}_2$.

\noindent
In the following four subcases, the achievability scheme presented above is used considering the following coding scheme: $\rho=0$, $\mu_1=0$, and $\mu_2=\frac{\INR_{12}^2\overleftarrow{\SNR}_1}{\left(\INR_{12}-1\right)\left(\INR_{12}\overleftarrow{\SNR}_1+\overrightarrow{\SNR}_1\right)}$. 

\hangindent=0.7cm \underline{Case 2.9}: $\overleftarrow{\SNR}_1 > \INR_{21}$, $\overleftarrow{\SNR}_2 \leqslant \INR_{12}$, $\INR_{12}\INR_{21} >  \overrightarrow{\SNR}_1$ and $\INR_{12}\INR_{21} >  \overrightarrow{\SNR}_2$.

\hangindent=0.7cm \underline{Case 2.10}: $\overleftarrow{\SNR}_1 >  \INR_{21}$, $\overleftarrow{\SNR}_2\INR_{21} \leqslant \overrightarrow{\SNR}_2$, $\INR_{12}\INR_{21} >  \overrightarrow{\SNR}_1$ and $\INR_{12}\INR_{21} <  \overrightarrow{\SNR}_2$.

\hangindent=0.7cm \underline{Case 2.11}: $\overleftarrow{\SNR}_1\INR_{12} > \overrightarrow{\SNR}_1$, $\overleftarrow{\SNR}_2 \leqslant \INR_{12}$, $\INR_{12}\INR_{21} <  \overrightarrow{\SNR}_1$ and $\INR_{12}\INR_{21} >  \overrightarrow{\SNR}_2$.

\hangindent=0.7cm \underline{Case 2.12}: $\overleftarrow{\SNR}_1\INR_{12} > \overrightarrow{\SNR}_1$, $\overleftarrow{\SNR}_2\INR_{21} \leqslant \overrightarrow{\SNR}_2$, $\INR_{12}\INR_{21} <  \overrightarrow{\SNR}_1$ and $\INR_{12}\INR_{21}$ $<$ $\overrightarrow{\SNR}_2$.

\noindent
\underline{Case 3}: $\INR_{12}>\overrightarrow{\SNR}_1$ and $\INR_{21}\leqslant\overrightarrow{\SNR}_2$. This case corresponds to the scenario in which transmitter-receiver pair $1$ is in HIR and transmitter-receiver pair $2$ is in LIR. There are four subcases that must be studied separately.

\noindent
In the following two subcases, the achievability scheme presented above is used considering the following coding scheme: $\rho=0$, $\mu_1=0$, and $\mu_2=0$. 

\hangindent=0.7cm \underline{Case 3.1}: $\overleftarrow{\SNR}_2 \leqslant \INR_{12}$ and $\INR_{12}\INR_{21} >  \overrightarrow{\SNR}_2$.

\hangindent=0.7cm \underline{Case 3.2}: $\overleftarrow{\SNR}_2\INR_{21} \leqslant  \overrightarrow{\SNR}_2$ and $\INR_{12}\INR_{21} <  \overrightarrow{\SNR}_2$.

\noindent
In the following two subcases, the achievability scheme presented above is used considering the following coding scheme: $\rho=0$, $\mu_1=1$, and $\mu_2=0$. 

\hangindent=0.7cm \underline{Case 3.3}: $\overleftarrow{\SNR}_2 > \INR_{12}$ and $\INR_{12}\INR_{21} >  \overrightarrow{\SNR}_2$.

\hangindent=0.7cm \underline{Case 3.4}: $\overleftarrow{\SNR}_2\INR_{21} >  \overrightarrow{\SNR}_2$ and $\INR_{12}\INR_{21} <  \overrightarrow{\SNR}_2$.

\noindent
The following is the calculation of the gap $\delta$ in Case $1.1$. 

\noindent
\begin{enumerate}
\item \underline {Calculation of $\delta_{R_1}$.}
\noindent
From \eqref{EqdeltaR1} and considering the corresponding coding scheme for the achievable region ($\rho=0$, $\mu_1=0$ and $\mu_2=0$), it follows that  
\begin{IEEEeqnarray} {rcl}
\label{EqdeltaR1proofgap}
\delta_{R_1}&\leqslant& \min\Big(\kappa_{1,1}(\rho'),\kappa_{2,1}(\rho'),\kappa_{3,1}(\rho')\Big)-\min\Big(a_{6,1}(0,0), a_{1,1}+a_{4,2}(0,0)\Big), 
\end{IEEEeqnarray} 
where the exact value of $\rho'$ is chosen to provide at least an outer bound for \eqref{EqdeltaR1proofgap}.

\noindent
Note that in this case:
\begin{subequations}
\label{EqprofGapcHIRbb}
\begin{IEEEeqnarray}{rcl}
\nonumber
 \kappa_{1,1}(\rho')  &=&  \frac{1}{2}\log \Big(b_{1,1}(\rho')+1\Big) \\
\nonumber
&\stackrel{(a)}{\leqslant} &   \frac{1}{2}\log \! \left( \! \overrightarrow{\SNR}_{1} \! + \! 2\sqrt{\overrightarrow{\SNR}_{1}\INR_{12}} \! + \! \INR_{12} \! + \! 1 \!\right) \\
\nonumber
&  \stackrel{(b)}{\leqslant} &  \frac{1}{2}\log \left(2\overrightarrow{\SNR}_{1}+2\INR_{12}+1\right) \\
\label{EqprofGapcHIRbb1}
&  \leqslant &  \frac{1}{2}\log \left(\overrightarrow{\SNR}_{1}+\INR_{12}+1\right)+\frac{1}{2}, \\
\nonumber
\kappa_{2,1}(\rho')  &=&  \frac{1}{2} \log \left(1+b_{4,1}(\rho')+b_{5,2}(\rho')\right) \\
\label{EqprofGapcHIRbb2}
& \leqslant & \frac{1}{2} \log \left(\overrightarrow{\SNR}_{1}+\INR_{21}+1\right), \\
\nonumber
\kappa_{3,1}(\rho') &=&   \frac{1}{2}\log\Big(b_{4,1}(\rho')+1\Big)+\frac{1}{2}\log \! \left(\frac{\overleftarrow{\SNR}_2\left(b_{4,1}(\rho')+b_{5,2}(\rho')+1\right)}{\left(b_{1,2}(1) \! + \! 1\right)\left(b_{4,1}(\rho') \! + \! 1\right)} \! + \! 1\right) \\
\nonumber
&\stackrel{(c)}{\leqslant} & \frac{1}{2}\log\left(\overrightarrow{\SNR}_1+1\right) \!+ \! \frac{1}{2} \! \log \! \left( \! \frac{\overleftarrow{\SNR}_2\left(\overrightarrow{\SNR}_1+\INR_{21}+1\right)}{\left(\overrightarrow{\SNR}_{2} \! +\! \INR_{21} \! + \! 1\right) \! \left( \! \overrightarrow{\SNR}_1 \! + \! 1\right)} \! + \! 1 \! \right)\\
\label{EqprofGapcHIRbb3}
&=&\! \frac{1}{2} \! \log \! \left( \! \frac{\overleftarrow{\SNR}_2 \! \left( \! \overrightarrow{\SNR}_1 \! + \! \INR_{21} \! + \! 1\right)}{\overrightarrow{\SNR}_{2}+\INR_{21}+1} \! + \! \overrightarrow{\SNR}_1 \! + \!1 \! \right) \! ,
\end{IEEEeqnarray}
\end{subequations}
where
(a) follows from the fact that $0\leqslant\rho'\leqslant1$; 
(b) follows from the fact that 
\begin{equation}
\label{Eqgapb}
\left(\sqrt{\overrightarrow{\SNR}_{1}}-\sqrt{\INR_{12}}\right)^2 \geqslant 0; 
\end{equation}
and
(c) follows from the fact that $\kappa_{3,1}(\rho')$ is a monotonically decreasing function of $\rho'$. 

\noindent
Note also that the achievable bound $a_{1,1}+a_{4,2}(0,0)$ can be lower bounded as follows:

\begin{IEEEeqnarray}{rcl}
\nonumber
a_{1,1} \! + \! a_{4,2}(0,0) &=& \frac{1}{2} \! \log \! \left( \! \frac{\overrightarrow{\SNR}_{1}}{\INR_{21}} \! + \! 2\right) \! + \! \frac{1}{2}\log\Big(\INR_{21} \! + \! 1 \! \Big) \! \\
\nonumber
& & - \! 1 \\
\nonumber
&\geqslant& \frac{1}{2} \! \log \! \left( \! \frac{\overrightarrow{\SNR}_{1}}{\INR_{21}} \! + \! 2\right) \! + \! \frac{1}{2} \! \log \! \Big( \! \INR_{21} \! \Big) \! - \! 1 \\
\nonumber
&=& \frac{1}{2}\log\left(\overrightarrow{\SNR}_{1}+2\INR_{21}\right)-1\\
\nonumber
&=& \frac{1}{2}\log\left(\overrightarrow{\SNR}_{1}+\INR_{21}+\INR_{21}\right)-1 \\
\label{Eqgapv1}
&\geqslant& \frac{1}{2}\log\left(\overrightarrow{\SNR}_{1}+\INR_{21}+1\right)-1.
\end{IEEEeqnarray}

From \eqref{EqdeltaR1proofgap}, \eqref{EqprofGapcHIRbb} and \eqref{Eqgapv1}, assuming that ${a_{1,1}+a_{4,2}(0,0) < a_{6,1}(0,0)}$, it follows that 
\begin{IEEEeqnarray} {rcl}
\nonumber
\delta_{R_{1}}&\leqslant&  \min \!\Big( \! \kappa_{1,1}(\rho'),\kappa_{2,1}(\rho'),\kappa_{3,1}(\rho') \!\Big) \! - \! \Big( \! a_{1,1} \!+ \! a_{4,2}(0,0) \! \Big) \\
\nonumber
&\leqslant& \kappa_{2,1}(\rho')-\Big(a_{1,1}+a_{4,2}(0,0)\Big)\\
\label{EqdeltaR11}
&\leqslant& 1. 
\end{IEEEeqnarray} 

\noindent
Now, assuming that $a_{6,1}(0,0) < a_{1,1}+a_{4,2}(0,0)$, the following holds: 

\begin{IEEEeqnarray} {rcl}
\label{EqdeltaR12p}
\delta_{R_{1}}&\leqslant& \min\Big(\kappa_{1,1}(\rho'),\kappa_{2,1}(\rho'),\kappa_{3,1}(\rho')\Big)-a_{6,1}(0,0). \qquad
\end{IEEEeqnarray} 

To calculate an upper bound for \eqref{EqdeltaR12p}, the following cases are considered: 

\noindent
\underline{Case 1.1.1}: $\overrightarrow{\SNR}_{1} \geqslant \INR_{21} \land \overrightarrow{\SNR}_{2} < \INR_{12}$; 

\noindent
\underline{Case 1.1.2}: $\overrightarrow{\SNR}_{1} < \INR_{21} \land \overrightarrow{\SNR}_{2} \geqslant \INR_{12}$; and

\noindent
\underline{Case 1.1.3}: $\overrightarrow{\SNR}_{1} < \INR_{21} \land \overrightarrow{\SNR}_{2} < \INR_{12}$. 

\noindent
In Case $1.1.1$,  from \eqref{EqprofGapcHIRbb} and \eqref{EqdeltaR12p}, it follows that
\begin{IEEEeqnarray} {rcl}
\nonumber
\delta_{R_{1}} &\leqslant& \kappa_{2,1}(\rho')-a_{6,1}(0,0)\\
\nonumber
&\leqslant& \frac{1}{2} \log \left(\overrightarrow{\SNR}_{1}+\INR_{21}+1\right)-\frac{1}{2}\log\left(\overrightarrow{\SNR}_{1}+2\right)+\frac{1}{2}\\
\nonumber
& \leqslant & \frac{1}{2} \log \left(\overrightarrow{\SNR}_{1}+\overrightarrow{\SNR}_{1}+1\right)-\frac{1}{2}\log\left(\overrightarrow{\SNR}_{1}+2\right)+\frac{1}{2}\\
\label{EqdeltaR12}
&\leqslant& 1. 
\end{IEEEeqnarray} 

\noindent
In Case $1.1.2$, from \eqref{EqprofGapcHIRbb} and \eqref{EqdeltaR12p}, it follows that
\begin{IEEEeqnarray} {rcl}
\nonumber
\delta_{R_{1}} &\leqslant& \kappa_{3,1}(\rho')-a_{6,1}(0,0)\\
\nonumber
& \leqslant & \frac{1}{2}\log\left(\frac{\overleftarrow{\SNR}_2\left(\overrightarrow{\SNR}_1\!+\!\INR_{21}\!+\!1\right)}{\overrightarrow{\SNR}_{2}+\INR_{21}+1}\!+\!\overrightarrow{\SNR}_1\!+\!1\right)-\frac{1}{2}\log\left(\overrightarrow{\SNR}_{1}+2\right)+\frac{1}{2} \\
\nonumber
& \leqslant & \frac{1}{2}\log\left(\overleftarrow{\SNR}_2+\overrightarrow{\SNR}_1+1\right)-\frac{1}{2}\log\left(\overrightarrow{\SNR}_{1}+2\right)+\frac{1}{2}\\
\nonumber
& \leqslant & \frac{1}{2}\log\left(\overrightarrow{\SNR}_1+\overrightarrow{\SNR}_1+1\right)-\frac{1}{2}\log\left(\overrightarrow{\SNR}_{1}+2\right)+\frac{1}{2}\\
\label{EqdeltaR13}
&\leqslant& 1. 
\end{IEEEeqnarray} 
\noindent
In Case $1.1.3$ two additional cases are considered:

\noindent
\underline{Case 1.1.3.1}: $\overrightarrow{\SNR}_{1} \geqslant \overrightarrow{\SNR}_{2}$; and 

\noindent
\underline{Case 1.1.3.2}: $\overrightarrow{\SNR}_{1} < \overrightarrow{\SNR}_{2}$.

\noindent
In  Case $1.1.3.1$, from \eqref{EqprofGapcHIRbb} and \eqref{EqdeltaR12p}, it follows that
\begin{IEEEeqnarray} {rcl}
\nonumber
\delta_{R_{1}} &\leqslant& \kappa_{3,1}(\rho')-a_{6,1}(0,0)\\
\nonumber
& \leqslant & \frac{1}{2}\log\left(\frac{\overleftarrow{\SNR}_2\left(\overrightarrow{\SNR}_1\!+\!\INR_{21}\!+\!1\right)}{\overrightarrow{\SNR}_{2}+\INR_{21}+1}\!+\!\overrightarrow{\SNR}_1\!+\!1\right)-\frac{1}{2}\log\left(\overrightarrow{\SNR}_{1}+2\right)+\frac{1}{2} \\
\nonumber
& = & \frac{1}{2}\log\left(\overrightarrow{\SNR}_1+1\right)+\frac{1}{2}\log\left(\frac{\overleftarrow{\SNR}_2\left(\overrightarrow{\SNR}_1+\INR_{21}+1\right)}{\left(\overrightarrow{\SNR}_{2}\!+\!\INR_{21}\!+\!1\right)\left(\overrightarrow{\SNR}_1\!+\!1\right)}\!+\!1\right)\\
\nonumber
& & -\frac{1}{2}\log\left(\overrightarrow{\SNR}_{1}+2\right)+\frac{1}{2} \\
\nonumber
& \leqslant & \frac{1}{2}\log\left(\frac{\overrightarrow{\SNR}_1\left(\INR_{21}+\INR_{21}+\INR_{21}\right)}{\INR_{21}\overrightarrow{\SNR}_1}+1\right)+\frac{1}{2} \\
\label{EqdeltaR14}
&=& \frac{3}{2}. 
\end{IEEEeqnarray} 

\noindent
In  Case $1.1.3.2$, from \eqref{EqprofGapcHIRbb} and \eqref{EqdeltaR12p}, it follows that
\begin{IEEEeqnarray} {rcl}
\nonumber
\delta_{R_{1}} &\leqslant& \kappa_{3,1}(\rho')-a_{6,1}(0,0)\\
\nonumber
& \leqslant & \frac{1}{2}\log\left(\frac{\overleftarrow{\SNR}_2\left(\overrightarrow{\SNR}_1\!+\!\INR_{21}\!+\!1\right)}{\overrightarrow{\SNR}_{2}+\INR_{21}+1}\!+\!\overrightarrow{\SNR}_1\!+\!1\right)-\frac{1}{2}\log\left(\overrightarrow{\SNR}_{1}+2\right)+\frac{1}{2} \\
\nonumber
& \leqslant & \frac{1}{2}\log\left(\overleftarrow{\SNR}_2+\overrightarrow{\SNR}_1+1\right)-\frac{1}{2}\log\left(\overrightarrow{\SNR}_{1}+2\right)+\frac{1}{2}\\
\nonumber
& \leqslant & \frac{1}{2}\log\left(\overrightarrow{\SNR}_1+\overrightarrow{\SNR}_1+1\right)-\frac{1}{2}\log\left(\overrightarrow{\SNR}_{1}+2\right)+\frac{1}{2} \\
\label{EqdeltaR15}
& \leqslant & 1. 
\end{IEEEeqnarray} 
Then, from \eqref{EqdeltaR11}, \eqref{EqdeltaR12}, \eqref{EqdeltaR13}, \eqref{EqdeltaR14}, and \eqref{EqdeltaR15}, it follows that in Case $1.1$: 
\begin{IEEEeqnarray}{rcl}
\label{EqGapR1}
\delta_{R_1} &\leqslant& \frac{3}{2}. 
\end{IEEEeqnarray} 

The same procedure holds to calculate $\delta_{R_2}$ and it yields:

\begin{IEEEeqnarray}{rcl}
\label{EqGapR2}
\delta_{R_2} &\leqslant& \frac{3}{2}. 
\end{IEEEeqnarray} 

\item \underline {Calculation of $\delta_{2R}$}. From \eqref{Eqdelta2R} and considering the corresponding coding scheme for the achievable region ($\rho=0$, $\mu_1=0$ and $\mu_2=0$),  it follows that 
\begin{IEEEeqnarray} {rcl}
\nonumber
\delta_{2R}&\leqslant&\min\Big(\kappa_{4}(\rho'),\kappa_{5}(\rho'),\kappa_{6}(\rho')\Big)-\min\Big(a_{2,1}(0)+a_{1,2},  a_{1,1}+a_{2,2}(0), a_{5,1}(0,0)+a_{5,2}(0,0)\Big)\\
\label{Eqdelta2Rproofgap}
&\leqslant&\min\Big(\kappa_{4}(\rho'),\kappa_{5}(\rho')\Big)-\min\Big(a_{2,1}(0)+a_{1,2}, a_{1,1}+a_{2,2}(0), a_{5,1}(0,0)+a_{5,2}(0,0)\Big).
\end{IEEEeqnarray} 

Note that 

\begin{subequations}
\label{EqprofGapcHIRsumratebb}
\begin{IEEEeqnarray}{rcl}
\nonumber
\kappa_{4}(\rho')  & = & \frac{1}{2}\log \left(1+\frac{b_{4,1}(\rho')}{1+b_{5,2}(\rho')}\right)+\frac{1}{2}\log \bigg(b_{1,2}(\rho')+1\bigg) \\
\nonumber
& \leqslant & \frac{1}{2}\log \left(1+\frac{b_{4,1}(\rho')}{b_{5,2}(\rho')}\right)+\frac{1}{2}\log \bigg(b_{1,2}(\rho')+1\bigg) \\
\nonumber
& = & \frac{1}{2}\log \left(1+\frac{\overrightarrow{\SNR}_{1}}{\INR_{21}}\right)+\frac{1}{2}\log \bigg(b_{1,2}(\rho')+1\bigg) \\
\nonumber
& \stackrel{(h)}{\leqslant} & \frac{1}{2}\log \left(1+\frac{\overrightarrow{\SNR}_{1}}{\INR_{21}}\right)+\frac{1}{2}\log \left(2\overrightarrow{\SNR}_{2}+2\INR_{21}+1\right) \\
\nonumber
& \leqslant & \frac{1}{2}\log \left(1+\frac{\overrightarrow{\SNR}_{1}}{\INR_{21}}\right)+\frac{1}{2}\log \left(\overrightarrow{\SNR}_{2}+\INR_{21}+1\right)+\frac{1}{2}\\
\label{EqprofGapcHIRsumratebb1}
& \leqslant & \frac{1}{2}\log \left(2+\frac{\overrightarrow{\SNR}_{1}}{\INR_{21}}\right)+\frac{1}{2}\log \left(\overrightarrow{\SNR}_{2}+\INR_{21}+1\right)+\frac{1}{2}, 
\end{IEEEeqnarray}
and
\begin{IEEEeqnarray}{rcl}
\nonumber
\kappa_{5}(\rho') & = & \frac{1}{2}\log \left(1+\frac{b_{4,2}(\rho')}{1+b_{5,1}(\rho')}\right)+\frac{1}{2}\log \bigg(b_{1,1}(\rho')+1\bigg) \\
\nonumber
& \leqslant & \frac{1}{2}\log \left(1+\frac{b_{4,2}(\rho')}{b_{5,1}(\rho')}\right)+\frac{1}{2}\log \bigg(b_{1,1}(\rho')+1\bigg) \\
\nonumber
& = & \frac{1}{2}\log \left(1+\frac{\overrightarrow{\SNR}_{2}}{\INR_{12}}\right)+\frac{1}{2}\log \bigg(b_{1,1}(\rho')+1\bigg) \\
\nonumber
&\stackrel{(i)}{\leqslant}& \frac{1}{2}\log \left(1+\frac{\overrightarrow{\SNR}_{2}}{\INR_{12}}\right)+\frac{1}{2}\log \left(2\overrightarrow{\SNR}_{1}+2\INR_{12}+1\right) \\
\nonumber
&\leqslant& \frac{1}{2}\log \left(1+\frac{\overrightarrow{\SNR}_{2}}{\INR_{12}}\right)+\frac{1}{2}\log \left(\overrightarrow{\SNR}_{1}+\INR_{12}+1\right)+\frac{1}{2}, \\
\label{EqprofGapcHIRsumratebb2}
&\leqslant& \frac{1}{2}\log \left(2+\frac{\overrightarrow{\SNR}_{2}}{\INR_{12}}\right)+\frac{1}{2}\log \left(\overrightarrow{\SNR}_{1}+\INR_{12}+1\right)+\frac{1}{2}, 
\end{IEEEeqnarray}
\end{subequations}
where
(h)  follows from the fact that 
\begin{equation}
\label{Eqgapeqh}
\left(\sqrt{\overrightarrow{\SNR}_{2}}-\sqrt{\INR_{21}}\right)^2 \geqslant 0; 
\end{equation}
and (i)  follows from the fact that 
\begin{equation}
\label{Eqgapeqi}
\left(\sqrt{\overrightarrow{\SNR}_{1}}-\sqrt{\INR_{12}}\right)^2 \geqslant 0. 
\end{equation}

\noindent
From \eqref{Eqdelta2Rproofgap} and  \eqref{EqprofGapcHIRsumratebb}, assuming that $a_{2,1}(0)+a_{1,2} < \min\Big(a_{1,1}+a_{2,2}(0), a_{5,1}(0,0)+a_{5,2}(0,0)\Big)$, it follows that
\begin{IEEEeqnarray}{rcl}
\nonumber
\delta_{2R}&\leqslant& \min\Big(\kappa_{4}(\rho'),\kappa_{5}(\rho')\Big)-\Big(a_{2,1}(0)+a_{1,2}\Big) \\
\nonumber
&\leqslant& \kappa_{5}(\rho')-\Big(a_{2,1}(0)+a_{1,2}\Big)\\
\nonumber
&\leqslant& \frac{1}{2}\log \left(2+\frac{\overrightarrow{\SNR}_{2}}{\INR_{12}}\right)+\frac{1}{2}\log \left(\overrightarrow{\SNR}_{1}+\INR_{12}+1\right)+\frac{1}{2} - \frac{1}{2}\log\left(\overrightarrow{\SNR}_{1}+\INR_{12}+1\right)\\
\nonumber
& & -\frac{1}{2}\log\left(\frac{\overrightarrow{\SNR}_{2}}{\INR_{12}}+2\right)+1 \\
\label{Eqgapsumrate1}
&=& \frac{3}{2}. 
\end{IEEEeqnarray} 
From \eqref{Eqdelta2Rproofgap} and  \eqref{EqprofGapcHIRsumratebb}, assuming that $a_{1,1}+a_{2,2}(0) < \min\Big(a_{2,1}(0)+a_{1,2}, a_{5,1}(0,0)+a_{5,2}(0,0)\Big)$, it follows that 
\begin{IEEEeqnarray}{rcl}
\nonumber
\delta_{2R}&\leqslant& \min\Big(\kappa_{4}(\rho'),\kappa_{5}(\rho')\Big)-\Big(a_{1,1}+a_{2,2}(0)\Big) \\
\nonumber
&\leqslant& \kappa_{4}(\rho')-\Big(a_{1,1}+a_{2,2}(0)\Big) \\
\nonumber
&\leqslant& \frac{1}{2}\log \left(2+\frac{\overrightarrow{\SNR}_{1}}{\INR_{21}}\right)+\frac{1}{2}\log \left(\overrightarrow{\SNR}_{2}+\INR_{21}+1\right)+\frac{1}{2}- \frac{1}{2}\log\left(\overrightarrow{\SNR}_{2}+\INR_{21}+1\right)\\
\nonumber
& & -\frac{1}{2}\log\left(\frac{\overrightarrow{\SNR}_{1}}{\INR_{21}}+2\right)+1 \\
\label{Eqgapsumrate2}
&=& \frac{3}{2}. 
\end{IEEEeqnarray} 

\noindent
Now,  assume that $a_{5,1}(0,0)+a_{5,2}(0,0) < \min(a_{2,1}(0)+a_{1,2}, a_{1,1}+a_{2,2}(0))$. In this case, the following holds: 
\begin{IEEEeqnarray}{rcl}
\label{Eqgapsumrate3p}
\delta_{2R}&\leqslant&  \min \!\Big( \!\kappa_{4}(\rho'),\kappa_{5}(\rho') \! \Big) \! - \! \Big( \! a_{5,1}(0,0) \! + \! a_{5,2}(0,0) \! \Big). \qquad
\end{IEEEeqnarray}
To calculate an upper bound for \eqref{Eqgapsumrate3p}, the cases $1.1.1$ - $1.1.3$ defined above are analyzed hereunder. 

\noindent
In Case $1.1.1$, $a_{5,1}(0,0)+a_{5,2}(0,0)$ can be lower bounded as follows:
\begin{IEEEeqnarray}{rcl}
\nonumber
a_{5,1}(0,0)+a_{5,2}(0,0) &=& \frac{1}{2} \! \log \! \left( \! \frac{\overrightarrow{\SNR}_{1}}{\INR_{21}} \! + \! \INR_{12} \! + \! 1 \! \right) \! + \! \frac{1}{2} \! \log \! \left( \! \frac{\overrightarrow{\SNR}_{2}}{\INR_{12}} \! + \! \INR_{21} \! + \! 1\right) \! - \! 1\\
\label{Eqgapv1v21}
&\geqslant& \frac{1}{2}\log\left(\INR_{12}+1\right)-1.
\end{IEEEeqnarray}
From \eqref{EqprofGapcHIRsumratebb}, \eqref{Eqgapsumrate3p}, and \eqref{Eqgapv1v21}, it follows that
\begin{IEEEeqnarray}{rcl}
\nonumber
\delta_{2R}&\leqslant& \! \min \!\Big( \!\kappa_{4}(\rho'),\kappa_{5}(\rho') \! \Big) \! - \! \Big( \! a_{5,1}(0,0) \! + \! a_{5,2}(0,0) \! \Big) \\
\nonumber
&\leqslant& \kappa_{5}(\rho')-\Big(a_{5,1}(0,0)+a_{5,2}(0,0)\Big) \\
\nonumber
&\leqslant& \frac{1}{2} \! \log \! \left( \! 2 \! + \! \frac{\overrightarrow{\SNR}_{2}}{\INR_{12}}\right) \! + \! \frac{1}{2}\log \left(\overrightarrow{\SNR}_{1} \! + \! \INR_{12} \! + \! 1\right) +\frac{1}{2}-\frac{1}{2}\log\left(\INR_{12}+1\right)+1 \\
\nonumber
& \leqslant & \frac{1}{2}\log \left(2+1\right)+\frac{1}{2}\log \left(\INR_{12}+\INR_{12}+1\right)-\frac{1}{2}\log\left(\INR_{12}+1\right)+\frac{3}{2} \\
\label{Eqgapsumrate3}
& \leqslant & \frac{1}{2}\log \left(3\right)+2.
\end{IEEEeqnarray}

\noindent
In Case $1.1.2$, $a_{5,1}(0,0)+a_{5,2}(0,0)$ can be lower bounded as follows:
\begin{IEEEeqnarray}{rcl}
\nonumber
a_{5,1}(0,0) \! + \! a_{5,2}(0,0) &=& \frac{1}{2} \!  \log \! \left( \! \frac{\overrightarrow{\SNR}_{1}}{\INR_{21}} \! + \! \INR_{12} \! + \! 1 \! \right) \! + \! \frac{1}{2} \! \log \! \left( \! \frac{\overrightarrow{\SNR}_{2}}{\INR_{12}} \! + \! \INR_{21} \! + \! 1\right) \! - \! 1  \\
\label{Eqgapv1v22}
& \geqslant & \frac{1}{2}\log\left(\INR_{21}+1\right)-1.
\end{IEEEeqnarray}
From \eqref{EqprofGapcHIRsumratebb}, \eqref{Eqgapsumrate3p}, and \eqref{Eqgapv1v22}, it follows that
\begin{IEEEeqnarray}{rcl}
\nonumber
\delta_{2R}&\leqslant&  \min \Big( \!\kappa_{4}(\rho'),\kappa_{5}(\rho') \! \Big) \! - \! \Big( \! a_{5,1}(0,0) \! + \! a_{5,2}(0,0) \! \Big) \\
\nonumber
&\leqslant& \kappa_{4}(\rho')-\Big(a_{5,1}(0,0)+a_{5,2}(0,0)\Big) \\
\nonumber
& \leqslant & \frac{1}{2} \! \log \! \left( \! 2 \! + \! \frac{\overrightarrow{\SNR}_{1}}{\INR_{21}}\right) \! + \! \frac{1}{2} \! \log \! \left(\! \overrightarrow{\SNR}_{2} \!+ \! \INR_{21} \!+ \! 1\right)+\frac{1}{2}-\frac{1}{2}\log\left(\INR_{21}+1\right)+1 \\
\nonumber
& \leqslant & \frac{1}{2}\log \left(2+1\right)+\frac{1}{2}\log \left(\INR_{21}+\INR_{21}+1\right)-\frac{1}{2}\log\left(\INR_{21}+1\right)+\frac{3}{2} \\
\label{Eqgapsumrate4}
& \leqslant & \frac{1}{2}\log \left(3\right)+2.
\end{IEEEeqnarray}
\noindent
In Case $1.1.3$, from  \eqref{EqprofGapcHIRsumratebb}, \eqref{Eqgapsumrate3p}, and \eqref{Eqgapv1v21}, it follows that
\begin{IEEEeqnarray}{rcl}
\nonumber
\delta_{2R}&\leqslant& \! \min \!\Big( \!\kappa_{4}(\rho'),\kappa_{5}(\rho') \! \Big) \! - \! \Big( \! a_{5,1}(0,0) \! + \! a_{5,2}(0,0) \! \Big) \\
\nonumber
&\leqslant& \kappa_{5}(\rho')-\Big(a_{5,1}(0,0)+a_{5,2}(0,0)\Big) \\
\nonumber
&\leqslant& \frac{1}{2} \! \log \! \left( \! 2 \! + \! \frac{\overrightarrow{\SNR}_{2}}{\INR_{12}}\right) \! + \! \frac{1}{2}\log \left(\overrightarrow{\SNR}_{1} \! + \! \INR_{12} \! + \! 1\right)+\frac{1}{2}-\frac{1}{2}\log\left(\INR_{12}+1\right)+1 \\
\nonumber
& \leqslant & \frac{1}{2}\log \left(2+1\right)+\frac{1}{2}\log \left(\INR_{12}+\INR_{12}+1\right) -\frac{1}{2}\log\left(\INR_{12}+1\right)+\frac{3}{2}\\
\nonumber
\label{Eqgapsumrate5}
& \leqslant & \frac{1}{2}\log \left(3\right)+2. 
\end{IEEEeqnarray}
Then, from \eqref{Eqgapsumrate1}, \eqref{Eqgapsumrate2}, \eqref{Eqgapsumrate3}, \eqref{Eqgapsumrate4}, and \eqref{Eqgapsumrate5}, it follows that in Case $1.1$: 
\begin{IEEEeqnarray}{rcl}
\label{EqGapR1R2}
\delta_{2R} &\leqslant& 2+\frac{1}{2}\log \left(3\right). 
\end{IEEEeqnarray} 

\item \underline {Calculation of $\delta_{3R_1}$}. From \eqref{Eqdelta3R1} and considering the corresponding coding scheme for the achievable region ($\rho=0$, $\mu_1=0$ and $\mu_2=0$), it follows that  
\begin{IEEEeqnarray} {rcl}
\label{Eqdelta3Rproofgap}
\delta_{3R_1} &\leqslant&\kappa_{7,1}(\rho')-\Big(a_{1,1}+a_{7,1}(0,0,0)+a_{5,2}(0,0)\Big). \qquad
\end{IEEEeqnarray} 

The sum $a_{1,1}+a_{7,1}(0,0,0)+a_{5,2}(0,0)$ can be lower bounded as follows:
\begin{IEEEeqnarray}{lcl}
\nonumber
a_{1,1}+a_{7,1}(0,0,0)+a_{5,2}(0,0) &=& \frac{1}{2}\log\left(\frac{\overrightarrow{\SNR}_{1}}{\INR_{21}}+2\right)+\frac{1}{2}\log\left(\overrightarrow{\SNR}_{1}+\INR_{12}+1\right)\\
\nonumber
& & +\frac{1}{2}\log\left(\frac{\overrightarrow{\SNR}_{2}}{\INR_{12}}+\INR_{21}+1\right)-\frac{3}{2} \\
\nonumber
&\geqslant& \frac{1}{2} \! \log \! \left( \! \frac{\overrightarrow{\SNR}_{1}}{\INR_{21}} \! + \! 2\right) \! + \! \frac{1}{2}\log\left(\overrightarrow{\SNR}_{1} \! + \! \INR_{12} \! + \! 1\right)\\
\label{Eqgap2v1v2}
& & +\frac{1}{2}\log\left(\INR_{21}+1\right)-\frac{3}{2}.
\end{IEEEeqnarray}
If the term $\kappa_{7,1}(\rho')$ is active in the converse region, this can be upper bounded by the sum $\kappa_{1,1}(\rho')+\kappa_{4}(\rho')$, which corresponds to the sum of the single rate and sum-rate outer bounds respectively, and this can be upper bounded as follows:
\begin{IEEEeqnarray}{rcl}
\nonumber
\kappa_{7,1}(\rho')  & \leqslant & \kappa_{1,1}(\rho')+\kappa_{4}(\rho') \\
\nonumber
& \leqslant & \frac{1}{2} \! \log \!  \left( \! \overrightarrow{\SNR}_{1} \! +  \! \INR_{12} \! +  \! 1\right) \! +  \! \frac{1}{2}  \! \log  \! \left( \! 2+  \! \frac{\overrightarrow{\SNR}_{1}}{\INR_{21}}  \! \right)+\frac{1}{2}\log \left(\overrightarrow{\SNR}_{2}+\INR_{21}+1\right)+1 \\
\nonumber
& \leqslant & \frac{1}{2} \! \log \!  \left( \! \overrightarrow{\SNR}_{1} \! +  \! \INR_{12} \! +  \! 1\right) \! +  \! \frac{1}{2}  \! \log  \! \left( \! 2+  \! \frac{\overrightarrow{\SNR}_{1}}{\INR_{21}}  \! \right)+\frac{1}{2}\log \left(\INR_{21}+\INR_{21}+1\right)+1 \\
\nonumber
& \leqslant & \frac{1}{2} \! \log \!  \left( \! \overrightarrow{\SNR}_{1} \! +  \! \INR_{12} \! +  \! 1\right) \! +  \! \frac{1}{2}  \! \log  \! \left( \! 2+  \! \frac{\overrightarrow{\SNR}_{1}}{\INR_{21}}  \! \right)\\
\label{EqprofGapcHIRweightedsumratebb1}
& & +\frac{1}{2}\log \left(\INR_{21}+1\right)+\frac{3}{2}. 
\end{IEEEeqnarray}

From \eqref{Eqdelta3Rproofgap}, \eqref{Eqgap2v1v2} and  \eqref{EqprofGapcHIRweightedsumratebb1}, it follows that in Case $1.1$:
\begin{IEEEeqnarray} {rcl}
\nonumber
\delta_{3R_1} &\leqslant&  \frac{1}{2} \! \log \!  \left( \! \overrightarrow{\SNR}_{1} \! +  \! \INR_{12} \! +  \! 1\right) \! +  \! \frac{1}{2}  \! \log  \! \left( \! 2+  \! \frac{\overrightarrow{\SNR}_{1}}{\INR_{21}}  \! \right)+\frac{1}{2}\log \left(\INR_{21} \! + \! 1\right) \! + \! \frac{3}{2} \! \\
\nonumber
& & - \!  \frac{1}{2}\log\left(\frac{\overrightarrow{\SNR}_{1}}{\INR_{21}} \! + \! 2\right) -\frac{1}{2} \! \log \! \left(\overrightarrow{\SNR}_{1} \! + \! \INR_{12} \! + \! 1\right) \! - \! \frac{1}{2} \! \log \! \left(\INR_{21} \! + \! 1\right)+\frac{3}{2} \\
\label{Eqgapweightedsumrate}
&=& 3. 
\end{IEEEeqnarray} 

The same procedure holds in the calculation of $\delta_{3R_2}$ and it yields:
\begin{IEEEeqnarray} {rcl}
\label{Eqgapweightedsumrate2}
\delta_{3R_2} &\leqslant& 3. 
\end{IEEEeqnarray} 

Therefore, in Case $1.1$, from \eqref{Eqdelta}, \eqref{EqGapR1}, \eqref{EqGapR2}, \eqref{EqGapR1R2}, \eqref{Eqgapweightedsumrate} and \eqref{Eqgapweightedsumrate2} it follows that
\begin{IEEEeqnarray} {lcl}
\label{EqdeltaHIRfinal}
\delta&=&\max\left(\delta_{R_1},\delta_{R_2},\frac{\delta_{2R}}{2},\frac{\delta_{3R_1}}{3}, \frac{\delta_{3R_2}}{3}\right) \leqslant \frac{3}{2}.
\end{IEEEeqnarray}      

\end{enumerate}

\noindent
This completes the calculation of the gap in Case$1.1$. Applying the same procedure to all the other cases listed above yields that $\delta \leqslant 4.4$ bits. 

\end{appendices}

\clearpage

\bibliographystyle{IEEEtran}
\bibliography{IT-GT}

\end{document}